\documentclass[12pt]{article}
\usepackage{pslatex,palatino,avant,graphicx,color}
\usepackage{colortbl}
\usepackage{fullpage}
\usepackage{mathtools}
\usepackage{natbib}
\usepackage{amsmath}
\usepackage{caption}
\usepackage{subcaption}
\usepackage{bm}
\usepackage{wrapfig}
\usepackage{enumerate}
\usepackage{rotating}
\usepackage{multirow}
\usepackage{tabularx}
\usepackage{tikz}
\usepackage{pgfplots}

\newcommand{\blind}{0}

\usepackage[titles,subfigure]{tocloft} % Alter the style of the Table of Contents

\usepackage{bbm}

\newcommand{\indicator}[1]{\mathbbm{1}\left( #1 \right) }

\newcommand{\hb}{\hat{b}}
\newcommand{\ha}{\hat{a}}
\newcommand{\htheta}{\hat{\theta}}
\newcommand{\halpha}{\hat{\alpha}}
\newcommand{\hmu}{\hat{\mu}}
\newcommand{\hsigma}{\hat{\sigma}}

\newcommand{\htau}{\hat{\tau}}

\newcommand{\E}[1]{\mbox{E}\left[#1\right]}
\newcommand{\Var}[1]{\mbox{Var}\left[#1\right]}

\newcommand{\dNormal}[3]{ N\left( #1 \left| #2, #3 \right. \right) }

\newcommand{\expo}[1]{ \exp\left\{ #1 \right\}}
\newcommand{\tauSquareDelta}{\htau^2
  \left(\frac{1-\expo{-2\htheta\Delta}}{2\htheta} \right)}

\newcommand{\InvGam}[2]{\mbox{Inv-Gamma}\left( #1, #2 \right) }

\newbox{\legendLineOne}
\savebox{\legendLineOne}{
	(\mbox{\protect\includegraphics[scale=1]{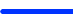}})
}

\newbox{\legendLineTwo}
\savebox{\legendLineTwo}{
	(\mbox{\protect\includegraphics[scale=1]{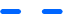}})
}

\newbox{\legendLineThree}
\savebox{\legendLineThree}{
	(\mbox{\protect\includegraphics[scale=1]{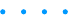}})
}

\newbox{\legendLineFour}
\savebox{\legendLineFour}{
	(\mbox{\protect\includegraphics[scale=1]{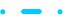}})
}

\newbox{\legendLineFive}
\savebox{\legendLineFive}{
	(\mbox{\protect\includegraphics[scale=1]{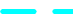}})
}

\newbox{\legendBlueTriangle}
\savebox{\legendBlueTriangle}{
	(\mbox{\protect\includegraphics[scale=1]{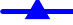}})
}

\newbox{\legendBlackRectangle}
\savebox{\legendBlackRectangle}{
	(\mbox{\protect\includegraphics[scale=1]{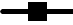}})
}

\begin{document}

\def\spacingset#1{\renewcommand{\baselinestretch}%
{#1}\small\normalsize} \spacingset{1}

\if0\blind
{
  \title{\bf Bayesian stochastic volatility models for high-frequency data}
  \author{Georgi Dinolov, Abel Rodriguez, and Hongyun Wang\\
    {\small Department of Applied Mathematics and Statistics, University of California, Santa Cruz, CA 95064, USA}}
  \maketitle
} \fi

\if1\blind
{
  \bigskip
  \bigskip
  \bigskip
  \begin{center}
    {\LARGE\bf Bayesian stochastic volatility models for high-frequency data}
\end{center}
  \medskip
} \fi

\bigskip

\abstract{We formulate a discrete-time Bayesian stochastic volatility model for high-frequency stock-market data that directly accounts for microstructure noise, and outline a Markov chain Monte Carlo algorithm for parameter estimation. The methods described in this paper are designed to be coherent across all sampling timescales, with the goal of estimating the latent log-volatility signal from data collected at arbitrarily short sampling periods. In keeping with this goal, we carefully develop a method for eliciting priors. The empirical results derived from both simulated and real data show that directly accounting for microstructure in a state-space formulation allows for well-calibrated estimates of the log-volatility process driving prices.}

\vspace{1cm}

\spacingset{1.00} % DON'T change the spacing! was set to 1.45
\section{Introduction}\label{se:introduction}

Estimating asset price volatilities is a common problem in finance; for example, accurate estimates of volatility paths play a key role in both option pricing and portfolio design.  Traditionally, financial models have used low-frequency returns (e.g., daily, weekly or monthly returns) to investigate price volatility.  Early attempts at incorporating higher-frequency information focused on using intra-period maximum and minimum prices (e.g., see \citealp{alizadeh2002range}, \citealp{brandt2003no-arb}, and \citealp{chou2010range}).  However, as high-frequency price data has become widely available, interest has turned to using all intra-period prices to generate high-resolution estimates of the volatility path and to improve estimates of the integrated volatility over higher frequencies.

A popular approach to estimating integrated volatilities from high-frequency price data is the realized variance estimator \citep{comte1998long,andersen2001distribution,barndorff2002estimating}.  The realized variance is defined as the sum of squared high-frequency log returns over the period of interest.  Under strict stationarity and some other weak regularity conditions for the volatility process, the realized variance converges in probability to the integrated variance of the true diffusion process as the sampling frequency increases.  An alternative to the realized variance estimator is to use standard parametric models such as the Generalized Autoregressive Conditionally Heteroscedastic (GARCH) model \citep{bollerslev1986,andersen1997intraday}.  However, the assumptions behind the classical GARCH model are not robust with respect to the specification of the sampling interval, and therefore the model is not invariant to temporal aggregation \citep{drost1993aggregation,andersen1997intraday,zumbach2000pitfalls}.  To address this issue some authors have turned to estimating low-frequency GARCH and stochastic volatility models using relevant summaries of the high-frequency prices.  For example, \cite{takahashi2009estimating} and \cite{shirota2014realized} use both high-frequency returns as well as the realized variance to estimate stochastic volatility models, while \cite{hansen2012realized} does the same for GARCH models.  Similarly, \cite{maneesoonthorn2014inference} use the realized volatility and the bipower variation estimators to estimate stochastic volatility models with jumps, while \cite{bollerslev2002estimating} use high order powers of the realized volatility as approximations to higher orders of integrated volatility.

A key challenge in working with high frequency prices is that they are often contaminated with microstructure noise. Indeed, as the sampling period shrinks down to the transaction-by-transaction frequency, irregular spacing between transactions, discreteness in transaction prices, and very short term temporal dependence become dominant features of the data \citep{stoll2000presidential}.  One consequence of the presence of microstructure noise is that the realized variance becomes a biased and inconsistent estimator of the true integrated variance \citep{zhou1996high}.  Possible solutions to this issue have been proposed by \cite{zhang2005tale}, who suggest sampling data sparsely at an optimally determined frequency and then averaging across the possible grids over the data, \cite{ait2011ultra}, who propose combining estimators based on subsampling data at different frequencies, and \cite{hansen2006realized} and \cite{barndorff2008designing}, who employ a class of kernel-based methods similar to those used for estimating the long-run variance of a stationary time-series in the presence of autocorrelation. In the context of model-based approaches it is common to assume that the summaries of the high-frequency returns used to estimate the model are noisy versions of the true realized volatilities (e.g., see \citealp{venter2012extended, shirota2014realized}).

This paper describes a Bayesian stochastic volatility model for high-frequency data that explicitly accounts for the presence of microstructure noise.  Unlike other approaches in the literature, we estimate our model directly using the high-frequency price data rather than summaries of the high-frequency returns.  To account for the effect of microstructure noise we introduce a hierarchical specification in which the observed high-frequency prices are noisy versions of the true unknown prices.  One appealing feature of our proposed model is that it is (approximately) coherent across all sampling frequencies, which is in line with previous efforts to validate the application of discrete-time models for volatility in high-frequency settings \citep{andersen1999forecasting}.  Coherency is achieved by starting with a continuous-time model and then carefully discretizing the exact solution to the stochastic differential equations for the price and volatility processes, and by carefully eliciting prior distributions for the parameters of the continuous-time model.

The remainder of the paper is structured as follows:  Section \ref{se:model_formulation} describes the continuous- and discrete-time version of the model. Section \ref{se:prior_elicitation} details the priors used and the method through which they were derived. Section \ref{se:computation} outlines the Bayesian Markov chain Monte Carlo (MCMC) algorithm used to fit the model. Section \ref{effect-mean-reverting-rate} examines the effect of certain model parameters on the posterior variance of the mean volatility level in our model. Finally, Section \ref{simulation-results} includes simulation results demonstrating the robustness of our inferential procedure to microstructure noise.

\section{Model Formulation}\label{se:model_formulation}

We begin with the continuous-time stochastic volatility model of \cite{hull1987pricing}, where the price $\hat{S}_t$ of an asset follows a Geometric Brownian motion and the time-varying log-volatility process $\log(\hat{\sigma}_t)$ follows a mean-reverting Ornstein-Uhlenbeck process,
\begin{align}
  d\log(\hat{S}_t) &= \hat{\mu}\, dt + \hat{\sigma}_t\, \sqrt{dt} \hat{\epsilon}_{t,1}  ,  \label{eq:price_evo} \\
  d\log( \hat{ \sigma }_t) &= -\hat{\theta} ( \log(\hat{\sigma}_t ) - \hat{\alpha} )\, dt + \hat{\tau}\, \sqrt{dt} \hat{\epsilon}_{t,2}  ,  \label{eq:vol_evo}
\end{align}
where $\hat{\epsilon}_{t,1}$ and $\hat{\epsilon}_{t,2}$ are dependent Weiner processes with instantaneous correlation $\rho$.  This model not only allows for the volatility to evolve over time, but also captures leverage effects though the correlation between $\hat{\epsilon}_{t,1}$ and $\hat{\epsilon}_{t,2}$ (e.g., see \citealp{black1976pricing}).

To generate a discretization of the model in \eqref{eq:price_evo} and \eqref{eq:vol_evo}, consider the (exact) solution of the Ornstein-Uhlenbeck process governing the evolution of the log-volatility in \eqref{eq:vol_evo},
\begin{align}\label{eq:sol_OU}
  \log(\hat{\sigma}_t) \sim N\left( \hat{\alpha} + \exp\left\{
      -\hat{\theta} t \right\} \left\{ \log(\hat{\sigma}_0) -
      \hat{\alpha} \right\}, \frac{\hat{\tau}^2}{2\hat{\theta}}\left\{
      1- \exp(-2\hat{\theta}t) \right\} \right)    ,
\end{align}
with stationary distribution
\begin{equation} \label{eq:stat-dist}
  \log(\hat{\sigma}_t) \sim N \left( \hat{\alpha} ,
    \frac{\hat{\tau}^2}{2\hat{\theta}} \right).
\end{equation}
For an arbitrary time interval $\Delta$ (which, for the purpose of this paper, we measure in milliseconds) we can use \eqref{eq:sol_OU} to generate the finite-difference equations
\begin{align*}
  \log(S_{j}) &= \log(S_{j-1}) + \mu(\Delta) + \sigma_{j} \, \epsilon_{j,1}   ,    \\
  \log(\sigma_{j+1}) &= \alpha(\Delta) + \theta(\Delta) \left\{ \log(\sigma_{j}) - \alpha(\Delta) \right\} + \tau(\Delta) \, \epsilon_{j,2}    ,
\end{align*}
where $j = 0, 1, \ldots, \left\lfloor T/\Delta \right\rfloor$ and 
\begin{align}
  \sigma_{j+1} &= \hat{\sigma}_{(j+1)\Delta}\sqrt{\Delta}, & S_j &= \hat{S}_{j\Delta},
\end{align}
\begin{align} \alpha(\Delta) &= \halpha + \frac{1}{2}\log(\Delta) , &
  \mu(\Delta) &= \hat{\mu} \Delta , & \theta(\Delta) &= \exp\left\{
    -\hat{\theta} \Delta \right\} , & \tau(\Delta) &= \hat{\tau}
  \sqrt{ \frac{1 - \exp \left\{ -2\hat{\theta} \Delta
      \right\}}{2\hat{\theta} } }, \label{eq:mu_sigma_tau}
\end{align}
and
\begin{align*}
  \left( \begin{matrix} \epsilon_{j,1} \\ 
      \epsilon_{j,2} \end{matrix} \right) &\sim
                                            N \left( \left(\begin{matrix} 0 \\
	                                          0 \end{matrix}
                                              \right) ,
  \left( \begin{matrix} 1 & \rho \\
      \rho & 1 \end{matrix} \right) \right) .
\end{align*}
We write $\alpha(\Delta)$, $\mu(\Delta)$, $\theta(\Delta)$, and $\tau(\Delta)$ to emphasize that we have a different set of parameters depending on the choice of $\Delta$.
%However, in the sequel we simplify notation by dropping the explicit dependence on $\Delta$.

Using the exact solution in \eqref{eq:sol_OU} to derive the finite difference equations allows us to take any step size $\Delta$ irrespective of the relative magnitude of the continuous-time model parameters. Indeed, the more standard forward-Euler discretization provides a poor approximation to the continuous-time model when $\Delta > 1/\hat{\theta}$, the timescale of inertia of the log-volatility process. Moreover, the inverse of the transformations in \eqref{eq:mu_sigma_tau} make it possible to meaningfully compare parameters inferred from different sampling frequencies and thereby check the coherency of our inferential procedure across different timescales.

In order to account for the effect of microstructure noise, we extend the previous model by differentiating between the \textit{true} log asset price $\log(S_j)$ and the \textit{discretely observed} log price $Y_j = \log(P_j)$. We treat these observed log prices as a noise-contaminated version of the true log price which is observed discretely only $n(\Delta) = \left\lfloor T/\Delta \right\rfloor$ times and whose index $j$ corresponds to $j\Delta$ in continuous-time.  More specifically, we let
\begin{align}\label{eq:microstructure_eq}
  Y_j &= \log (S_j) + \zeta_j,
\end{align}
where $\zeta_1, \zeta_2, \ldots$ are independent and identically distributed errors with mean zero and standard deviation $\xi$.  To motivate \eqref{eq:microstructure_eq}, consider one possible source of microstructure noise, the bid-ask spread. We can think of the idealized, ``true'' equilibrium price as evolving continuously in time by being moved by market supply and demand. Real-time order arrival and market friction makes it so that transaction prices are recorded at the highest bid or lowest ask levels only, thereby bounding the equilibrium market price in the bid-ask range. In this case it is natural to assume that $P_j = S_j + \nu_j$, where $\nu_j \sim U[ - D_p/2, D_p/2]$ and $D_p$ is the size of the bid-ask spread.  Using a first-order Taylor approximation then leads to $Y_j = \log(P_j) \approx \log(S_j) + \zeta_j$, where $\zeta_j = \frac{1}{S_j}\nu_j$.  A similar argument can be used to account for the effect of price discretization.

More generally, in order to account for the approximation error as well as for other sources of microstructure noise, we let $\zeta_t \sim N (0, \xi^2)$ where $\xi \approx \frac{D}{2Q}$, $D = \max\{ D_p, D_s\}$, $D_p$ represents a rough estimate of the bid-ask spread over the period of interest, $D_s$ represents another possible source of microstructure noise (such as price truncation to the nearest cent), and $Q$ is a rough guess of the average price of the asset over the period of interest.  Note that the distribution of $\xi$ is independent of the time scale $\Delta$ used for the discretization of the continuous-time process, and therefore independent of the frequency at which prices are observed.

To summarize, our hierarchical discrete-time stochastic volatility model reduces to  
\begin{align}
  Y_j &= \log(S_j) + \zeta_j  ,   \label{eq:mod1}   \\
  \log(S_{j}) &= \mu(\Delta) + \log(S_{j-1}) + \sigma_{j} \epsilon_{j,1}  ,  \label{eq:mod2}   \\
  \log(\sigma_{j+1}) &= \alpha(\Delta) + \theta(\Delta) \left\{ \log(\sigma_j) - \alpha(\Delta) \right\} + \tau(\Delta) \epsilon_{j,2}  ,   \label{eq:mod3}   
\end{align}
where 
\begin{align*}
  \zeta_j &\sim N(0, \xi^2)  ,  &
                                  \left( \begin{matrix} \epsilon_{j,1} \\
                                      \epsilon_{j,2} \end{matrix} \right) &\sim 
                                                                            N \left( \left( \begin{matrix} 0 \\
                                                                                  0 \end{matrix} \right),  
  \left( \begin{matrix} 1 & \rho \\
      \rho & 1 \end{matrix} \right) \right)  ,
\end{align*}
and initial conditions
\begin{align*}
  \log(\sigma_0) & \sim N \left( \alpha , \frac{\tau(\Delta)^2}{1 - \theta(\Delta)^2} \right)  ,  &   \log(S_0) & \sim N \left( \eta, \kappa^2 \right)  .
\end{align*}
Table \ref{ta:parameters} summarizes our notation:
\begin{table}[h!]
\begin{center}
%\begin{tabular}{c|c|l}
\begin{tabular}{c|c|p{10cm}}
&  Symbol   &   Interpretation  \\ \hline \hline
\multirow{2}{*}{\begin{sideways} Data \end{sideways}} &  $P_j$   &   Observed asset price at time $j\Delta$.  \\
&  $Y_j$   &   Logarithm of the observed asset price at time $j\Delta$.  \\ \hline
\multirow{10}{*}{\begin{sideways} Parameters \end{sideways}} &  $S_j$   &   True asset price at time $j\Delta$.  \\
&  $\sigma_j$   &   Discrete-time approximation of the volatility of the true asset price at time $j\Delta$.  \\
%&  $\gamma_j$  & Mixture indicator at time $j \Delta $. In MCMC sampling computation, $\log\left(\varepsilon_{j,1}^2\right)/2$ is approximated using a mixture of Gaussians (see Section \ref{se:computation}). \\
&  $\alpha(\Delta)$   &   Discrete-time approximation of the stationary mean of the log volatility.  \\
&  $\mu(\Delta)$   &  Discrete-time approximation to the mean asset return.  \\
&  $\theta(\Delta)$   &  Discrete-time approximation to the autocorrelation associated with the log volatility.  \\
&  $\tau(\Delta)$   &  Discrete-time approximation to the volatility of volatility.  \\
&  $\rho$   &  Correlation coefficient between volatility and price innovations.  \\
&  $\xi$     &   Standard deviation associated with the microstructure noise.  \\ \hline
\multirow{4}{*}{\begin{sideways} Hyper \end{sideways}} &  $\Delta$   &  Time step between observations (Fixed).  \\
&  $\eta$   &  Mean for the true asset price at time 0. (Fixed; no inference performed)  \\
&  $\kappa$   &  Standard deviation for the true asset at time 0. (Fixed; no inference performed) \\
\end{tabular}
\caption{Notational summary for our high-frequency stochastic volatility model, including data, parameters and hyperparameters.}\label{ta:parameters}
\end{center}
\end{table}

\section{Prior Elicitation}\label{se:prior_elicitation}

We approach the problem of estimation and prediction for the model described above using Bayesian methods. This requires that we elicit priors for the unknown parameters $\rho$, $\xi^2$, $\alpha(\Delta)$, $\mu(\Delta)$, $\theta(\Delta)$ and $\tau(\Delta)$.  Eliciting a prior for the correlation parameter $\rho$ and the microstructure variance $\xi^2$ is relatively straightforward since their value and interpretation are independent of the time step $\Delta$.  On the other hand, ensuring that the priors for $\alpha(\Delta)$, $\mu(\Delta)$, $\theta(\Delta)$ and $\tau(\Delta)$ are coherent across scales, i.e., that the priors provide the same information no matter what the time step $\Delta$ is, is non trivial.  To address this problem we proceed to elicit priors on the continuous-time parameters $\hat{\alpha}$, $\hat{\mu}$, $\hat{\theta}$, $\hat{\tau}$ and then use the formulas in \eqref{eq:mu_sigma_tau} to obtain the implied priors on $\alpha(\Delta)$, $\mu(\Delta)$, $\theta(\Delta)$ and $\tau(\Delta)$ for any time step $\Delta$.  Ideally, such priors would be invariant to the transformations in \eqref{eq:mu_sigma_tau}.  However, fully invariant priors are difficult to elicit and would, in any case, lead to computationally complicated models even using simulation-based methods such as Markov chain Monte Carlo algorithms.  Hence, we settle for the more modest goal of assigning priors that belong to families that are conditionally conjugate and therefore lead to computationally tractable models, but whose first two moments are (approximately) coherent across scales.
\begin{enumerate}[]

%%%%%%%%%%%%%%%%%%%%%%%

\item{\textbf{Prior for } $\boldsymbol{\rho}$:} We assign $(\rho + 1)/2$ a symmetric beta distribution with mean $1/2$ and precision $c$.  This prior ensures that $\rho \in [-1,1]$ as required and implies that $E(\rho) = 0$ a priori.  Furthermore, for large values of $c$, this means that we believe a priori that the  leverage effect is relatively small.

%%%%%%%%%%%%%%%%%%%%%%%

\item{\textbf{Prior for } $\boldsymbol{\mu}(\boldsymbol{\Delta})$:} For the mean of the asset returns a prior in the normal family leads to a simple full conditional distribution for MCMC sampling.  If we let $\E{ \hat{\mu} } = \hat{a}_{\hat{\mu}}$ and $\Var{ \hat{\mu} } = \hat{b}^2_{\hat{\mu}}$, then $\mu(\Delta) = \hat{\mu} \Delta$ leads to $\E{ \mu(\Delta) } = \Delta \hat{a}_{\hat{\mu}}$ and $\Var{ \mu(\Delta) } = \Delta^2 \hat{b}^2_{\hat{\mu}}$.  Hence, in our analysis we use the prior
$$
\mu(\Delta) \sim N(\Delta \hat{a}_{\hat{\mu}}, \Delta^2 \hat{b}^2_{\hat{\mu}})
$$
where values of $\hat{a}_{\hat{\mu}}$ and $\hat{b}^2_{\hat{\mu}}$ are elicited from historical data.

%%%%%%%%%%%%%%%%%%%%%%%

\item{\textbf{Prior for } $\boldsymbol{\theta}(\boldsymbol{\Delta})$:} The discrete-time autocorrelation coefficient $\theta(\Delta)$ of the volatility process is bounded above by 1 and below by 0 such that $\log(\sigma_j)$ is bounded as $j \to \infty$.  Hence, we employ a truncated normal prior for $\theta(\Delta)$,
$$
p(\theta(\Delta)) \propto N\left(a_{\theta}(\Delta), b^2_{\theta}(\Delta) \right) \indicator{\theta(\Delta) \in [0,1]},
$$
which leads again to a tractable computational algorithm. Note that because of the truncation,
\begin{align}
\E{\theta(\Delta)} &= a_{\theta}(\Delta) + \frac{ \phi\left( -\frac{a_{\theta}(\Delta)}{b_{\theta}(\Delta)} \right) - \phi\left(\frac{1- a_{\theta}(\Delta)}{b_{\theta}(\Delta)} \right)}{\Phi\left( \frac{1- a_{\theta}(\Delta) }{b_{\theta}(\Delta)} \right) - \Phi\left( -\frac{a_{\theta}(\Delta)}{b_{\theta}(\Delta)} \right)}  b_{\theta}(\Delta)  \label{eq:meantheta1} \\
%%%%%%
\Var{\theta(\Delta)} &= b^2_{\theta}(\Delta)  \left[ 1 + \frac{ -\frac{a_{\theta}(\Delta)}{b_{\theta}(\Delta)}\phi\left( -\frac{a_{\theta}(\Delta)}{b_{\theta}(\Delta)} \right) - \frac{1- a_{\theta}(\Delta)}{b_{\theta}(\Delta)}\phi\left(\frac{1- a_{\theta}(\Delta)}{b_{\theta}(\Delta)} \right)}{\Phi\left( \frac{1- a_{\theta}(\Delta) }{b_{\theta}(\Delta)} \right) - \Phi\left( -\frac{a_{\theta}(\Delta)}{b_{\theta}(\Delta)} \right)}   \right.  \nonumber \\ &  \;\;\;\;\;\;\;\;\;\;\;\;\;\;\;\;\;\;\;\;\;\;\;\;\;\;\;\;\;\;\;\;\;\;\;\;\;\;\;\;\;\;\;\;\;\;\;\;\;\;\;\;\;\;\;\;\;\;\;\;\;\;\;\;\;\;\;\;\;\;\;\;\;\;\;    \left. 
+\left\{ \frac{ \phi\left( -\frac{a_{\theta}(\Delta)}{b_{\theta}(\Delta)} \right) - \phi\left(\frac{1- a_{\theta}(\Delta)}{b_{\theta}(\Delta)} \right)}{\Phi\left( \frac{1- a_{\theta}(\Delta) }{b_{\theta}(\Delta)} \right) - \Phi\left( -\frac{a_{\theta}(\Delta)}{b_{\theta}(\Delta)} \right)} \right\}^2    \label{eq:vartheta1} 
\right]  
\end{align}
where $\phi(\cdot)$ and $\Phi(\cdot)$ denote the density and the cumulative distribution functions of the standard normal distribution.  Now, given the prior mean $\ha_{\htheta}$ and variance $\hb^2_{\htheta}$ for $\htheta$, we choose the values of $a_{\theta}(\Delta)$ and $b_{\theta}(\Delta)$ so that the mean and variance of $\theta(\Delta)$ above are approximately equal to the mean and variance of $\exp\left\{ -\htheta \Delta \right\}$. To simplify calculation of the moments of $\exp\left\{ -\htheta \Delta \right\}$ we use a second-order Taylor expansion of $\exp\left\{ -\hat{\theta} \Delta \right\}$ to approximate the first two moments of $\theta(\Delta)$ in terms of $\ha_{\htheta}$ and $\hb^2_{\theta}$, an approach known as the Delta-Method (e.g., see \citealp{casella2002statistical}):
\begin{align}
\E{\expo{-\htheta \Delta}} &\approx \exp \left(-\ha_{\htheta} \Delta \right) \left( 1 + \frac{1}{2} \hb_{\htheta}^2 \Delta^2 \right) \label{eq:e-theta-delta}, \\
\E{\expo{-2\htheta \Delta}} &\approx \exp \left(-2\ha_{\htheta} \Delta \right) \left( 1 + 2 \hb_{\htheta}^2 \Delta^2 \right) \label{eq:var-theta-delta}.
\end{align}
Using \eqref{eq:meantheta1}, \eqref{eq:vartheta1}, \eqref{eq:e-theta-delta}, and \eqref{eq:var-theta-delta}, and by setting $\E{\theta(\Delta)}  = \E{\expo{-\htheta \Delta}}$ and $\Var{\theta(\Delta)} = \Var{\expo{-\htheta \Delta}}$, we obtain a system of two equations with two unknowns that can be solved numerically to find the values of $a_\theta(\Delta)$ and $b^2_\theta(\Delta)$ in terms of $\ha_{\htheta}$, $\hb^2_{\htheta}$, and $\Delta$.

To elicit $\ha_{\htheta}$ and $\hb^2_{\htheta}$, recall that $\hat{\theta}$ is the inverse of the time scale of inertia for $\log(\hsigma_t)$ in the continuous-time formulation, which can be thought of as the characteristic time length, or unit of time, over which the process for the diffusion of $\log(\hsigma_t)$ ``forgets'' about an endogenous shock. The two hyper-parameters can be chosen so that the prior probability mass for $\htheta$ permits a reasonable range for the timescale of inertia.

%%%%%%%%%%%%%%%%%%%%%%%

\item{\textbf{Prior for } $\boldsymbol{\alpha}(\boldsymbol{\Delta})$:} For the mean log-volatility level $\halpha$, we once again use a computationally convenient prior in the normal family.  Letting $\E{\halpha} = \hat{a}_{\halpha}$ and $\Var{\halpha} = \hat{b}^2_{\halpha}$, and recalling that $\alpha(\Delta) = \hat{\alpha} + \frac{1}{2}\log(\Delta)$, we have
$$
\alpha(\Delta) \sim N\left(\hat{a}_{\halpha} + \frac{1}{2} \log(\Delta), \hat{b}^2_{\halpha}\right).
$$
To elicit the values of $\hat{a}_{\halpha}$ and $\hat{b}^2_{\halpha}$, recall that $\hat{\alpha}$ is the stationary (long-term) median of the volatility process.  Hence, for most assets these parameters could be elicited by looking at the time series of the asset's implied volatility (e.g., the VIX index if the asset is the S\&P500 index).

%%%%%%%%%%%%%%%%%%%%%%%

\item{\textbf{Prior for} $\boldsymbol{\tau^2}(\boldsymbol{\Delta})$:} We use a prior in the Inverse-Gamma family for $\tau^2(\Delta)$, so that
$$
\tau^2(\Delta) \sim \InvGam{ a_{\tau^2}(\Delta)}{b_{\tau^2}(\Delta) }. 
$$ 
To find the values of $a_{\tau^2}(\Delta)$ and $b_{\tau^2}(\Delta)$ recall that $\tau^2(\Delta) = \htau^2 \left( 1 - \exp\{ -2\htheta \Delta \} \right)/ \left( 2\htheta \right)$.  If we let $\E{\htau^2} = \hat{a}_{\hat{\tau}^2}$ and $\Var{\htau^2} = \hat{b}^2_{\hat{\tau}^2}$, and if we use the prior mean and variance of $\htheta$ as before, we can again apply the Delta-Method to approximate the prior first and second moments of $\tau^2(\Delta)$ by performing a second-order Taylor expansion of $\tau^2(\Delta)$ and $(\tau^2(\Delta))^2$ about the prior means of $\htau^2$ and $\htheta$, leading to

\begin{multline} \label{eq:htau-first-moment}
\E{\htau^2 \left( \frac{1 - \exp\{ -2\htheta \Delta  \}}{2\htheta} \right)} \approx  \left. \tauSquareDelta \right|_{\htau^2 =  \ha_{\htau^2}, \htheta = \ha_{\htheta}}   \\  
+  \hb^2_{\htau^2}  \left. \frac{\partial^2}{\partial^2 \htau^2} \left[ \tauSquareDelta \right] \right|_{\htau^2 = \ha_{\htau^2}, \htheta = \ha_{\htheta}}   \\
+  \hb^2_{\htheta}  \left. \frac{\partial^2}{\partial^2 \htheta} \left[ \tauSquareDelta \right] \right|_{\htau^2 = \ha_{\htau^2}, \htheta = \ha_{\htheta}}  
\end{multline}
and
\begin{multline}  \label{eq:htau-second-moment}
\E{\left\{\htau^2 \left( \frac{1 - \exp\{ -2\htheta \Delta \}}{2\htheta} \right) \right\}^2} \approx  \left. \left\{ \tauSquareDelta\right\}^2 \right|_{\htau^2 = \ha_{\htau^2}, \htheta = \ha_{\htheta}}   \\
+  \hb^2_{\htau^2}   \left. \frac{\partial^2}{\partial^2 \htau^2} \left\{ \tauSquareDelta \right\}^2 \right|_{\htau^2 = \ha_{\htau^2}, \htheta = \ha_{\htheta}}  \\
+  \hb^2_{\htheta}   \left. \frac{\partial^2}{\partial^2 \htheta} \left\{ \tauSquareDelta \right\}^2 \right|_{\htau^2 = \ha_{\htau^2}, \htheta = \ha_{\htheta}}    .
\end{multline}
The right sides of \eqref{eq:htau-first-moment} and \eqref{eq:htau-second-moment} are functions of $a_{\tau^2}(\Delta)$ and $b_{\tau^2}(\Delta)$, so that the above system of equations can be solved numerically to find $a_{\tau^2}(\Delta)$ and $b_{\tau^2}(\Delta)$ in terms of the other known prior hyperparameters. Finally, to elicit $\ha_{\htau^2}$ and $\hb_{\htau^2}$,we have to recall that the ratio $\frac{\htau^2}{2\htheta}$ represents the long-run variance of the log-volatility process $\log(\hat{\sigma})$. The prior for $\htau^2$ can therefore be elicited from market-traded approximations of the volatility process, such as the VIX.

%%%%%%%%%%%%%%%%%%%%%%%

\item{\textbf{Prior for} $\boldsymbol{\xi^2}$:}  For computational convenience, the variance of the microstructure noise is assigned an inverse Gamma with shape parameter $a_{\xi}$ and rate parameter $b_{\xi}$.  The mean of the prior can be elicited from information about the bid-ask spread, the tick size and the average price of the stock as discussed in Section \ref{se:model_formulation}, while its standard deviation is selected so that we stay within an order of magnitude (above and below) of the mean.
\end{enumerate}

\section{Computation}\label{se:computation}

The posterior distribution of high-frequency stochastic volatility is analytically intractable, so we perform parameter inference and prediction using a Markov chain Monte Carlo (MCMC) algorithm.  Our sampler extends the ideas introduced in \cite{omori2007stochastic}, which used a mixture of normals approximation to the distribution of a log $\chi^2$ distribution.  More specifically, our algorithm alternates between 1) sampling the true asset prices $S_0, S_1, \ldots, S_{n(\Delta)}$ from their joint full conditional distribution using a Forward-Backward algorithm \citep{carter1994gibbs,fruhwirth1994data}, 2) jointly sampling the mixture indicators $\gamma_1, \ldots, \gamma_{n(\Delta)}$ (to be introduced below) given all other parameters, 3) jointly sampling the volatilities $\sigma_1, \ldots, \sigma_{n(\Delta)}, \sigma_{n(\Delta) + 1}$ using a second Forward-Backward algorithm, and 4) sampling from each model parameter given all other parameters.

For Step 1 in our inferential procedure, note that, given the mean return $\mu(\Delta)$, the microstructure variance $\xi^2$ and the volatilities $\sigma_1, \ldots, \sigma_{n(\Delta)}, \sigma_{n(\Delta) + 1}$, equations \eqref{eq:mod1} and \eqref{eq:mod2} define a linear state-space model with state variable $x_j = \log (S_j)$ and Gaussian innovations.  Hence, using a Forward-Backward algorithm to sample the true asset prices is straightforward.  For Step 2, we note that
$$
\log\left\{ | \log(S_j / S_{j -1}) - \mu(\Delta) | \right\} = \log(\sigma_j) + \log( \epsilon^2_{j,1} )/2 .
$$
Following \cite{omori2007stochastic} we approximate the error term using a mixture of Gaussian distributions,
$$
\log( \epsilon^2_{j,1} )/2 \sim \sum_{l=1}^{10} p_l N \left( \frac{m_l}{2}, \frac{v_l^2}{4} \right)
$$
(see Table \ref{ta:mixture_parameters} for the values of $\{ p_l \}$, $\{ m_l \}$ and $\{ v_l \}$).  The mixture can be rewritten by introducing auxiliary indicators $\gamma_1, \ldots, \gamma_{n(\Delta)}$ such that
\begin{align*}
\log( \epsilon^2_{j,1} )/2 \mid \gamma_j &\sim N \left( \frac{m_{\gamma_j}}{2}, \frac{v_{\gamma_j}^2}{4} \right)   ,   &   \Pr(\gamma_k = l) = p_l .
\end{align*}
The auxiliary indicators are sampled jointly conditional on all other parameters, and $\Pr(\gamma_k = l) = p_l$ is interpreted as the prior probability that observation $k$ belongs to mixture element $l$.

Conditionally on the true prices, the indicators $\gamma_{1}, \ldots, \gamma_{n(\Delta)}$, and the hyperparameters $\mu(\Delta)$, $\alpha(\Delta)$, $\theta(\Delta)$, $\tau(\Delta)$ and $\rho$, we have again a linear state-space model with Gaussian innovations, so the volatilities can be sampled using another Forward-Backward algorithm for Step 3.  In Step 4, for the prior distributions discussed in Section \ref{se:prior_elicitation} the full conditional distributions for each of the parameters given the volatilities, prices, mixture indicators, and other hyperparameters follow standard distributions such as Gaussians, truncated Gaussians and inverse-Gammas.  Details of the algorithm are given in Appendix \ref{ap:mcmc}.
\begin{table}
\begin{center}
\begin{tabular}{c|ccc}
Component   &   $p_l$   &   $m_l$   &   $v^2_l$  \\ \hline
1 & 0.00609 & 1.92677 & 0.11265 \\
2 & 0.04775 & 1.34744 & 0.17788 \\
3 & 0.13057 & 0.73504 & 0.26768 \\
4 & 0.20674 & 0.02266 & 0.40611 \\
5 & 0.22715 & -0.85173 & 0.62699 \\
6 & 0.18842 &-1.97278 & 0.98583 \\
7 & 0.12047 & -3.46788  & 1.57469 \\
8 & 0.05591 & -5.55246 & 2.54498 \\
9 & 0.01575 & -8.68384 & 4.16591 \\
10 & 0.00115 & -14.65000 & 7.33342
\end{tabular}
\caption{Parameters of the mixture representation of the log Chi-squared distribution, provided in \cite{omori2007stochastic}.}\label{ta:mixture_parameters}
\end{center}
\end{table}

Once the algorithm has converged and the burn-in samples have been discarded, point and interval estimates can be easily obtained using empirical estimates.  For example, given a sample of the volatility path $\left( \sigma^{(b)}_{1}, \sigma^{(b)}_{2}, \ldots,  \sigma^{(b)}_{n(\Delta)} \right)$ for $b=1, \ldots, B$, a sample of the in-sample (approximate) integrated variance $IV = \int_0^T \hsigma^2_t dt$ can be obtained as
$$
IV^{(b)} \approx \sum_{j=1}^{n(\Delta)} \left( \sigma^{(b)}_{j} \right)^2   .
$$
A similar approach can be used to make out-of-sample predictions of the integrated volatility.

\section{The effect of the mean-reverting rate $\hat{\theta}$ and the observational duration on the posterior variance of the mean log-volatility $\hat{\alpha}$}\label{effect-mean-reverting-rate}

When estimating model parameters, the common intuition is that an increase in sample size leads to a decrease in posterior uncertainty. When dealing with the estimation of stochastic volatility models for high-frequency data, one may be prone to apply this thinking when the sample size is increased by obtaining move frequent price path samples for a fixed observational period. However, in the case where the volatility process has a finite non-zero mean-reversion timescale (as is the case for the Ornstein-Uhlenbeck process), an increase in the number of intraperiod observations does not add information about the mean-level of the process. Rather, the posterior uncertainty for this model parameter can only be decreased by increasing how long we observe the process.  To demonstrate this feature of the model, we study analytically the relationship between the mean-reverting rate $\hat{\theta}$, the time duration of observation $T$, and the posterior variance of mean log-volatility $\hat{\alpha}$.  To proceed analytically, we consider a simplified inference problem described by the following assumptions:  1) the mean log-volatility $\hat{\alpha}$ is the only parameter to be inferred -- all other parameters are known; 2) the prior distribution for the mean log-volatility $\hat{\alpha}$ is normal and is denoted by $N(\hat{a}_{\hat{\alpha}}, \hat{b}^2_{\hat{\alpha}}) $ (previously there was no parametric assumption made on the on the prior for $\halpha$); c) the log-volatility $\log( \hat{ \sigma }_t)$ is observed exactly (without error) on a uniform grid $\{0, \Delta, 2\Delta, \ldots , N\Delta \} $ in time duration $[0, T]$ where $\Delta $ is the sampling period and $N=T/\Delta $.

The exact solution of the Ornstein-Uhlenbeck process \eqref{eq:vol_evo} is given in \eqref{eq:sol_OU}.  Applying the exact solution \eqref{eq:sol_OU} to the time interval $[j\Delta, (j+1)\Delta]$, we obtain
$$
  \log( \hat{ \sigma }_{(j+1)\Delta}) = \theta(\Delta) \log( \hat{ \sigma}_{j\Delta}) + (1-\theta(\Delta)) \hat{\alpha} +  \tau(\Delta) \varepsilon_j \;\; , \hspace{0.5cm} 0\le j \le N-1
$$
where $\varepsilon_j \sim N(0, 1)$, and $\theta(\Delta)$ and $\tau(\Delta)$ are given in \eqref{eq:mu_sigma_tau}.  Recall the stationary distribution of the continuous-time log-volatility process in \eqref{eq:stat-dist}
$$
  \log(\hat{\sigma}_t) \sim N \left( \hat{\alpha} , \tau(\infty)^2 \right), \hspace*{1cm} \tau(\infty)^2 = \frac{\hat{\tau}^2}{2\hat{\theta}}.
$$
Here we denote the stationary variance as $\tau(\infty)^2$ for mathematical convenience.  The likelihood of $\hat{\alpha} $ given the observation $\{\log(\hat{\sigma}_0), \log(\hat{\sigma}_1), \ldots, \log(\hat{\sigma}_N)\}$ is
\begin{multline}\label{likelihood_alpha_n1}
L \left(\hat{\alpha} \left| \log(\hat{\sigma}_0), \log(\hat{\sigma}_1), \ldots, \log(\hat{\sigma}_N) \right. \right) \propto \exp\left( \frac{-(\log(\hat{\sigma}_0)-\hat{\alpha})^2}{2 \tau(\infty)^2} \right)  \\
 \times \prod_{j-0}^{N-1} \exp\left( \frac{-\left(\log\left(\hat{\sigma}_{(j+1)\Delta}\right) - \theta(\Delta) \log(\hat{\sigma}_{j\Delta})-(1-\theta(\Delta)) \hat{\alpha}\right)^2}{2 \tau(\Delta)^2} \right).
\end{multline}
Since we assume $N(\hat{a}_{\hat{\alpha}}, \hat{b}^2_{\hat{\alpha}})$ as the prior for $\hat{\alpha}$, the posterior distribution of $\hat{\alpha}$ is normal and the reciprocal of the posterior variance of $\hat{\alpha}$ has the expression
\begin{align}
  \frac{1}{\Var{\hat{\alpha}} } &= \frac{1}{\hat{b}^2_{\hat{\alpha}}}  + \frac{1}{\tau(\infty)^2} + \sum_{j=0}^{N-1} \frac{(1-\theta(\Delta))^2}{\tau(\Delta)^2}  %\nonumber \\
                                            = \frac{1}{\hat{b}^2_{\hat{\alpha}} }+ \frac{2\hat{\theta} }{\hat{\tau}^2} \left(1 + N \cdot \tanh\left(\frac{\hat{\theta} \Delta}{2} \right) \right).  \label{var_alpha_n1}
\end{align}
In the above, we have used the expressions of $\tau(\Delta )$ and $\theta(\Delta )$ given in \eqref{eq:mu_sigma_tau}.

Now, using the linear approximation $\tanh\left(\frac{x}{2} \right) \approx \frac{x}{2}$ and setting $T=N \Delta $, we can write \eqref{var_alpha_n1} as
\begin{equation}
\frac{1}{\Var{\hat{\alpha}} } \approx \frac{1}{\hat{b}^2_{\hat{\alpha}} } + \frac{2\hat{\theta} }{\hat{\tau}^2}  \left(1 + \frac{\hat{\theta} T}{2} \right).  \label{var_alpha_n2}
\end{equation}
This expression is valid for $\hat{\theta} \Delta \le 1$, i.e. when the timescale of inertia of the log-volatility process is \textit{greater} than the spacing between observations. The important consequence of \eqref{var_alpha_n2} is that decreasing $\Delta$ does not decrease the posterior variance of $\halpha$. In other words, an increase in the number of intraperiod observations does not add information about $\halpha$. Rather, the posterior uncertainty for $\halpha$ can only be decreased by increasing $T$ (increasing how long we observe the process) or increasing $\htheta$ (on average, increasing the number of reversions to the mean). The rate of information increase for $\halpha$ with respect to $T$ and $\htheta$ is examined under two conditions.

When $\hat{\tau}$ is fixed, $1/\Var{\hat{\alpha}} $ increases linearly with the time duration $T$ and increases quadratically with the mean-reverting rate $\hat{\theta} $.  The quadratic increase of $1/\Var{\hat{\alpha}} $ with respect to $\hat{\theta} $ is the combined result from two contributions: i) for larger $\hat{\theta} $, the variance of $\log(\hat{\sigma}_t) $ is smaller and consequently each data point is a more accurate approximation to $\hat{\alpha }$; and ii) for larger $\hat{\theta} $, the time duration $[0, T]$ covers more rounds of $\log(\hat{\sigma}_t) $ fluctuating away from $\hat{\alpha }$ and relaxing back toward $\hat{\alpha }$.

When $\frac{\hat{\tau}^2}{2\hat{\theta} } $ (the stationary variance of log-volatility) is fixed, $1/\Var{\hat{\alpha}} $ increases linearly with $\hat{\theta} T$. In this case, if the prior is wider than the stationary distribution ($\hat{b}^2_{\hat{\alpha}} \ge \frac{\hat{\tau}^2}{2\hat{\theta} } $) and the time duration is much larger than the time scale of inertia ($T \gg 1/\hat{\theta}$), then the posterior variance of $\hat{\alpha} $ is inversely proportional to the time duration:
\begin{equation}
\Var{\hat{\alpha}} \approx \frac{\hat{\tau}^2}{\hat{\theta} } \cdot \frac{1}{\hat{\theta} T}. \label{var_alpha_n3}
\end{equation}

When $\htheta \Delta \gg 1$ (i.e. when the timescale of inertia of the log-volatility process is much \textit{smaller} than the spacing between
observations), the linear approximation 
$\tanh(\htheta \Delta/2) \approx \htheta \Delta/2$ is invalid. Instead, we have
the approximation $\tanh(\htheta \Delta/2) \approx 1 $, which leads to 
$$ \frac{1}{\Var{\hat{\alpha}} } \approx \frac{1}{\hat{b}^2_{\hat{\alpha}} }+ 
\frac{2\hat{\theta} }{\hat{\tau}^2} (1 + N), \hspace*{1cm} \mbox{for }
\; \htheta \Delta \gg 1 $$
Under this regime, $1/\Var{\halpha}$ is approximately proportional to the number of observations, $N$, provided that $\htheta \Delta \gg 1$ is preserved as $N$ is increased.  This occurs when the spacing between observations, $\Delta $, is fixed and the increase in $N$ comes from extending the observational duration $T$.  When $T$ is fixed, as $N$ increases $\Delta $ decreases, which eventually will carry the system from the regime of $\htheta \Delta \gg 1$ to that of $\htheta \Delta \le 1$.  The behavior of the posterior variance of $\halpha$ for $\htheta \Delta \leq 1$ with $N$ increasing through either lowering $\Delta$ or increasing $T$ is illustrated in Section \ref{se:effect-timescale} below.

The results above, which indicate that $\halpha$ cannot be estimated consistently under in-fill asymptotics, apply to any stochastic volatility model based on the Ornstein-Uhlenbeck process.  Note, however, that they do not contradict standard asymptotic results from the realized volatility literature, which focus on the integrated variance during a finite period of time and not on the long-term median volatility of the process. 

\section{Illustrations}\label{simulation-results}

\subsection{Effect of microstructure noise and sampling frequency on estimates: simulation studies}

In this section we examine the effects of microstructure noise and sampling frequency on our inference of the model parameters. We first consider a one-day (6.5 hours) simulated dataset in which the true log-prices were generated according to \eqref{eq:mod2} - \eqref{eq:mod3} with $\Delta = 10^{-3}$ seconds, but where the microstructure noise was incorporated by adding to each point (exponentiated to transform from log-price to price level) a uniformly distributed random variable between $-0.05$ and $0.05$, simulating a \$0.1 bid-ask spread. The price was rounded to the nearest 100th, then transformed back to the log scale.

The true parameters used in the simulations were set to be reasonably close to typical values on the S\&P500 market. The instantaneous return per millisecond $\hmu$ was set to $\hmu = 1.7 \cdot 10^{-12}/\mbox{millisecond}$, corresponding to an annual return of 1\%, based on 251 trading days per year, 6.5 trading hours per trading day, excluding jumps between trading sessions. Assuming a characteristic timescale of intertia of the continuous log-volatility process to be 30 min (measured in milliseconds), $\htheta$ was set to $\htheta = 5.6 \cdot 10^{-7}$/millisecond. The remaining parameters $\htau^2$ and $\halpha$ governing the behaviour of the log-volatility process were set using the publicly traded VIX index. The VIX is the square root of the risk-neutral market expectation of the S\&P 500 variance over the next 30 days on an annualized scale, such that
$$
\log \hsigma_t \approx \log\left( \log\left(1 + \mbox{VIX}_t / 100 \right) \right) - \frac{1}{2} \log(T_{\rm year})
$$
where $T_{\rm year} = 1 \mbox{ year} = 251 \times 6.5 \times 3.6 \cdot 10^6 \mbox{ ms}$. We can transform between $\mbox{VIX}_t$ and $\log(\hsigma_t)$, obtaining approximations of the historical log-volatility path. With $\htheta$ set, the parameters $\halpha$ and $\htau^2$ determine the stationary distribution of the log-volatility process, $\log(\hsigma_t) \sim N\left(\halpha, \frac{\htau^2}{2\htheta}\right)$.  Thus we set $\halpha$ to be the mean of this VIX-derived log-volatility path, and $\htau^2$ the variance thereof, multiplied by $2\htheta$.  A summary of the true model parameters used in this simulation is presented in Table \ref{ta:simulation-parameters}.
\begin{table}[h!]
\begin{center}
\begin{tabular}{c!{\color{black}\vrule}c!{\color{black}\vrule}p{8.5cm}}  
  Parameter & Value & Interpretation \\ \hline 
  $\hmu_T$ & $1.7 \cdot 10^{-12}/\mbox{ms}$ & Annual asset return of 1\%, based on 251 trading days per year, 6.5 trading hours per trading day, excluding jumps between trading sessions. \\ \arrayrulecolor{gray} \hline
  $\htheta_T$ & $5.6 \cdot 10^{-7}/\mbox{ms}$ & Timescale of inertia for the log-volatility process equal to 30 min. \\ \arrayrulecolor{gray} \hline
  $\halpha_T$ & $-13-\log(\sqrt{\mbox{ms}})$ & Average of daily closing VIX values from 1/2/1990 to 4/10/2015 (18.9\% a year), transformed to the $\log(\hsigma_t)$ scale with time measured in milliseconds. \\ \arrayrulecolor{gray} \hline
  $\htau^2_T$ & $1.3 \cdot 10^{-7}/\mbox{ms}$ & $2\htheta$ times the variance of daily closing VIX values from 1/2/1990 to 4/10/2015, transformed to the $\log(\hsigma_t)$ scale with time measured in milliseconds. In terms of the annualized volatility for the price, the corresponding distribution of annualized volatility has (1st, 10th, 50th, 90th, 99th) percentiles given by $( 8.16 \%, 11.8\%, 18.9\%, 30.7\%, 46.6\%)$\\ \arrayrulecolor{gray} \hline
  $\xi^2_T$ & $2.5 \cdot 10^{-7}$ & Bid-ask spread of \$0.1 for an average price of \$100. \\ \arrayrulecolor{gray} \hline
  $\rho_T$ & 0 & Innovations in the price and log-volatility process are independent, no leverage effect.
\end{tabular}
\caption{Summary for model parameters in simulation data, along with associated market interpretation of these parameters.}\label{ta:simulation-parameters}
\end{center}
\end{table}

We fit our model to the simulated data using priors whose means equal the true simulation values and and whose standard deviations are roughly one order of magnitude higher than the prior mean (recall that the prior specification procedure described in Section \ref{se:prior_elicitation} only requires the first two moments of each continuous-time parameter). For example, since $\hmu_T = 1.7 \cdot 10^{-12}$, we let $\E{\hmu} = \ha_{\hmu} = 1.7 \cdot 10^{-12}$ and $\Var{\hmu} = \hb_{\hmu}^2 = (1 \cdot 10^{-11})^2$. In this case, the prior region covered by three standard deviations to the left and right of the prior mean approximately corresponds to a range from -15\% to a 20\% annual return.  In general, specifying the prior standard deviation in this manner usually leads to a relatively wide but reasonable prior coverage of parameter model values. A summary of the prior means and standard deviations used to fit the model is provided in Table \ref{ta:priors}.
\begin{table}[h]
\begin{center}
\begin{tabular}{c|c|c}
  Parameter & Prior mean & Prior standard deviation \\ \hline
  $\hmu$ & $1.7 \cdot 10^{-12}$ & $1 \cdot 10^{-11}$  \\ 
  $\htheta$ & $5.6 \cdot 10^{-7}$ & $1 \cdot 10^{-6}$  \\ 
  $\halpha$ & $-13$ & $10$  \\ 
  $\htau^2$ & $1.3\cdot 10^{-7}$ & $1\cdot 10^{-6}$  \\ 
  $\xi^2$ & $2.5 \cdot 10^{-7}$ & $1\cdot 10^{-6}$ \\ 
  $\rho$ & 0 & 1 
\end{tabular}
\caption{Parameters of the prior distributions used for inference with simulated data.}\label{ta:priors}
\end{center}
\end{table}

%{\color{red} We should note that we are using \textit{daily} observations of volatility (i.e. the VIX) over a long time scale (years) to center our priors for $\halpha$ and $\htau^2$ governing the log-volatility process that a priori is active on a much shorter time scale (minutes; the prior mean for $\htheta$ corresponds to a timescale of inertia of 30 min). We recognize the possible importance of this distinction, but modeling multiple volatility timescales is beyond the scope of this work. Further, as long as the priors are sufficiently diffuse, we hope that the VIX and other daily indicators normally used to measure daily volatility can provide a reasonable first-order approximation for model parameters at shorter timescales.  {\bf Moved this paragraph (which deals with the prior) after the prior is described).  However, not sure what the point of this paragraph is, I would remove it.}  }

We fit three slightly different versions of our model to the simulated dataset.  In the first version, the microstructure noise parameter was set to $\xi^2 = 0$, so our model does not take into account microstructure noise and reduces to the standard SV models used in the literature. In the second version, $\xi^2$ was fixed to $2.5\cdot 10^{-7}$, a level of microstructure noise roughly consistent with the true level of microstructure noise added in the data.  Finally, the third version corresponds to our full model where $\xi^2$ is estimated from the data by assigning it a Gamma prior with mean $2.5\cdot 10^{-7}$ and standard deviation $1\cdot 10^{-6}$ (see Table \ref{ta:priors}).  The dataset was analyzed assuming sampling periods of 300, 30, 15, and 5 seconds (note that, because the size of the microstructure noise is assumed to be the same at every sampling scale, we use the same prior for $\xi^2$ for all sampling periods).

\begin{figure}
	\centering
	\begin{tabular}{m{0.25cm}ccc}
		 & Inference with & Inference with & Inference with \\
		 & $\xi^2 = 0$ & $\xi^2 = 2.5 \cdot 10^{-7}$ & $\xi^2 \mbox{ estimated }$ \\
		\begin{sideways} $\Delta = 5$ min \end{sideways}
			& \begin{minipage}{0.25\textwidth}
				\centering
				\includegraphics[width=1\linewidth]{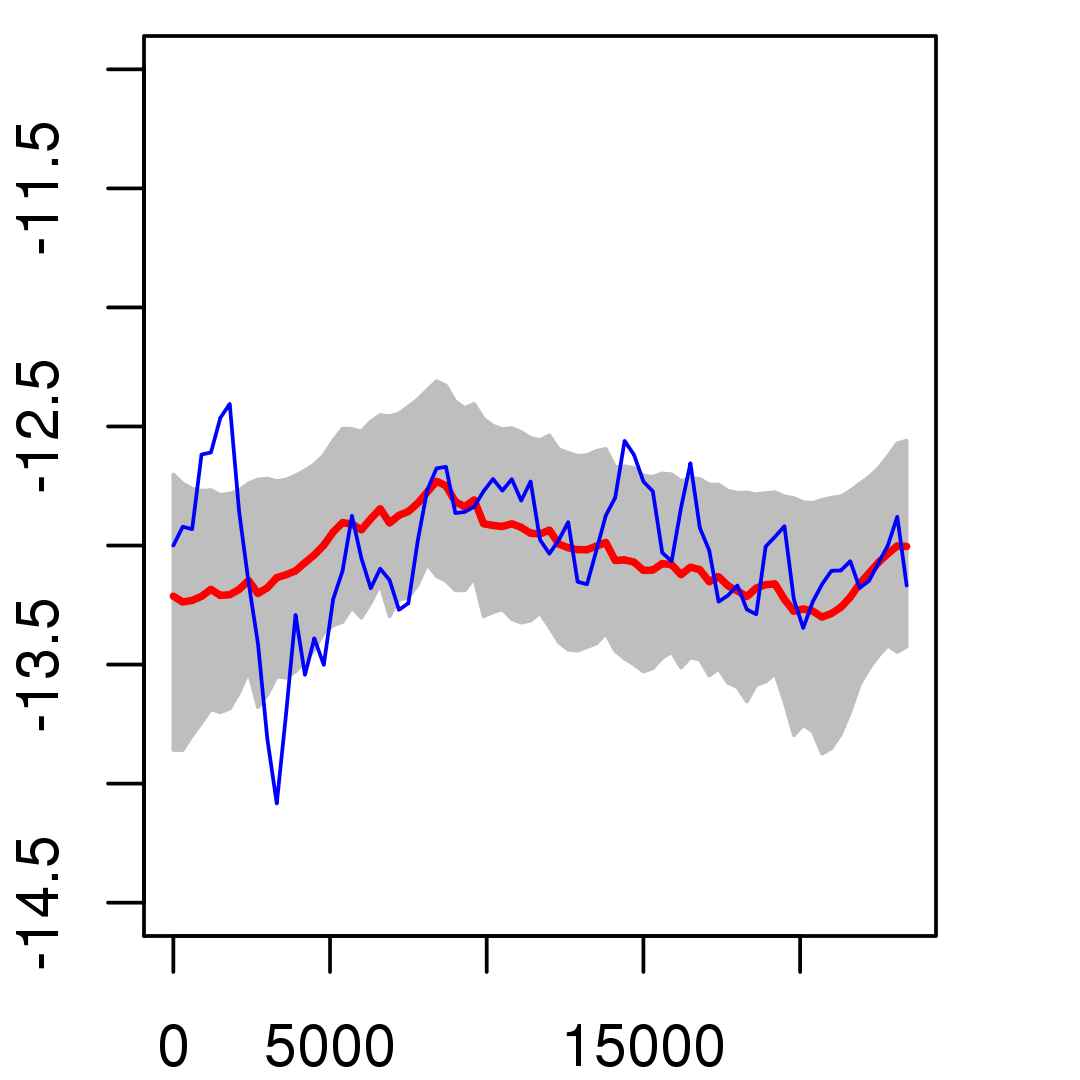}
				\end{minipage} 
			& \begin{minipage}{0.25\textwidth}
				\centering
				\includegraphics[width=1\linewidth]{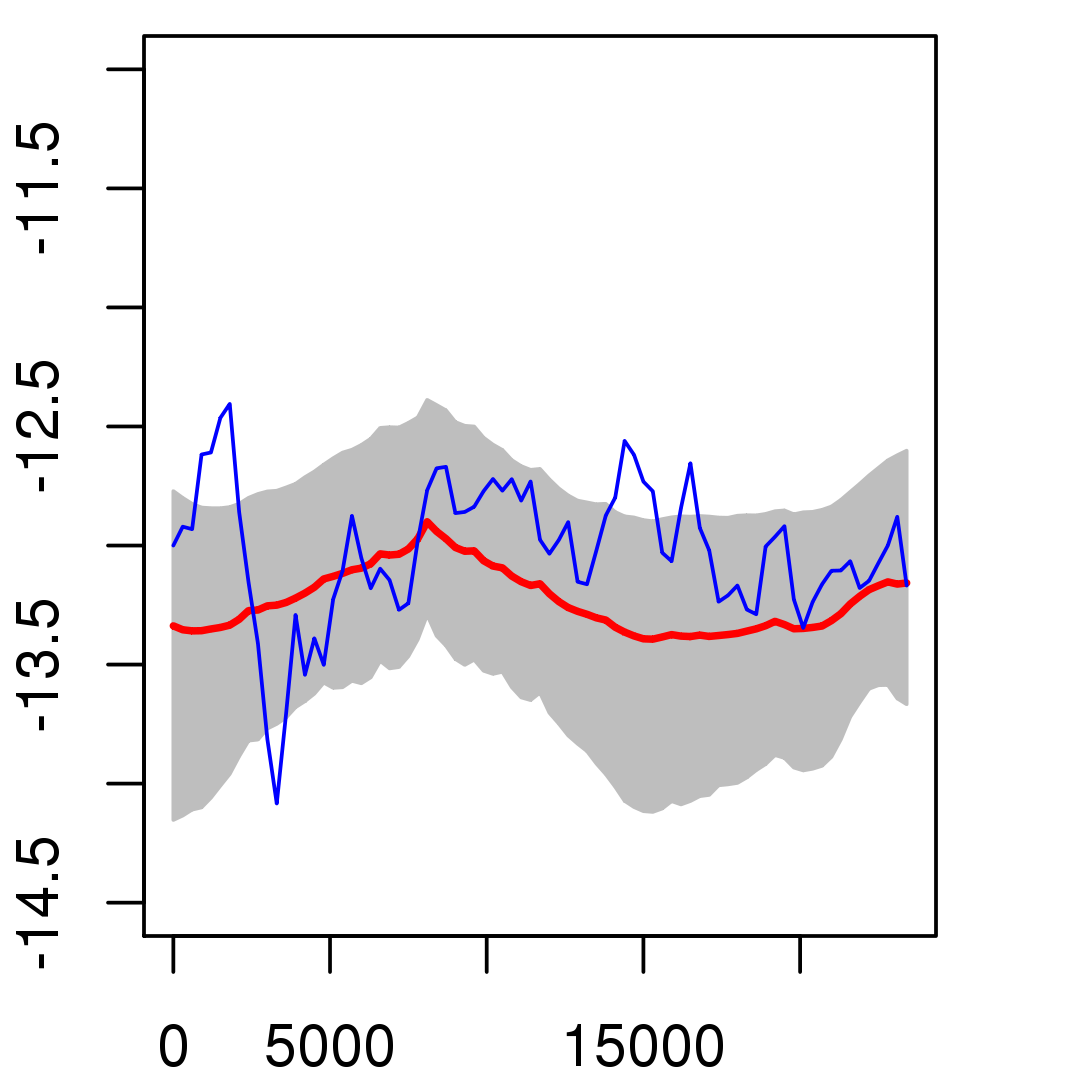}
				\end{minipage} 
			& \begin{minipage}{0.25\textwidth}
				\centering
				\includegraphics[width=1\linewidth]{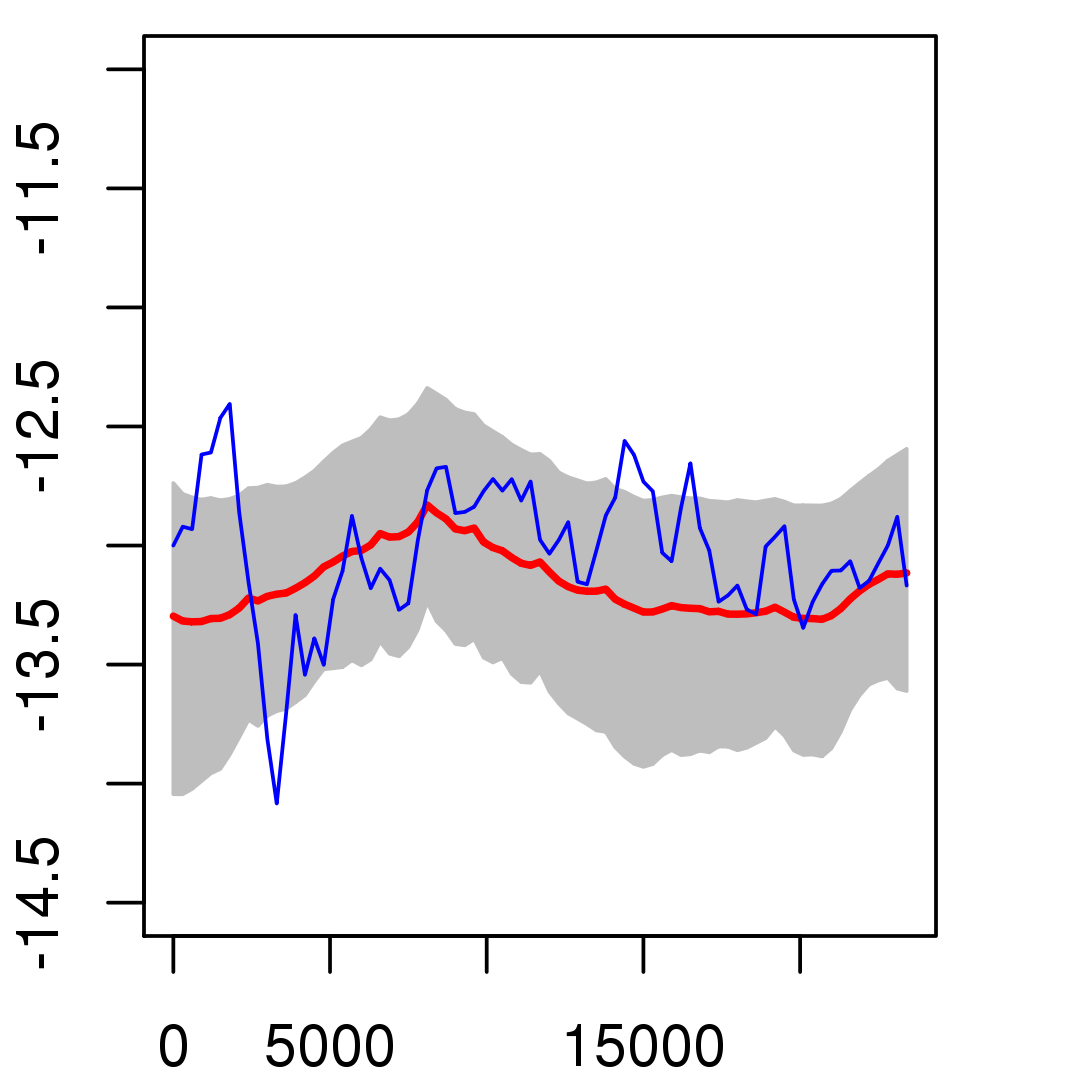}
				\end{minipage}  \\
		\begin{sideways} $\Delta = 15$ sec \end{sideways} 
			& \begin{minipage}{0.25\textwidth}
				\centering
				\includegraphics[width=1\linewidth]{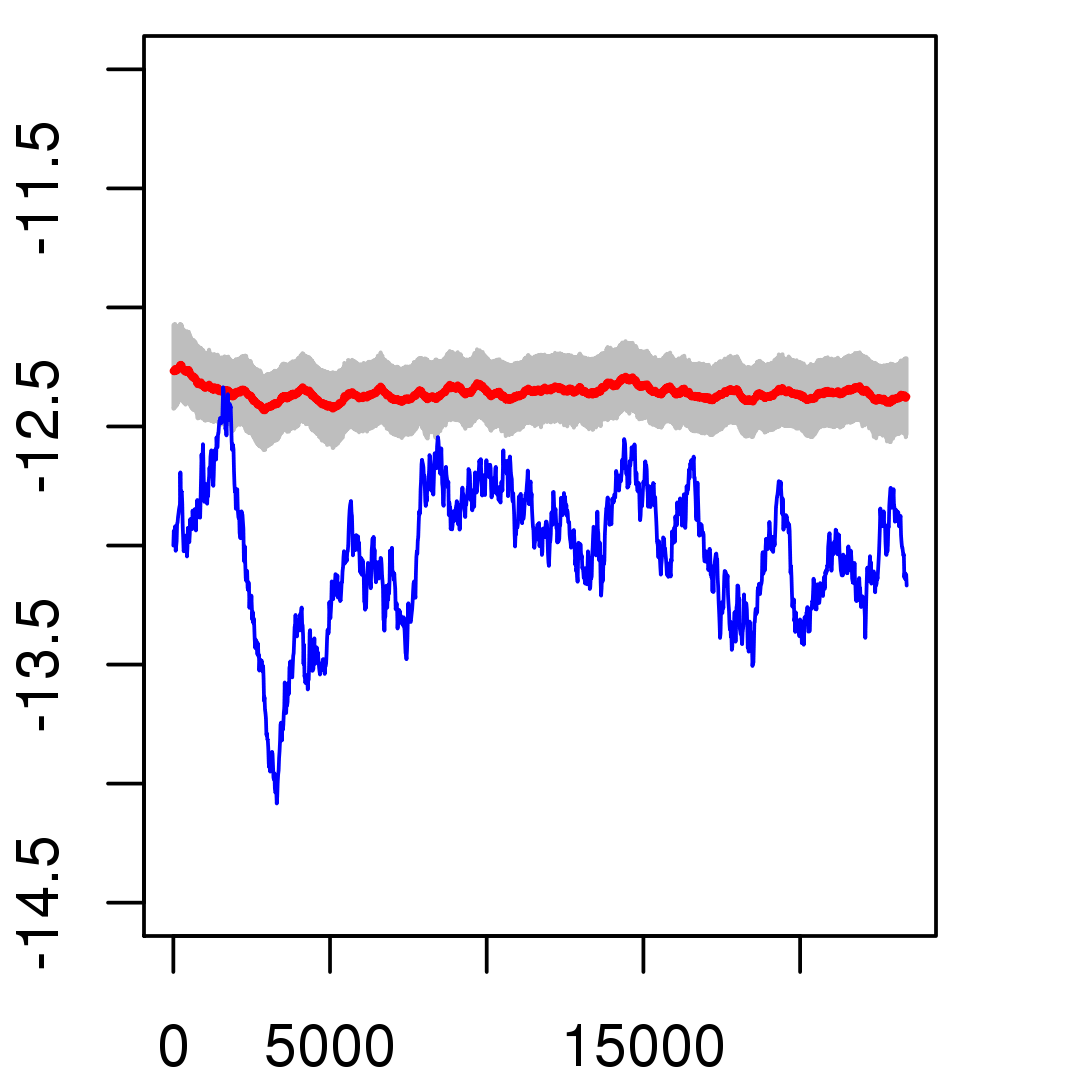}
				\end{minipage} 
			& \begin{minipage}{0.25\textwidth}
				\centering
				\includegraphics[width=1\linewidth]{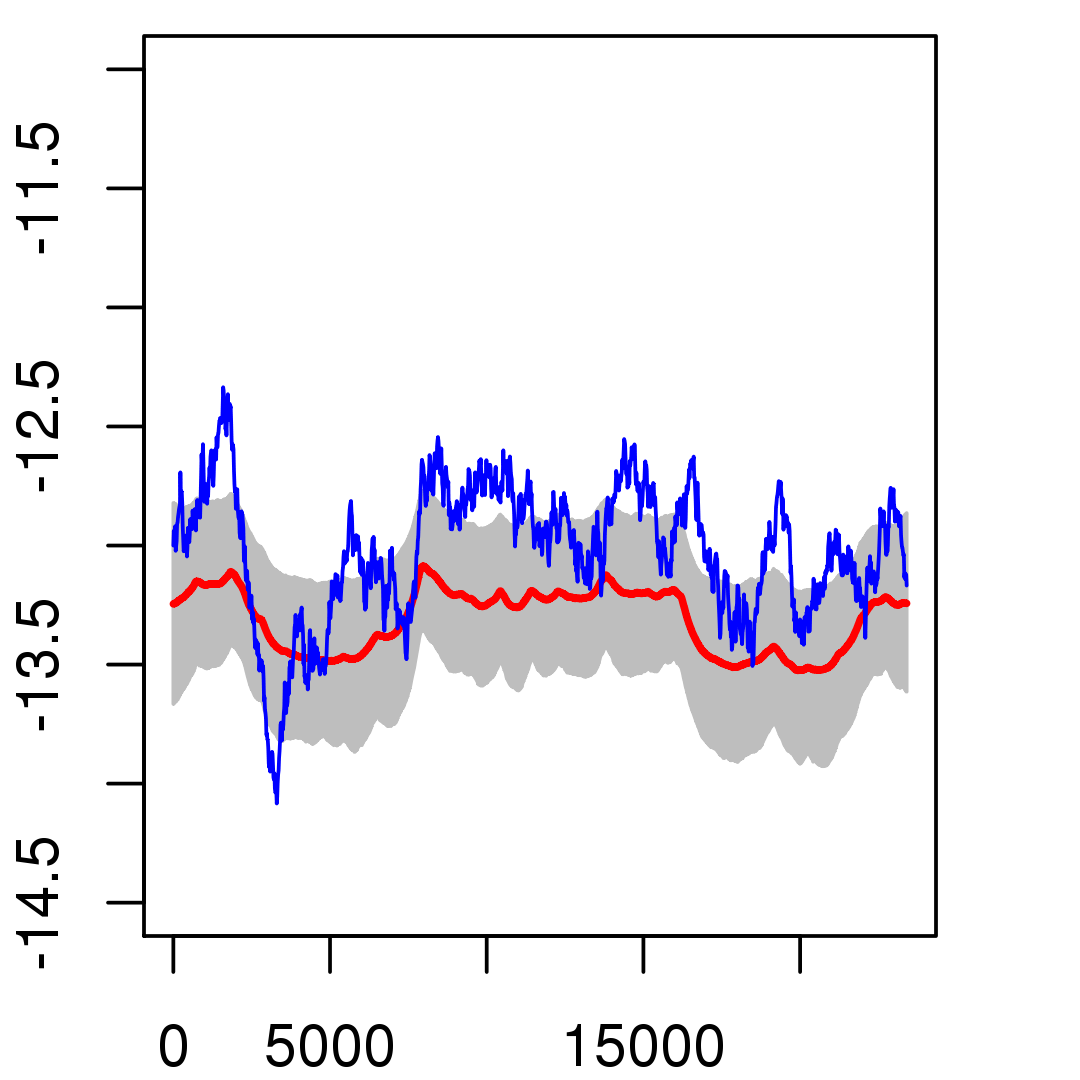}
				\end{minipage} 
			& \begin{minipage}{0.25\textwidth}
				\centering
				\includegraphics[width=1\linewidth]{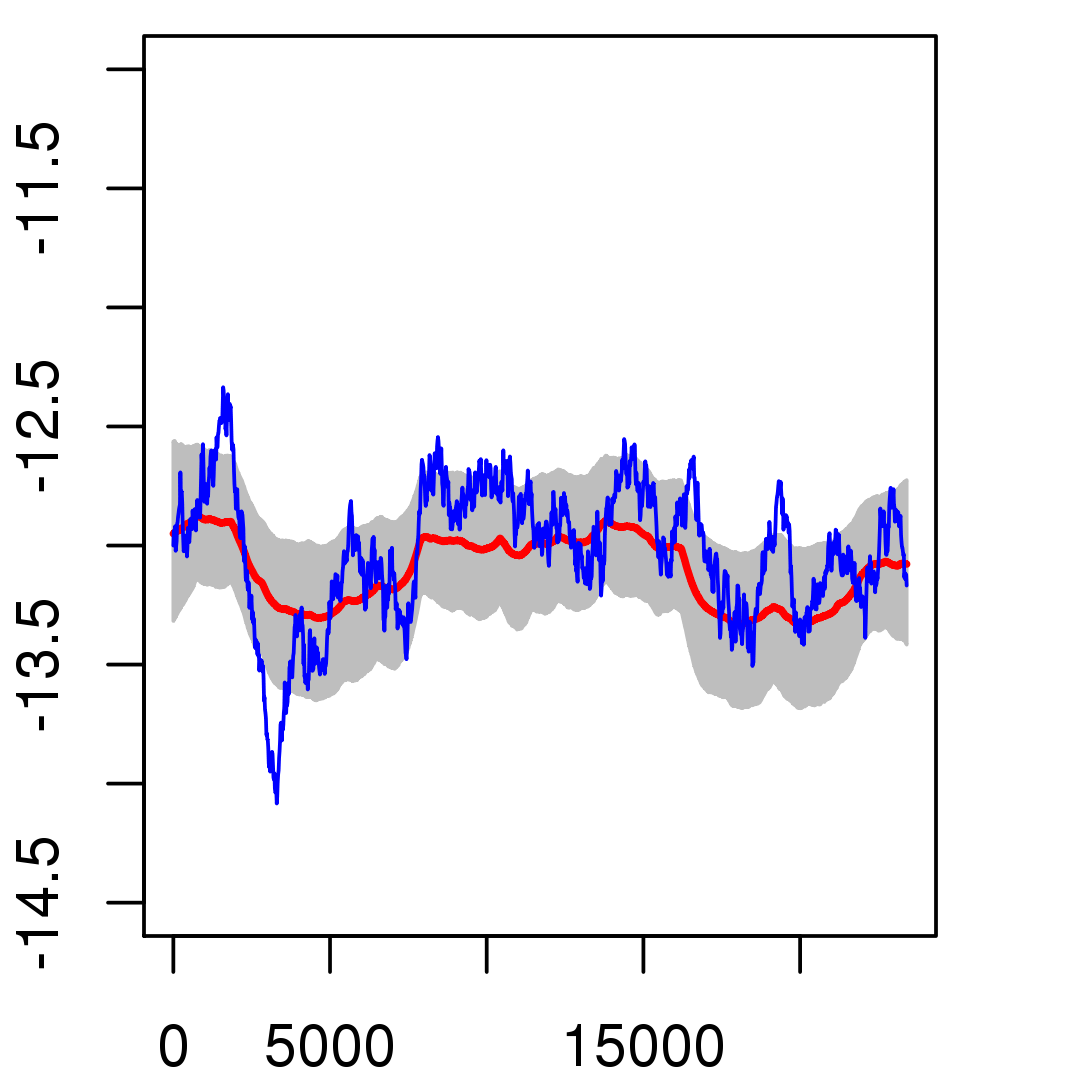}
				\end{minipage}  \\
			\begin{sideways} $\Delta = 5$ sec \end{sideways} 
			& \begin{minipage}{0.25\textwidth}
				\centering
                                \includegraphics[width=1\linewidth]{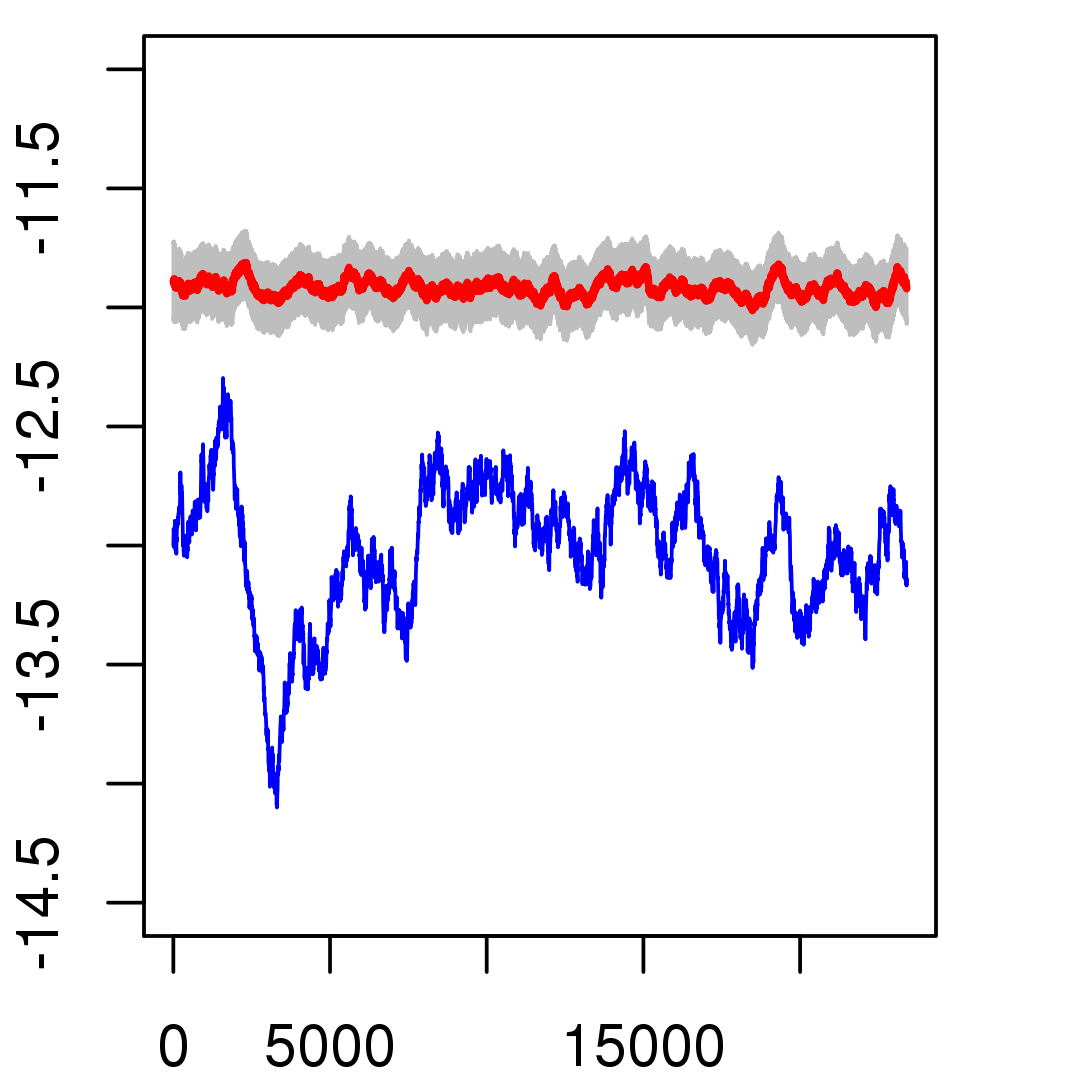}
				\end{minipage} 
			& \begin{minipage}{0.25\textwidth}
				\centering
                                \includegraphics[width=1\linewidth]{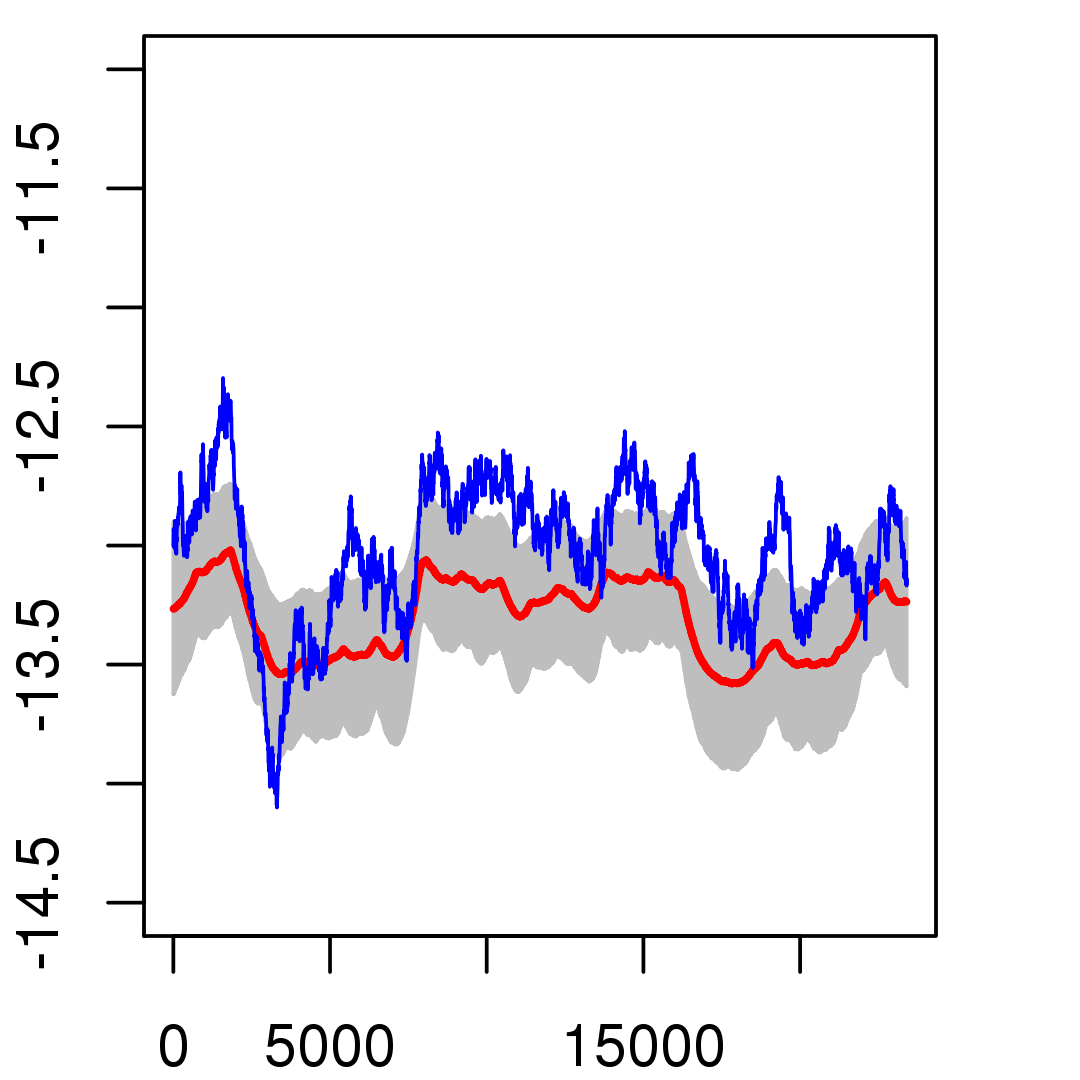}
				\end{minipage} 
			& \begin{minipage}{0.25\textwidth}
				\centering
				\includegraphics[width=1\linewidth]{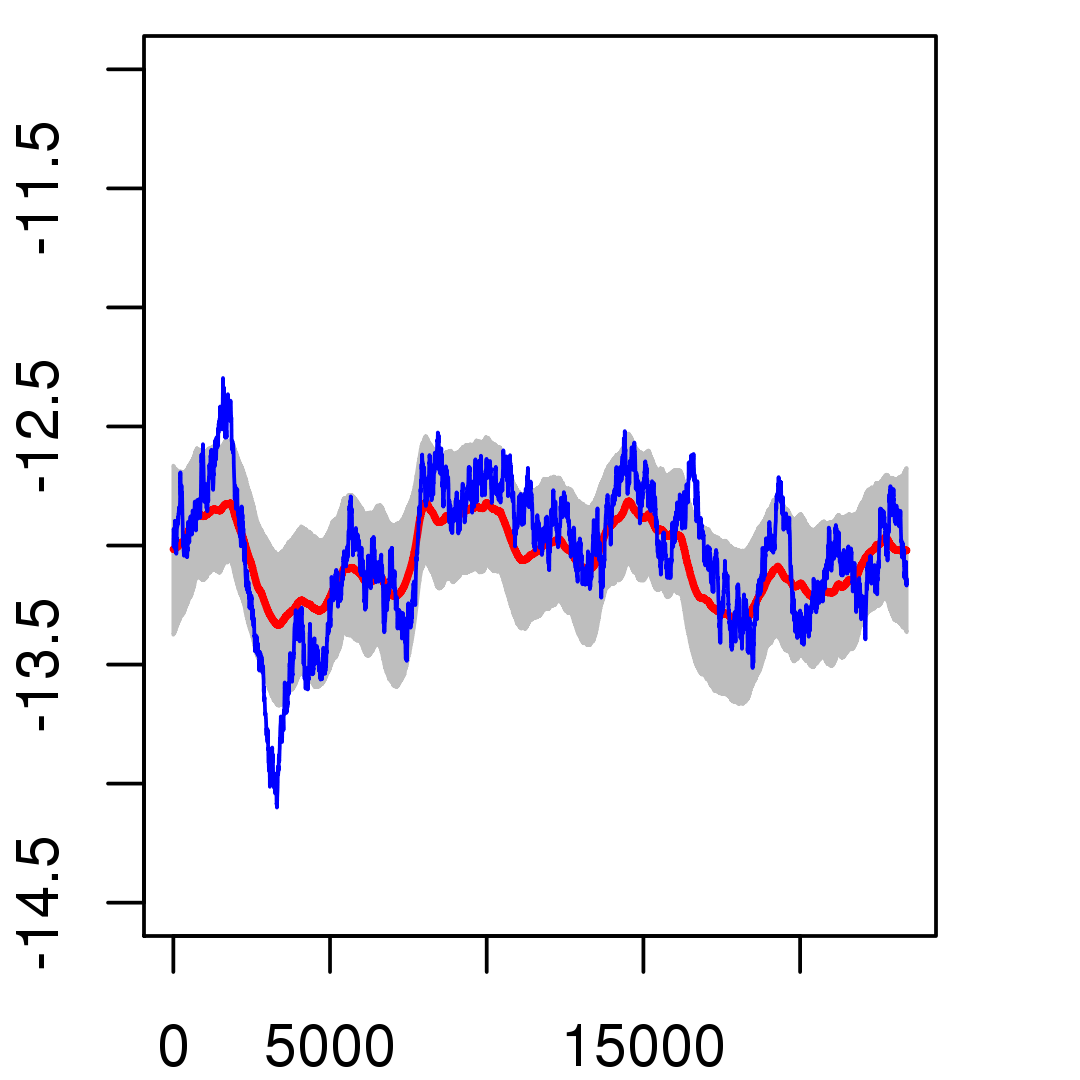}
				\end{minipage}
	\end{tabular}
\caption{Log-volatility paths for simulated data. All three inference approaches are applied to the same data set that contains microsructure noise. The miscrostructure noise added in the simulated data is approximately at the level of $\xi^2 = 2.5\cdot 10^{-7}$. Red denotes the posterior mean of the paths, while the gray region denotes the posterior 95\% probability for the log-volatility value. Blue is the true log-volatility signal. We see that when microstructure is ignored ($\xi^2 = 0$), we fail to recover the true signal.}  \label{fig:log-vol-simulation}
\end{figure}

The plots for the posterior mean and 95\% probability bands for the estimated log-volatility paths are given in Figure \ref{fig:log-vol-simulation}, along with the true signal from the simulated data. As the sampling period decreases, we see that the first version of the model (which ignores microstructure noise by fixing $\xi^2 = 0$) fails to capture the latent signal. On the other hand, the other two versions of our models produce much better posterior estimates for the latent signal. In particular, we see that, as expected, the naive choice $\xi^2 = 0$ model overestimates the volatility signal for higher sampling frequencies where the time interval $\Delta$ diminishes enough so that the microstructure noise dominates for the volatility signal.  We can also see that, although the second and third versions of the model tend to smooth out the true volatility path, the reconstruction generated by the model that estimates $\xi^2$ from the data is somewhat more accurate.

%%%%% LOG-VOL SIMULATION %%%%%

\begin{figure}[h!]
\centering
\begin{tabular}{m{0.25cm}ccc}
		 & Inference with & Inference with & Inference with \\
		 & $\xi^2 = 0$ & $\xi^2 = 2.5 \cdot 10^{-7}$ & $\xi^2 \mbox{ estimated }$ \\
		\begin{sideways} $\halpha$ \end{sideways}
			& \begin{minipage}{0.20\textwidth}
				\centering
				\includegraphics[width=1\linewidth]{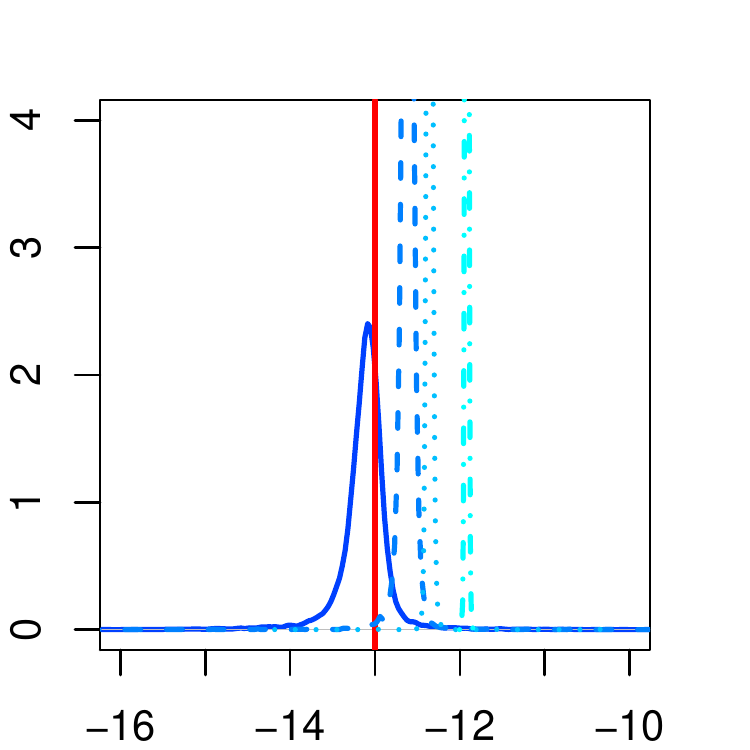}
				\end{minipage} 
			& \begin{minipage}{0.20\textwidth}
				\centering
				\includegraphics[width=1\linewidth]{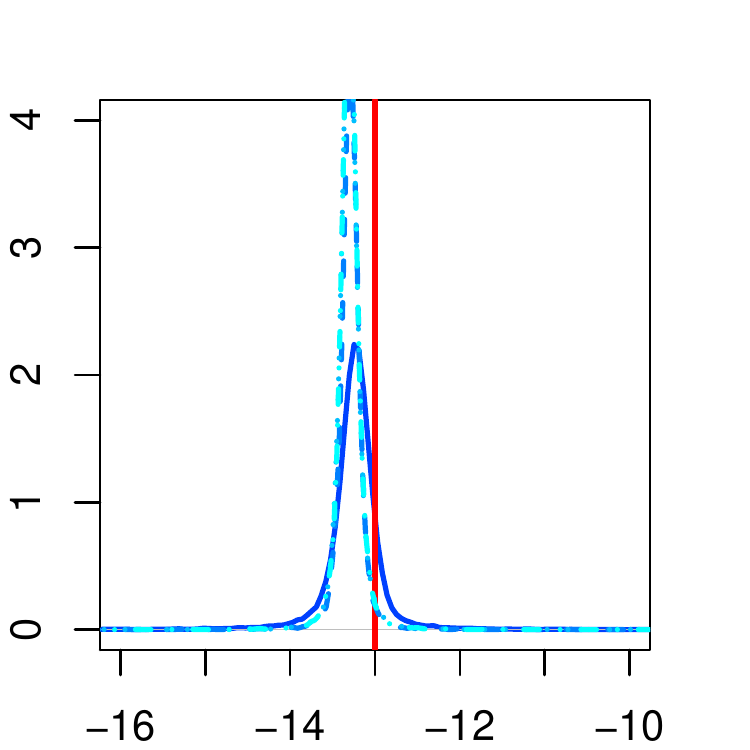}
				\end{minipage} 
			& \begin{minipage}{0.20\textwidth}
				\centering
				\includegraphics[width=1\linewidth]{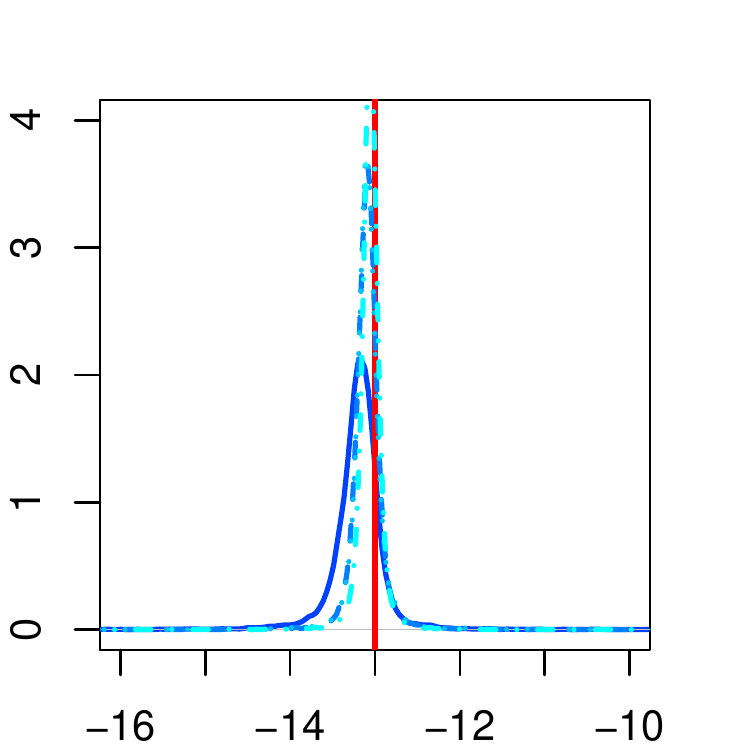}
				\end{minipage}  \\
		\begin{sideways} $\htau^2$ \end{sideways} 
			& \begin{minipage}{0.20\textwidth}
				\centering
				\includegraphics[width=1\linewidth]{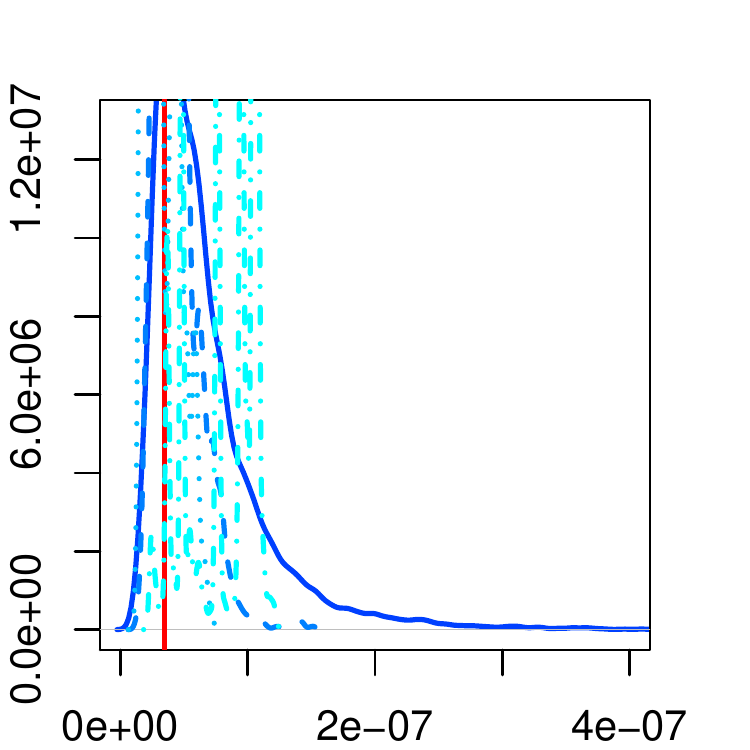}
				\end{minipage} 
			& \begin{minipage}{0.20\textwidth}
				\centering
				\includegraphics[width=1\linewidth]{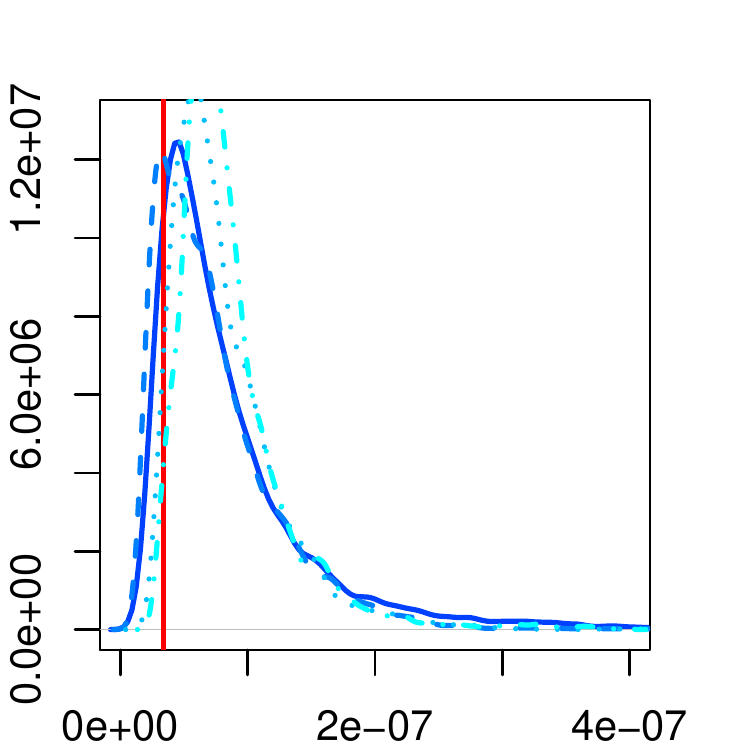}
				\end{minipage} 
			& \begin{minipage}{0.20\textwidth}
				\centering
				\includegraphics[width=1\linewidth]{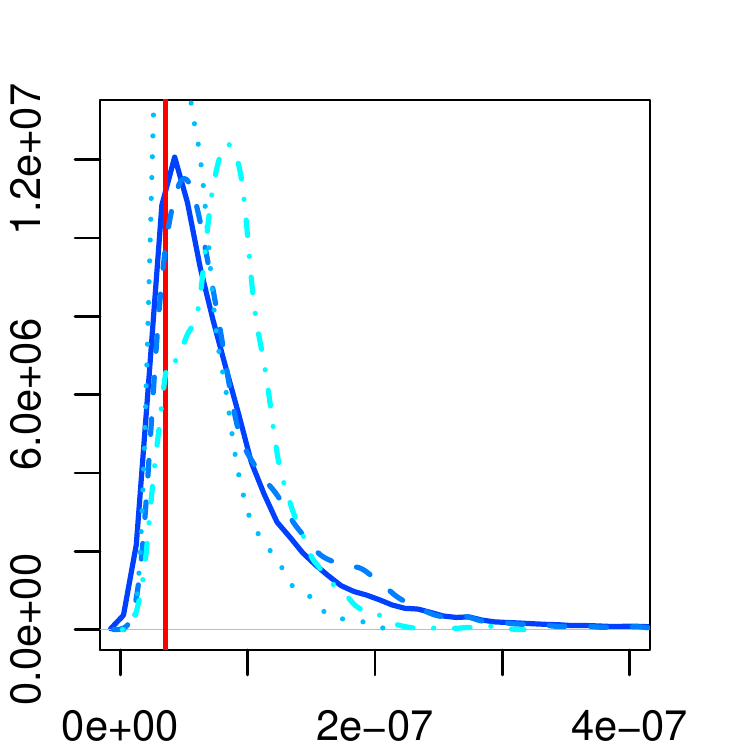}
				\end{minipage}  \\
			\begin{sideways} $\hat{\theta}$ \end{sideways} 
			& \begin{minipage}{0.20\textwidth}
				\centering
				\includegraphics[width=1\linewidth]{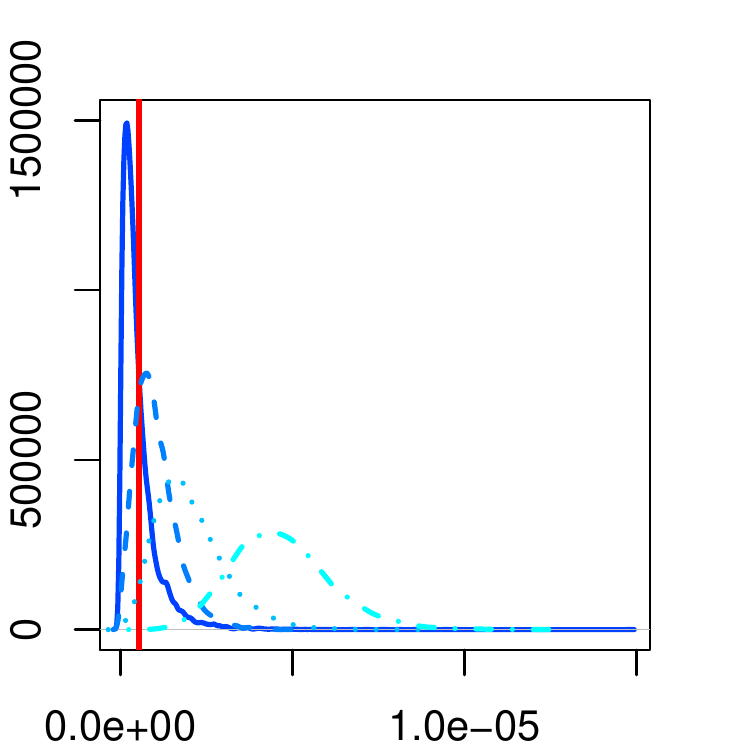}
				\end{minipage} 
			& \begin{minipage}{0.20\textwidth}
				\centering
				\includegraphics[width=1\linewidth]{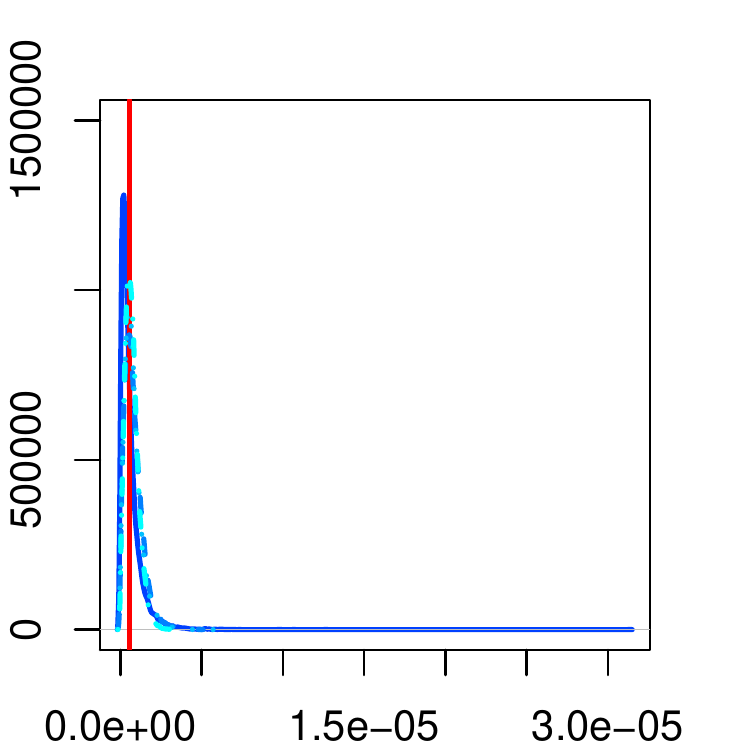}
				\end{minipage}  
			& \begin{minipage}{0.20\textwidth}
				\centering
				\includegraphics[width=1\linewidth]{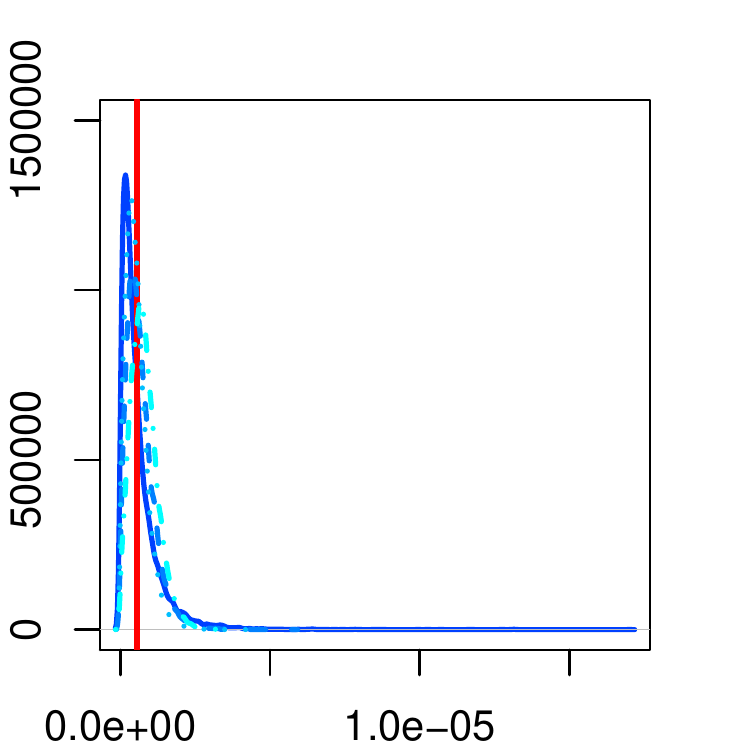}
				\end{minipage}  \\
			\begin{sideways} $\hat{\mu}$ \end{sideways} 
			& \begin{minipage}{0.20\textwidth}
				\centering
				\includegraphics[width=1\linewidth]{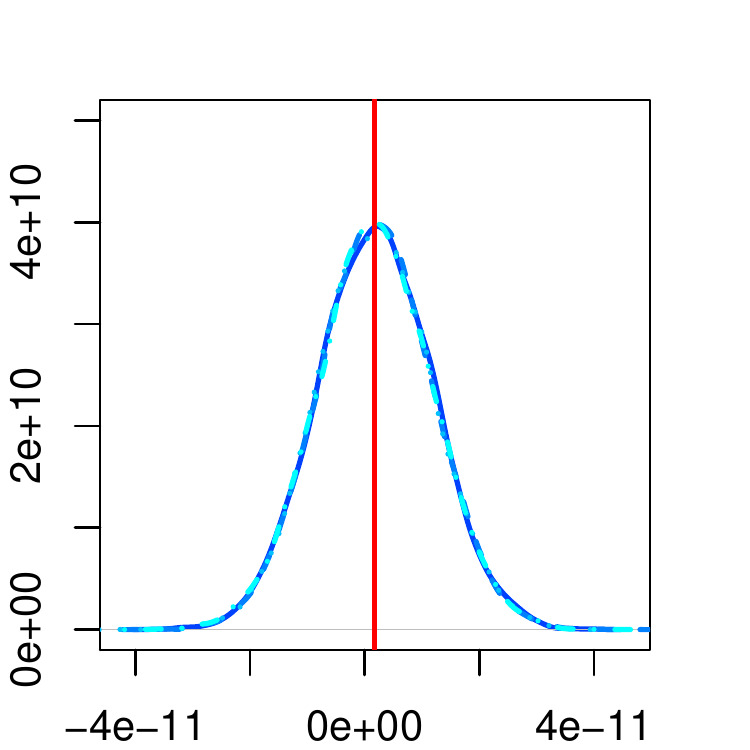}
				\end{minipage} 
			& \begin{minipage}{0.20\textwidth}
				\centering
				\includegraphics[width=1\linewidth]{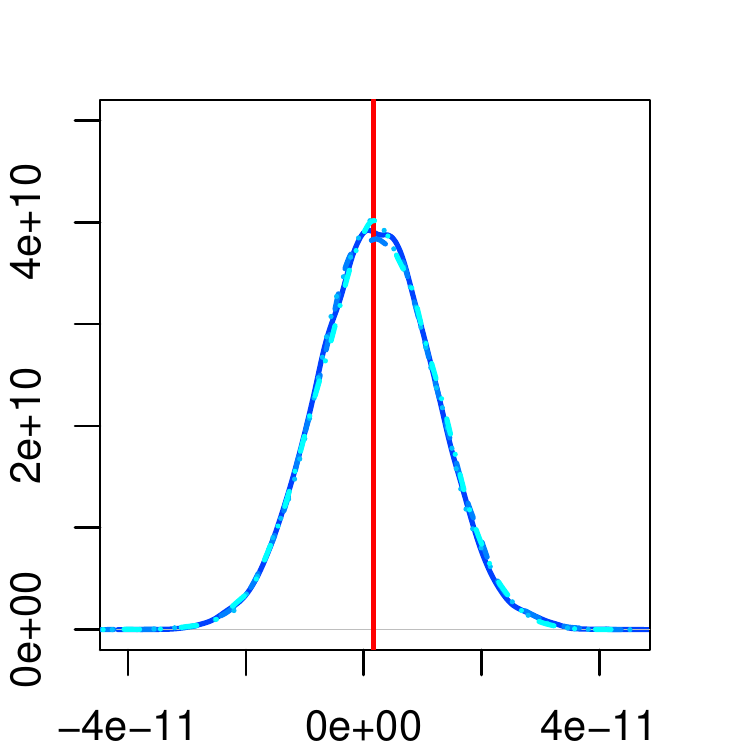}
				\end{minipage} 
			& \begin{minipage}{0.20\textwidth}
				\centering
				\includegraphics[width=1\linewidth]{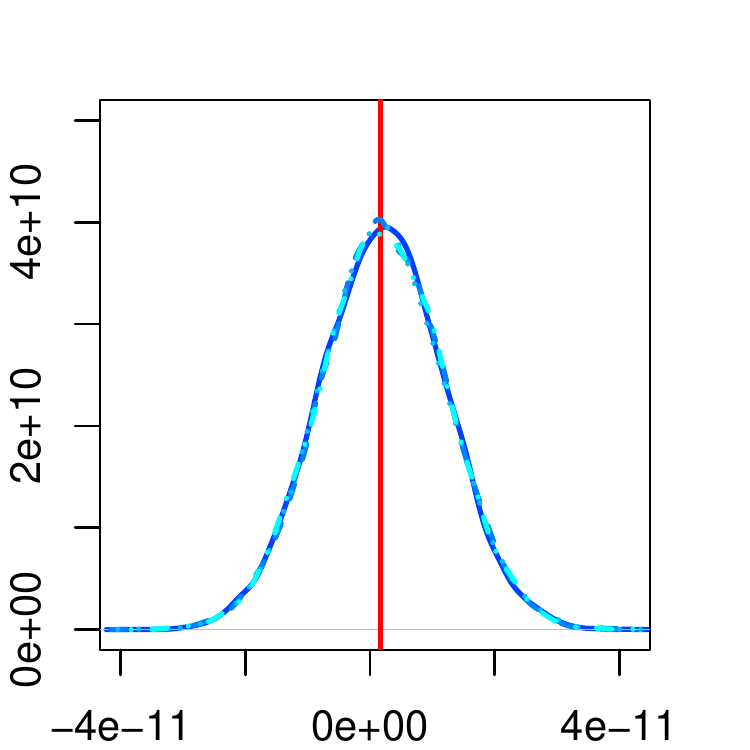}
				\end{minipage} \\
          \begin{sideways} $\xi^2$ \end{sideways} 
			& 
			& 
			& \begin{minipage}{0.20\textwidth}
				\centering
				\includegraphics[width=1\linewidth]{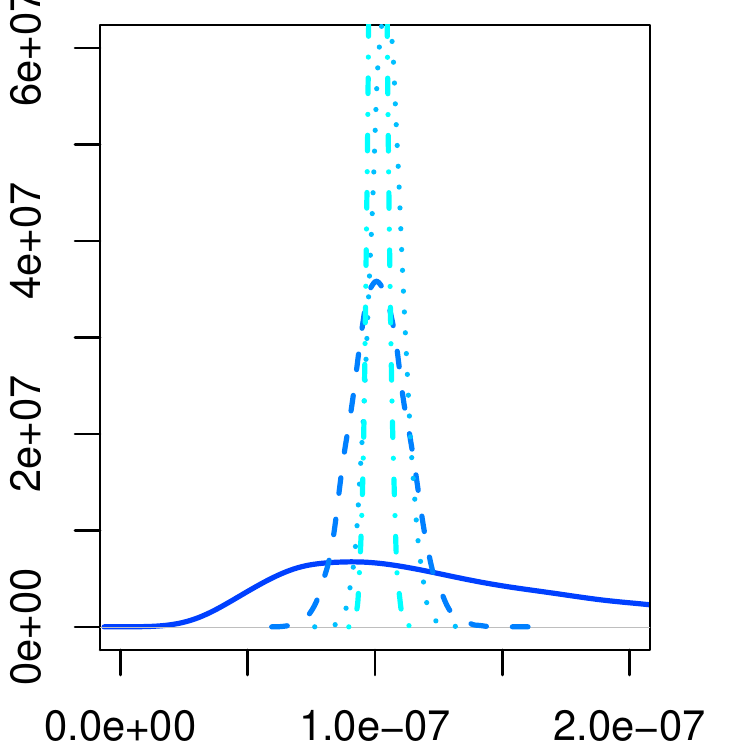}
				\end{minipage} 
\end{tabular}
\caption{Posterior density approximations of model parameters for
  simulated data. The red vertical line represents the true parameter
  value. The sampling periods used are: 5 minutes
  \usebox{\legendLineOne}, 30 seconds \usebox{\legendLineTwo}, 15
  seconds \usebox{\legendLineThree}, and 5 seconds
  \usebox{\legendLineFour}.}  \label{fig:posterior-parameters-simulated}
\end{figure}

In addition to estimating the volatility path, we also investigate the ability of the model to infer model parameters.  In particular, Figure \ref{fig:posterior-parameters-simulated} shows the posterior density estimates for the continuous-time parameters $\hat{\alpha}$, $\hat{\tau}^2$, $\hat{\theta}$ and $\hat{\mu}$ (which are comparable across scales), as well as the posterior distribution for $\xi^2$ (in the case of the third version of the model, which is the only one in which it is estimated from the data).  Note that, when the model is estimated with $\xi^2 = 0$ fixed, the posterior densities for $\halpha$, the mean level of log volatility, show a reduction in variance with increasing sampling frequency:  posterior draws become more centered around a wrong, overestimated value for mean log-volatility level. These results are consistent with those obtained for the volatility path and show that the model fails to capture the constant information content in the data regarding $\halpha$.  We also note that learning $\htau^2$ and $\htheta$ is difficult whether we do or do not include microstructure noise. However, due to the constant information in the data with respect to $\halpha$, the posterior uncertainty for $\htheta$ and $\htau^2$ seems to remain constant even with increasing sampling frequency when $\xi$ is not fixed.

\subsection{Estimating Integrated Variance} 

As described in Section \ref{se:computation}, the posterior draws for $\sigma^2_{j}$ allow us to approximate the posterior distributions for the integrated variance of the latent volatility process. In this Section we extend the previous simulation study to compare the 95\% intervals generated by the three versions of our model with those generated from a realized variance estimate.  The literature on realized volatility estimators for high-frequency data is vast (for a review, refer to \citealp{pigorsch2012volatility}), but the construction of confidence intervals for the realized volatility estimators can be challenging.  Here we compare the coverage properties of our model-based credible intervals against bootstrap-based confidence intervals of the the kernel-based realized variance estimator introduced by \cite{zhou1996high} and \cite{hansen2006realized}. The idea behind the bootstrapping method is to periodically extend the available data set and randomly reselect a new data set to construct a bootstrap sample (see \citealp{hwang2013stationary-bootstrap} for a full description of the procedure).

The results of the comparison are shown in Table \ref{ta:coverage}. The table is constructed using 300 simulated data sets, each corresponding to a single trading day. We compare the percentage of times the 95\% confidence/credible intervals for the integrated variance (IV) estimator covers the true integrated variance value.  A well-calibrated interval will produce a 95\% coverage on average, and we see that the estimator based on our approach where $\xi^2$ is fully estimated preforms very well, both when compared to $\xi^2 =0$, $\xi^2 = 2.5 \cdot 10^{-7}$, and when compared to the kernel-based estimator.

\begin{table}[h]
\begin{center}
\begin{tabular}{l|ccccc}
Sampling period   &   5 min  &	60 sec 	&   30 sec   &   15 sec & 5 sec  \\ \hline \hline
Inference with $\xi^2 = 0$  &  93  &  72  &   28  &	 3 & 0 \\
Inference with $\xi^2 = 2.5 \cdot 10^{-7}$ & 95 & 79 & 57 & 23 & 0 \\
Inference with $\xi^2$ estimated & 95 & 91 & 92 & 96 & 97  \\ \hline
Inference with kernel-based estimator &  53 & 51 & 48 & 59  & 76
\end{tabular}
\caption{Coverage table comparing the percentage of covered integrated variance levels by 95\% confidence/credible intervals of our estimator and a kernel-based estimator.}\label{ta:coverage}
\end{center}
\end{table}

\newpage
\subsection{ Effect of microstructure noise and sampling frequency on estimates: market data }

\begin{figure}
	\centering
	\begin{tabular}{m{0.25cm}ccc}
		 & Inference with & Inference with & Inference with \\
		 & $\xi^2 = 0$ & $\xi^2 = 2.5 \cdot 10^{-7}$ & $\xi^2 \mbox{ estimated }$ \\
		\begin{sideways} $\Delta = 5$ min \end{sideways}
			& \begin{minipage}{0.25\textwidth}
				\centering
				\includegraphics[width=1\linewidth]{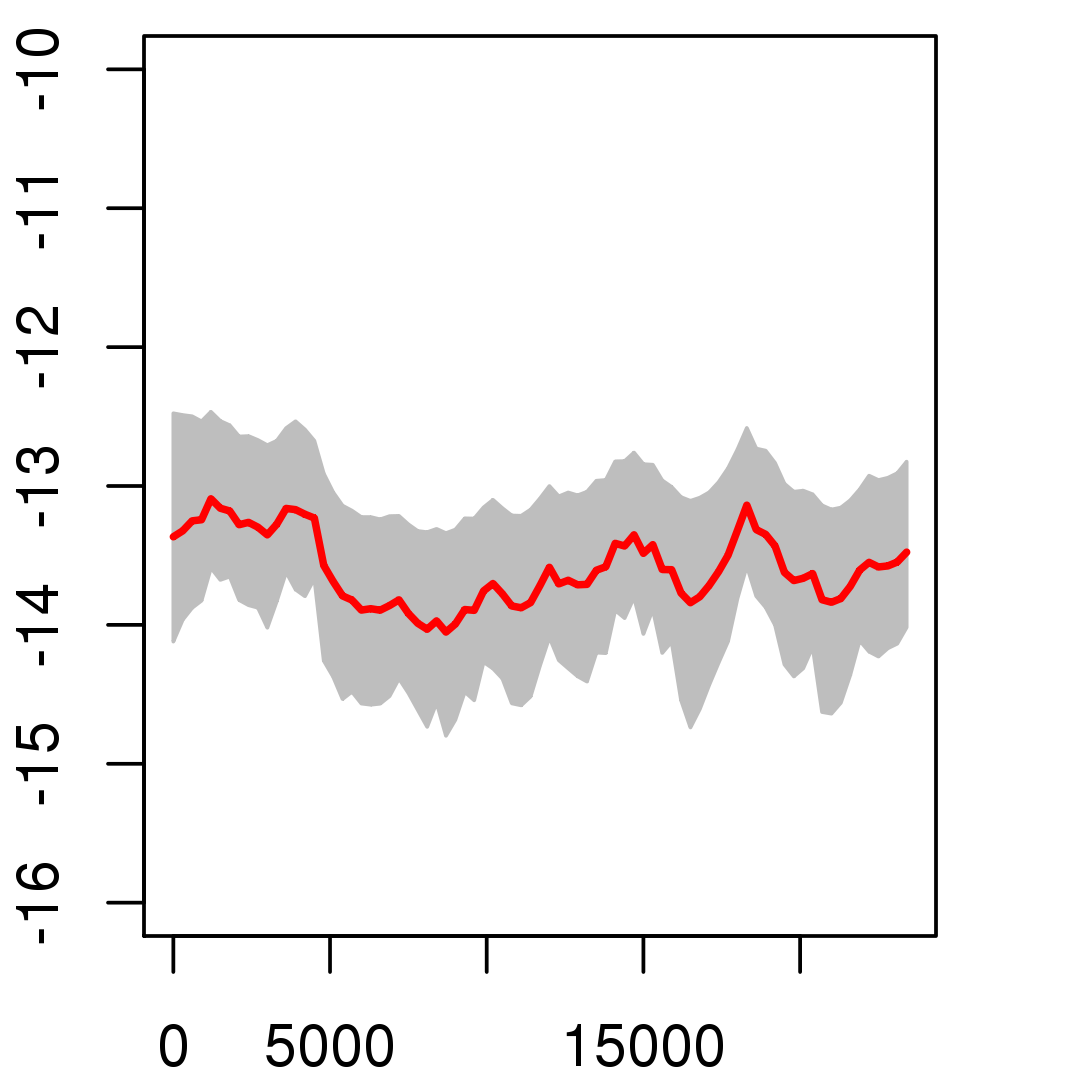}
				\end{minipage} 
			& \begin{minipage}{0.25\textwidth}
				\centering
				\includegraphics[width=1\linewidth]{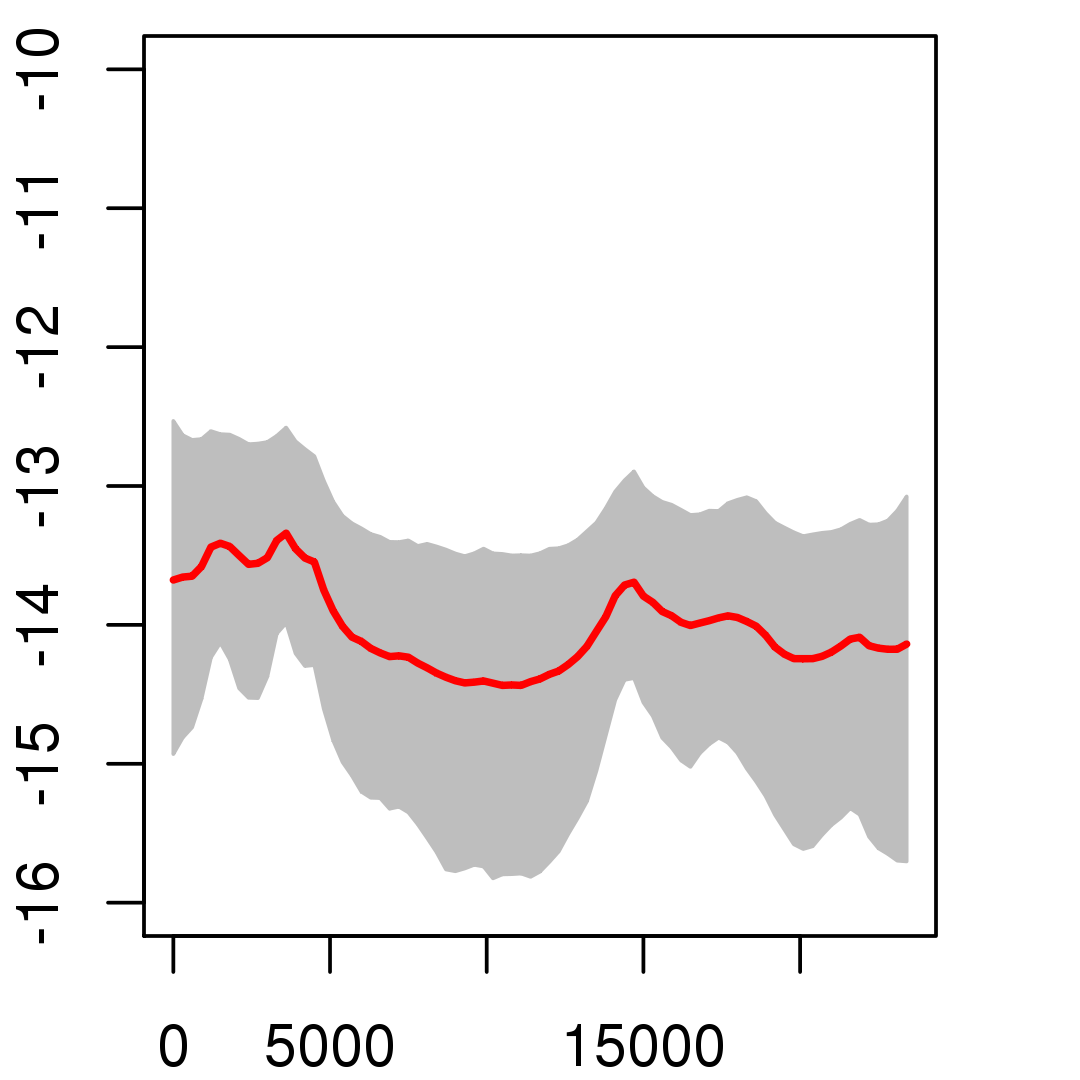}
				\end{minipage} 
			& \begin{minipage}{0.25\textwidth}
				\centering
				\includegraphics[width=1\linewidth]{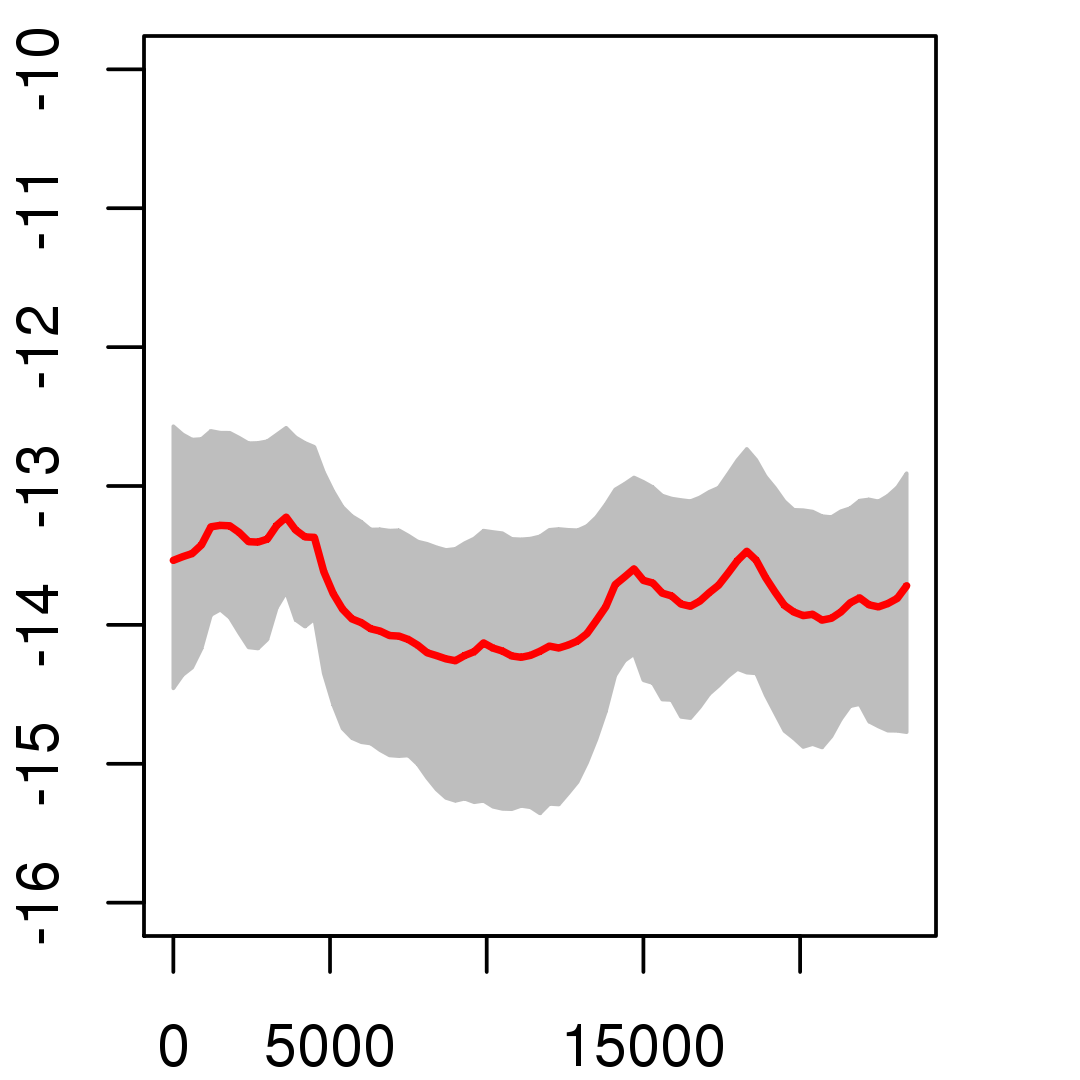}
				\end{minipage}  \\
		\begin{sideways} $\Delta = 15$ sec \end{sideways} 
			& \begin{minipage}{0.25\textwidth}
				\centering
				\includegraphics[width=1\linewidth]{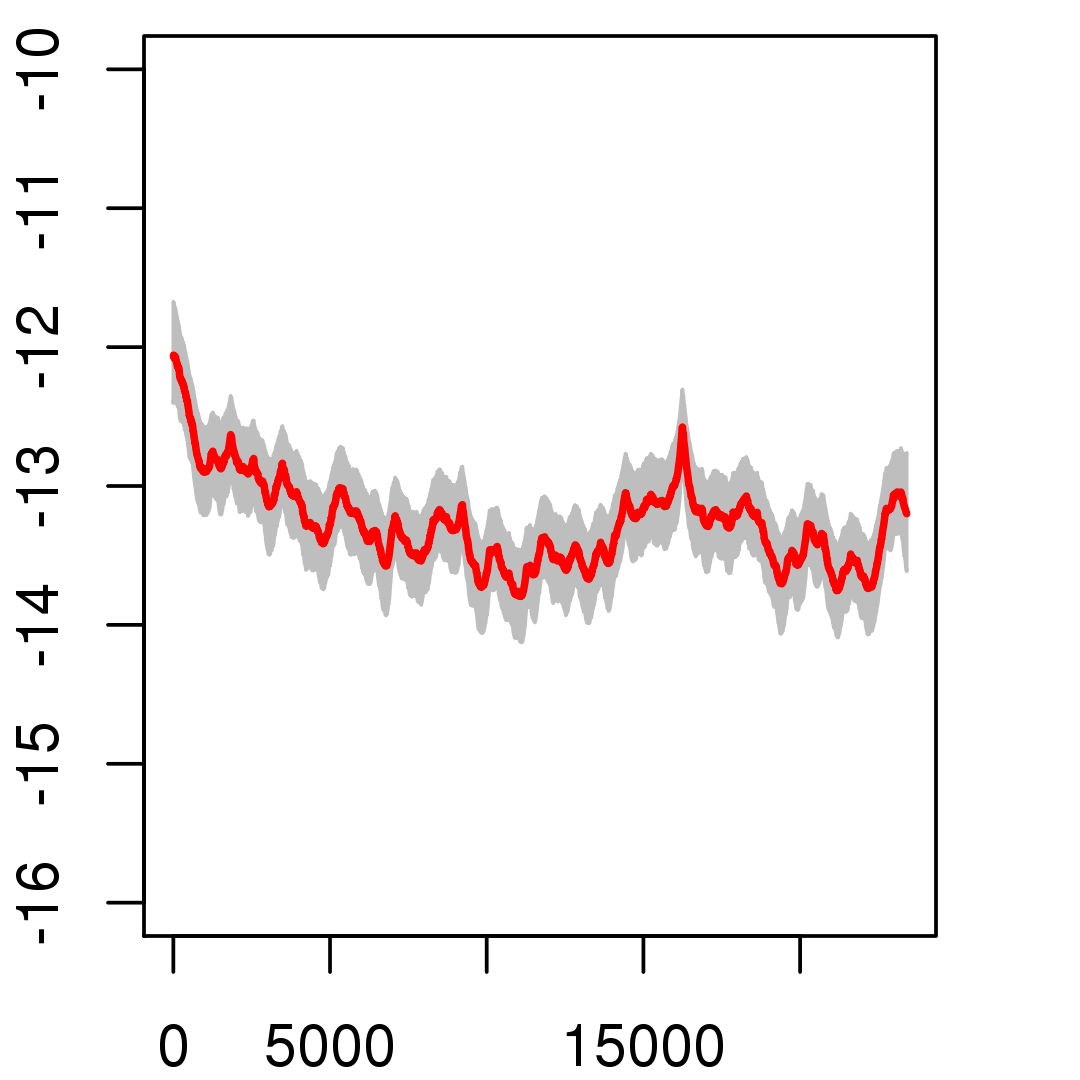}
				\end{minipage} 
			& \begin{minipage}{0.25\textwidth}
				\centering
				\includegraphics[width=1\linewidth]{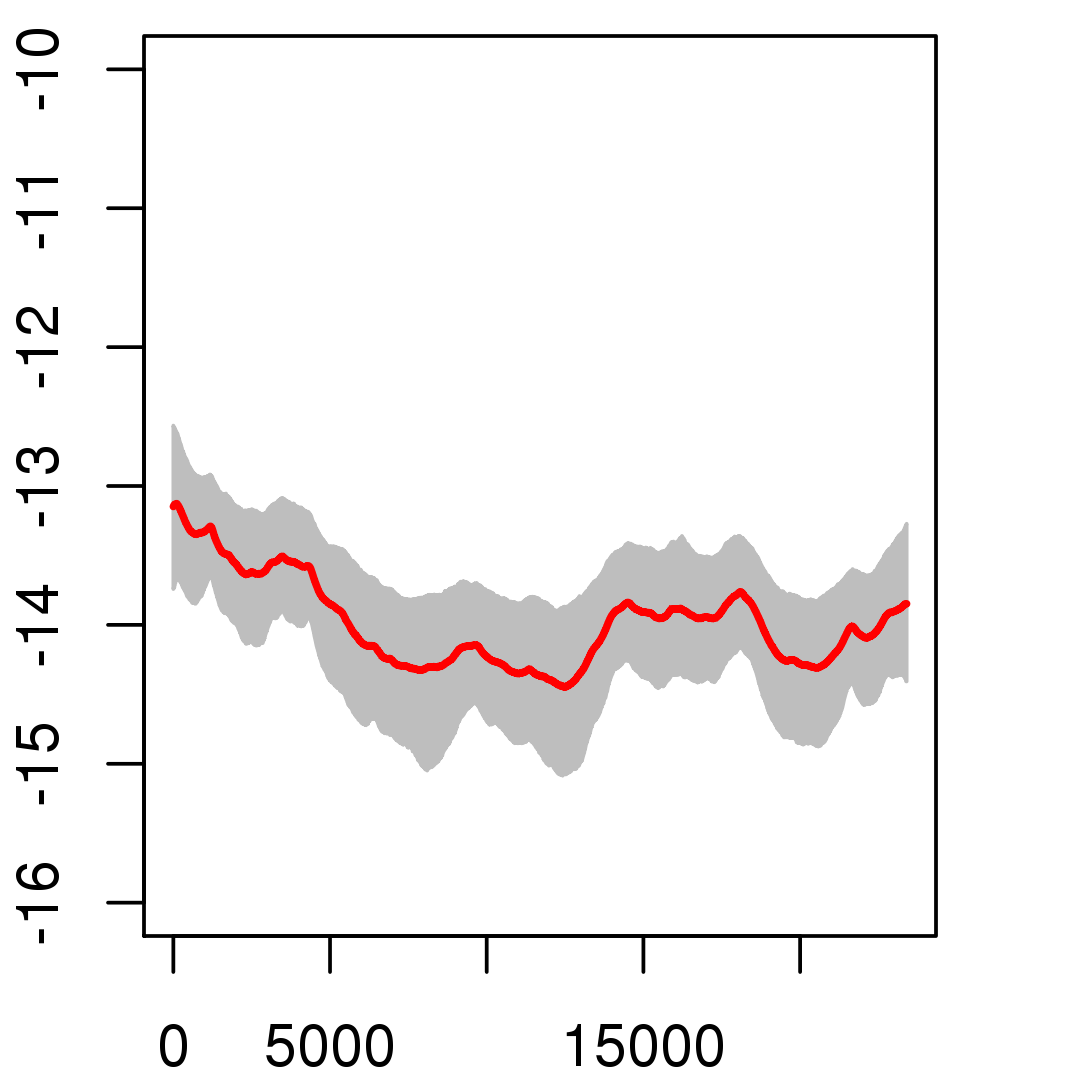}
				\end{minipage} 
			& \begin{minipage}{0.25\textwidth}
				\centering
				\includegraphics[width=1\linewidth]{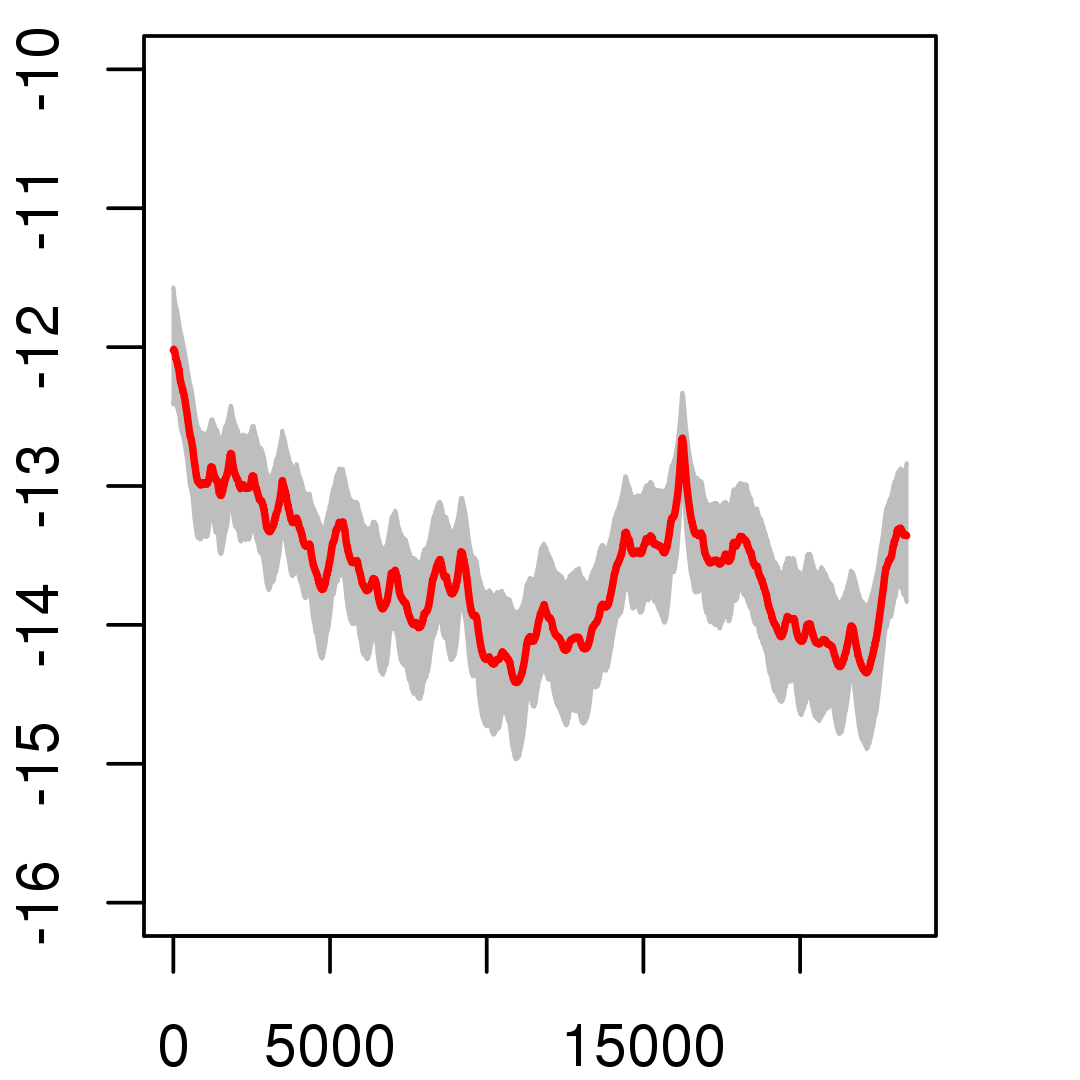}
				\end{minipage}  \\
			\begin{sideways} $\Delta = 5$ sec \end{sideways} 
			& \begin{minipage}{0.25\textwidth}
				\centering
				\includegraphics[width=1\linewidth]{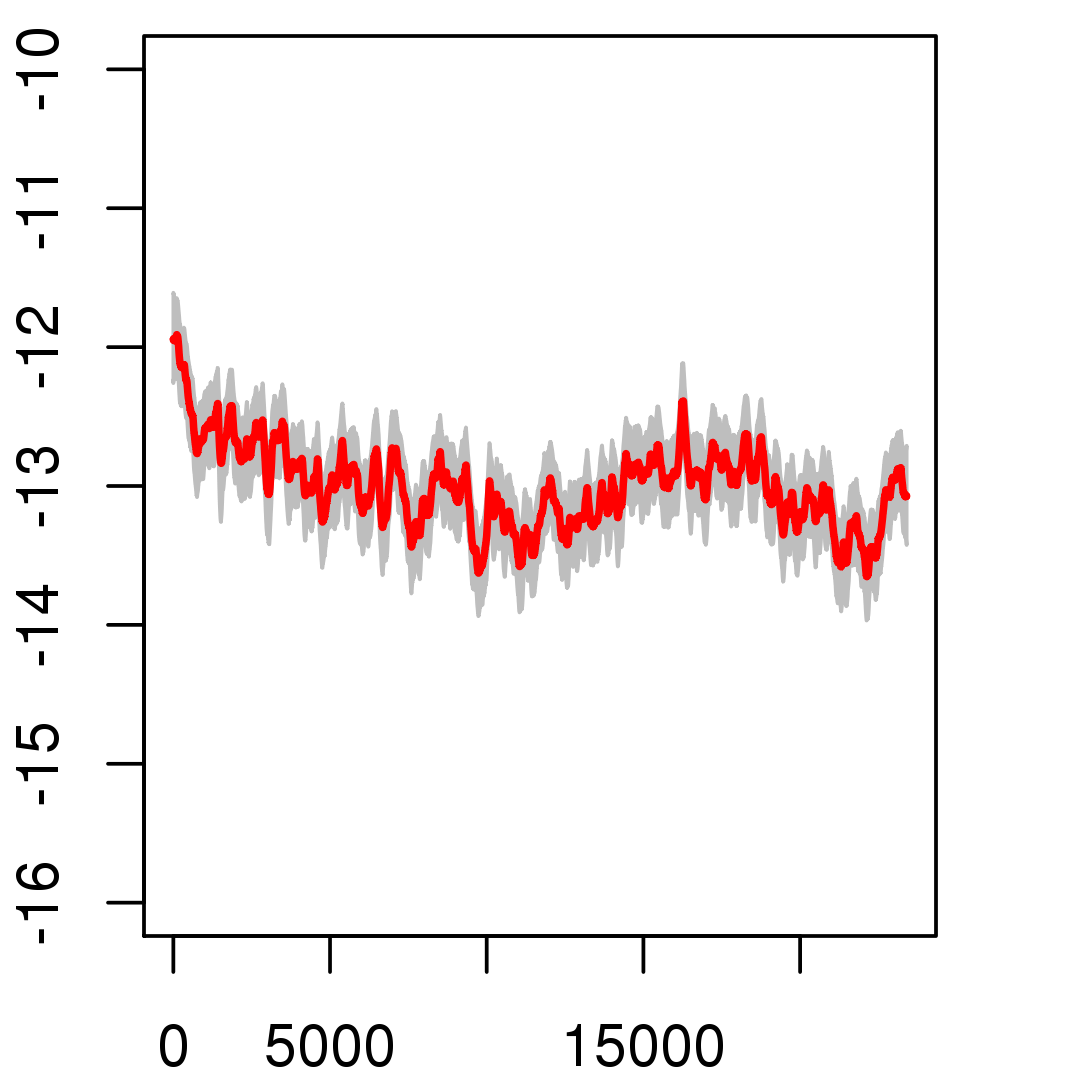}
				\end{minipage} 
			& \begin{minipage}{0.25\textwidth}
				\centering
				\includegraphics[width=1\linewidth]{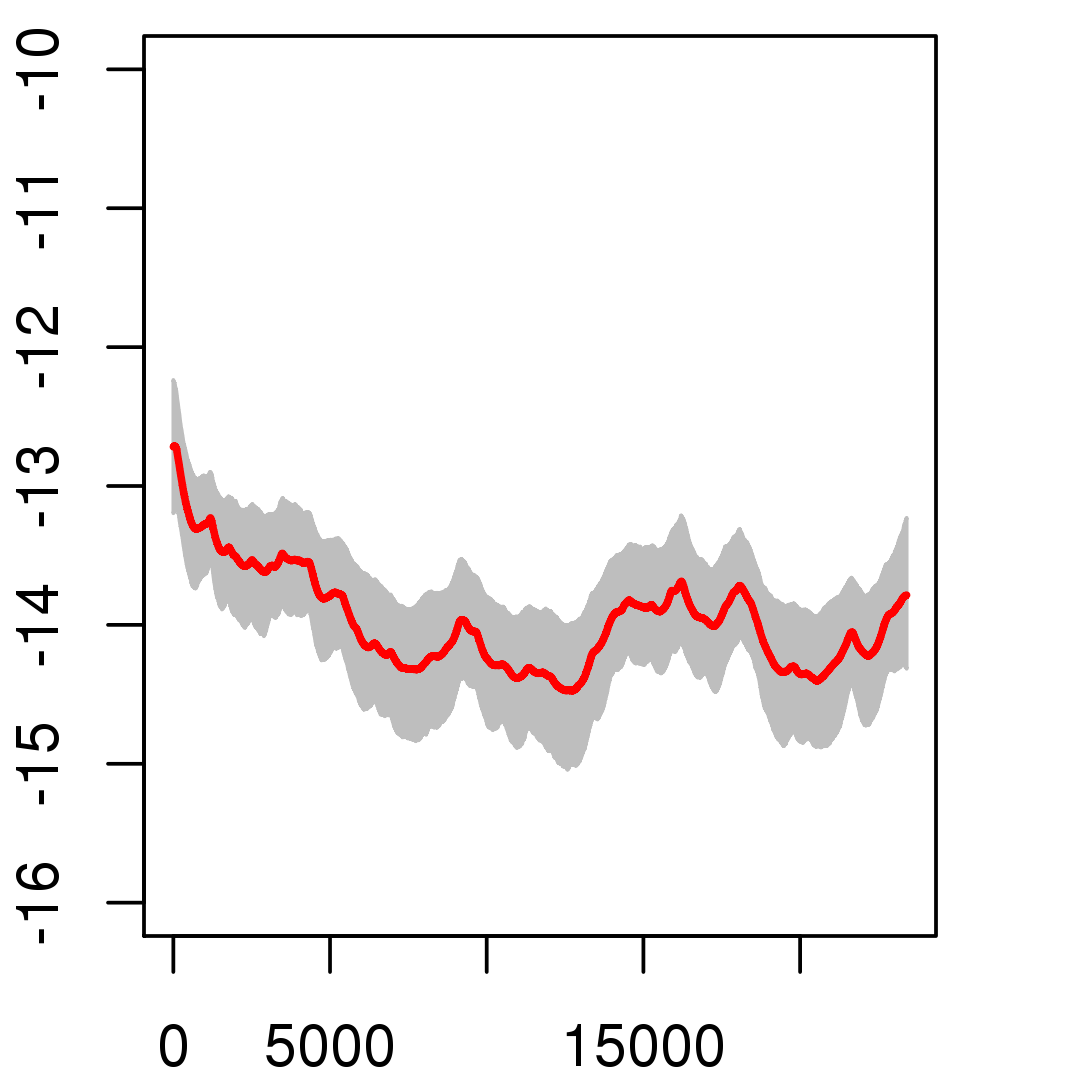}
				\end{minipage} 
			& \begin{minipage}{0.25\textwidth}
				\centering
				\includegraphics[width=1\linewidth]{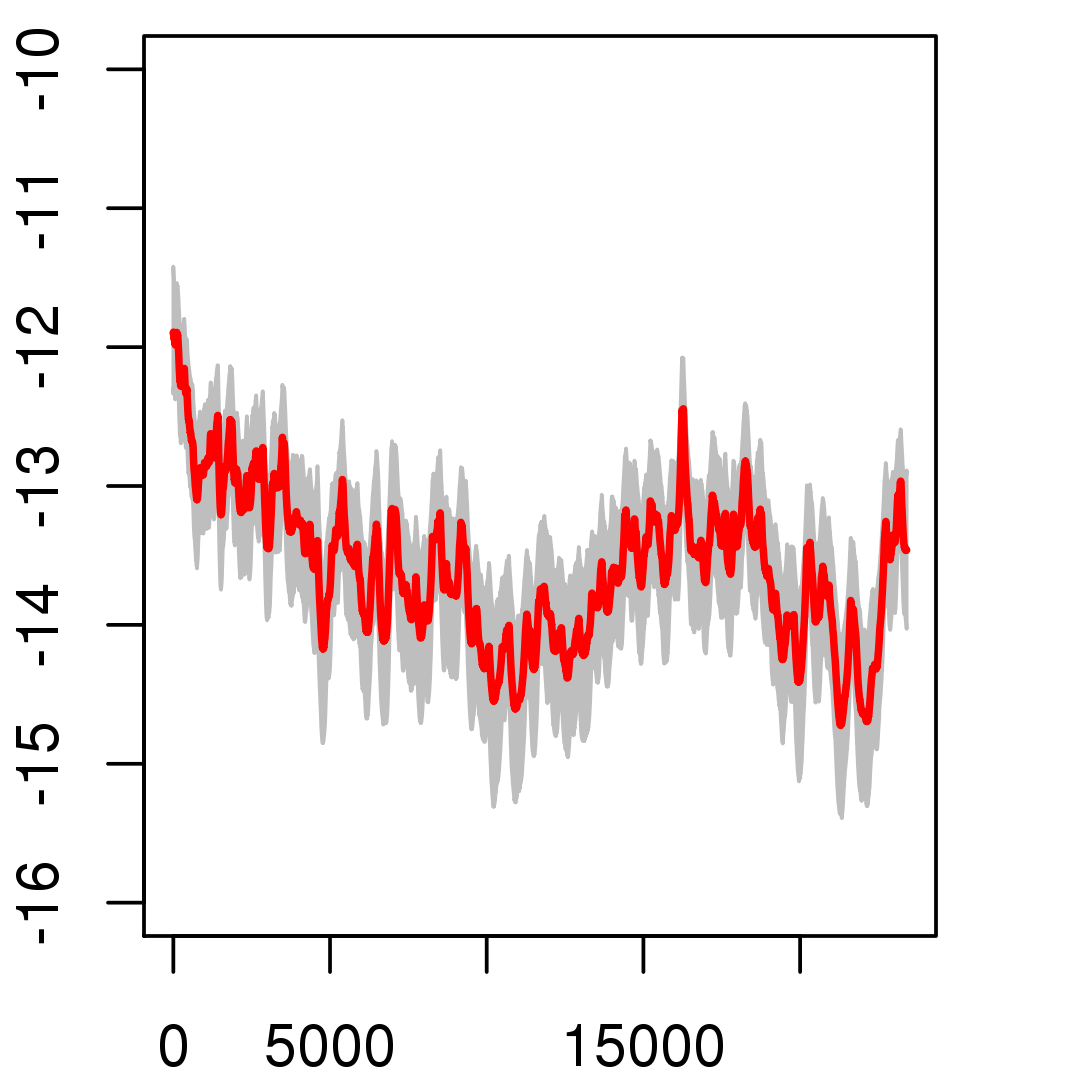}
				\end{minipage}  \\
%%
%			\begin{sideways} $\Delta = 1$ sec \end{sideways} 
%			& \begin{minipage}{0.25\textwidth}
%				\centering
%				\includegraphics[width=1\linewidth]{results/real-data/plots/VOL-PATHS/microstructure-VOL-PATHS-XI-0-dt-1000-SDs-0.png}
%				\end{minipage} 
%			& \begin{minipage}{0.25\textwidth}
%				\centering
%				\includegraphics[width=1\linewidth]{results/real-data/plots/VOL-PATHS/microstructure-VOL-PATHS-XI-1e-06-dt-1000-SDs-0.png}
%				\end{minipage} 
%			& \begin{minipage}{0.25\textwidth}
%				\centering
%				\includegraphics[width=1\linewidth]{results/real-data/plots/VOL-PATHS/microstructure-VOL-PATHS-XI-Inf-dt-1000-SDs-0.png}
%				\end{minipage} 
	\end{tabular}
	\caption{Log-volatility paths for the AAPL 03/06/2014 data. Red denotes the posterior mean of the paths, while the gray region denotes the posterior 95\% probability for the log-volatility value.}
	\label{fig:log-vol-real}
\end{figure}

\begin{figure}[h!]
	\centering
	\begin{tabular}{m{0.25cm}ccc}
		 & Inference with & Inference with & Inference with \\
		 & $\xi^2 = 0$ & $\xi^2 = 1 \cdot 10^{-6}$ & $\xi^2 \mbox{ estimated }$ \\
		\begin{sideways} $\halpha$ \end{sideways}
			& \begin{minipage}{0.20\textwidth}
				\centering
				\includegraphics[width=1\linewidth]{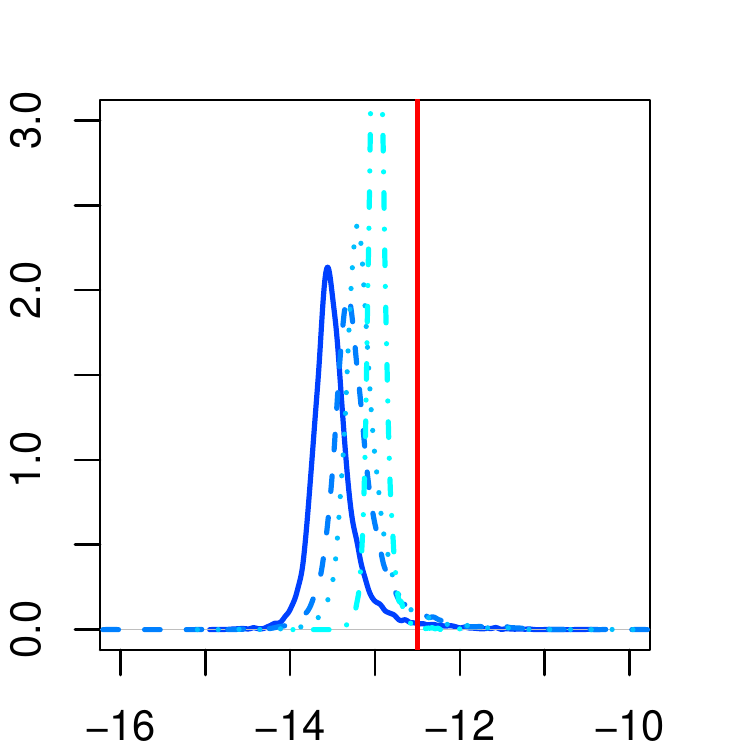}
				\end{minipage} 
			& \begin{minipage}{0.20\textwidth}
				\centering
				\includegraphics[width=1\linewidth]{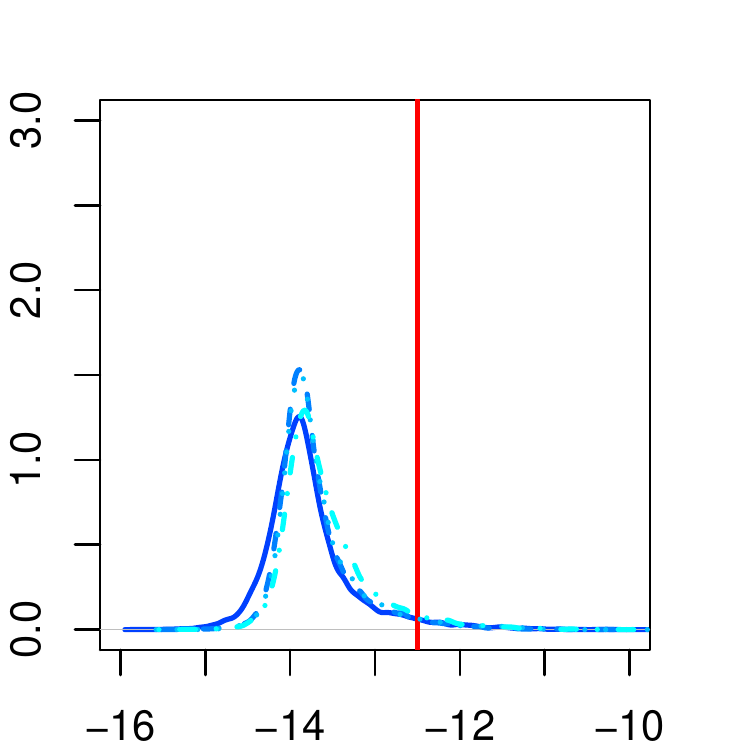}
				\end{minipage} 
			& \begin{minipage}{0.20\textwidth}
				\centering
				\includegraphics[width=1\linewidth]{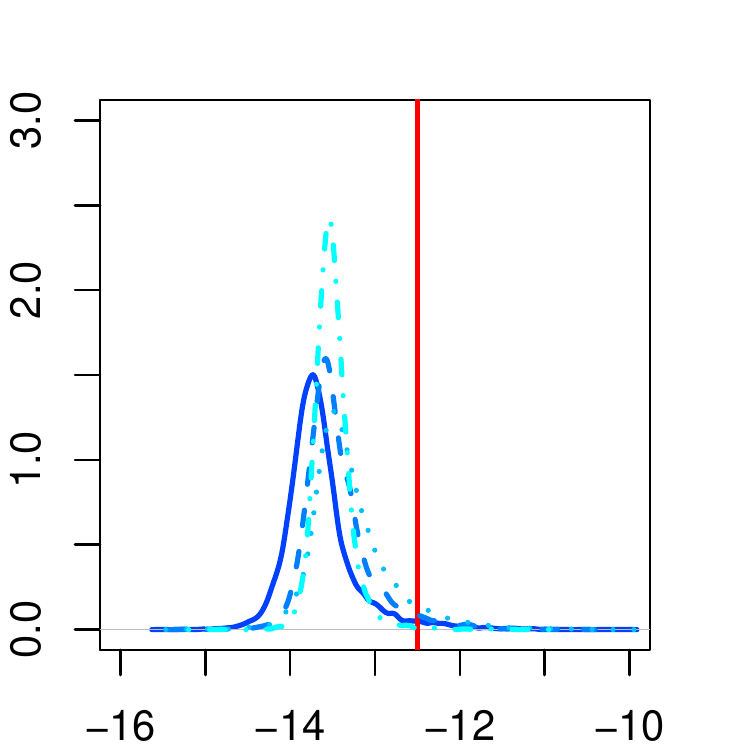}
				\end{minipage}  \\
		\begin{sideways} $\htau^2$ \end{sideways} 
			& \begin{minipage}{0.20\textwidth}
				\centering
				\includegraphics[width=1\linewidth]{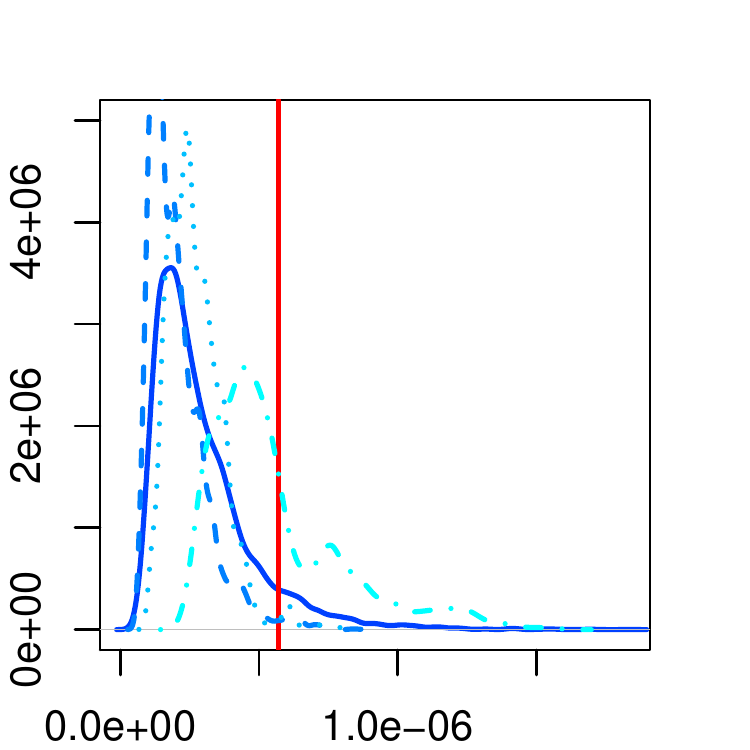}
				\end{minipage} 
			& \begin{minipage}{0.20\textwidth}
				\centering
				\includegraphics[width=1\linewidth]{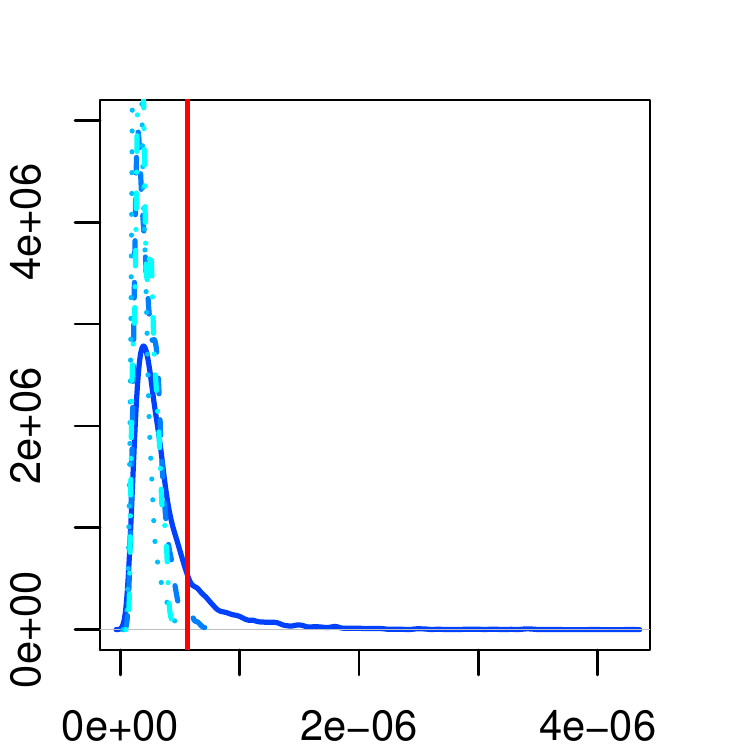}
				\end{minipage} 
			& \begin{minipage}{0.20\textwidth}
				\centering
				\includegraphics[width=1\linewidth]{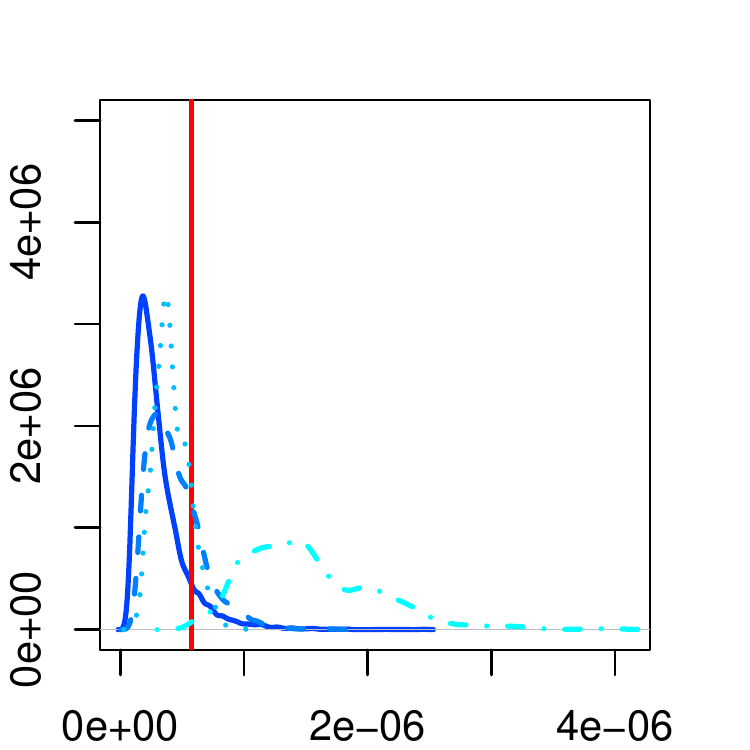}
				\end{minipage}  \\
			\begin{sideways} $\hat{\theta}$ \end{sideways} 
			& \begin{minipage}{0.20\textwidth}
				\centering
				\includegraphics[width=1\linewidth]{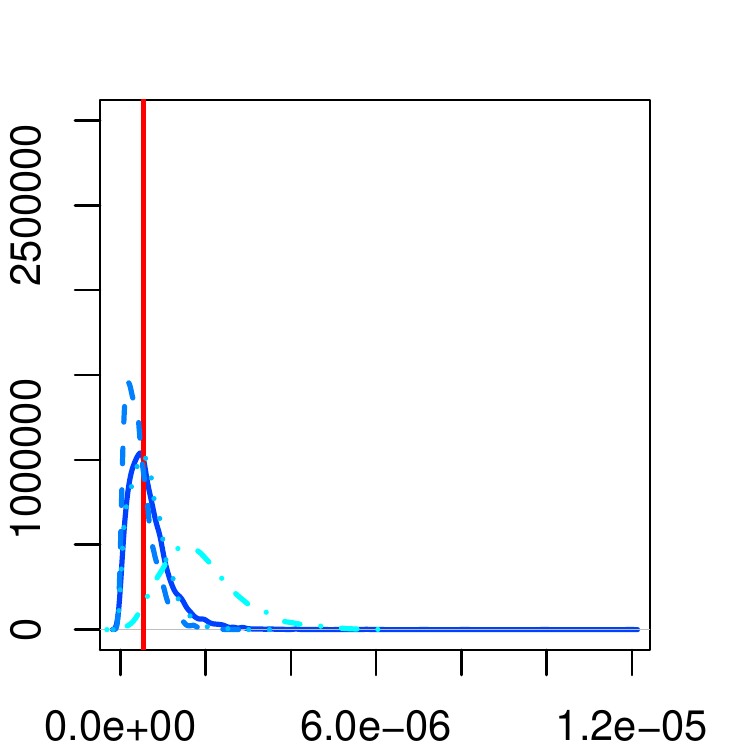}
				\end{minipage} 
			& \begin{minipage}{0.20\textwidth}
				\centering
				\includegraphics[width=1\linewidth]{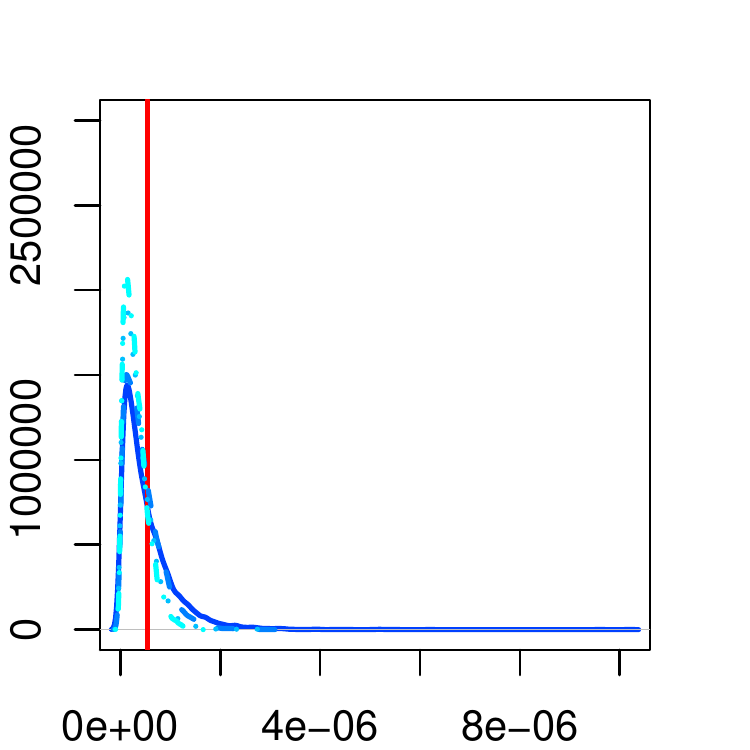}
				\end{minipage} 
			& \begin{minipage}{0.20\textwidth}
				\centering
				\includegraphics[width=1\linewidth]{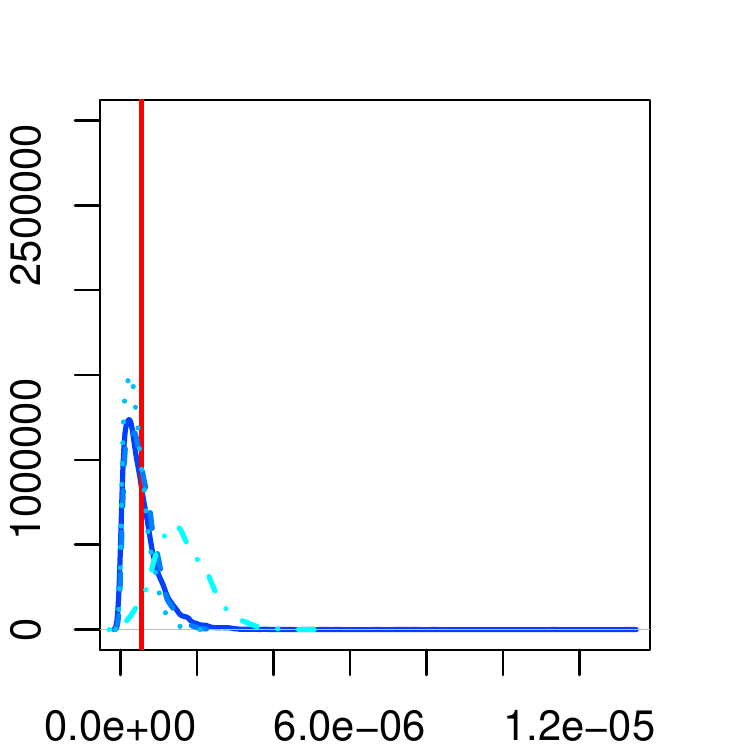}
				\end{minipage}  \\
			\begin{sideways} $\hat{\mu}$ \end{sideways} 
			& \begin{minipage}{0.20\textwidth}
				\centering
				\includegraphics[width=1\linewidth]{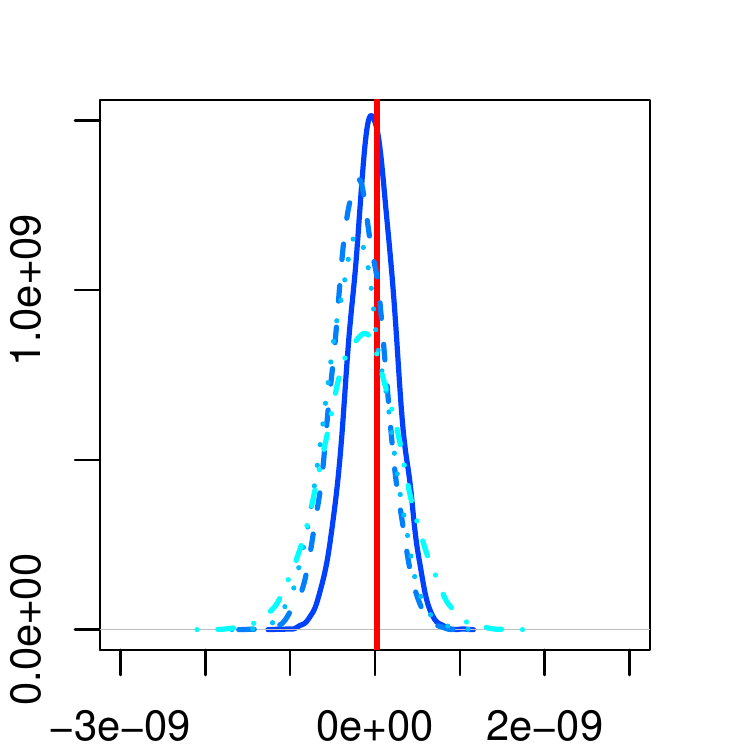}
				\end{minipage} 
			& \begin{minipage}{0.20\textwidth}
				\centering
				\includegraphics[width=1\linewidth]{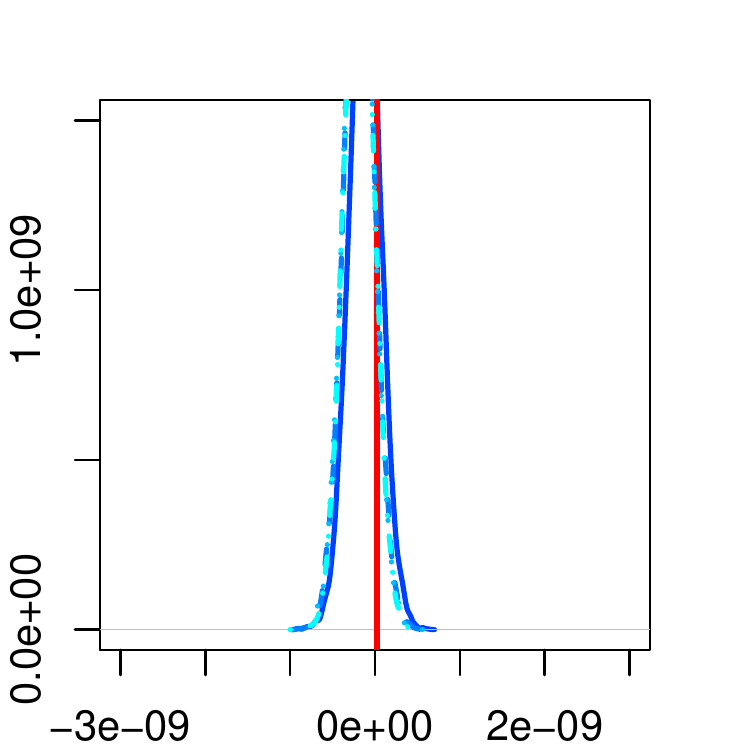}
				\end{minipage} 
			&\begin{minipage}{0.20\textwidth}
				\centering
				\includegraphics[width=1\linewidth]{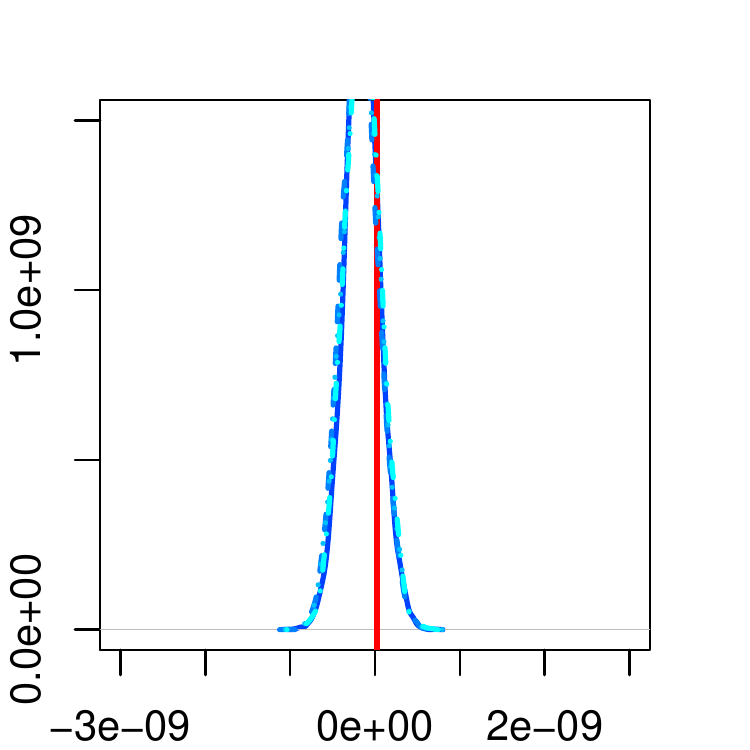}
				\end{minipage} \\
          \begin{sideways} $\xi^2$ \end{sideways} 
			& 
			& 
			&\begin{minipage}{0.20\textwidth}
				\centering
				\includegraphics[width=1\linewidth]{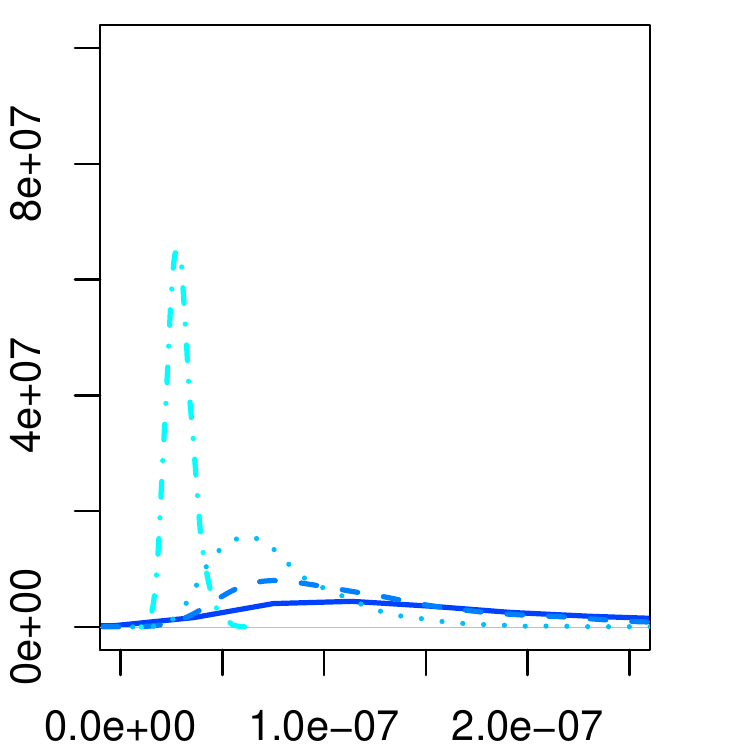}
				\end{minipage} 
	\end{tabular}
	\caption{Posterior density approximations of model parameters
          for the AAPL 03/06/2014 data. The sampling periods used are:
          5 minutes \usebox{\legendLineOne}, 30 seconds
          \usebox{\legendLineTwo}, 15 seconds
          \usebox{\legendLineThree}, 5 seconds
          \usebox{\legendLineFour}, and 1 second
          \usebox{\legendLineFive}}
	\label{fig:posterior-parameters-real}
\end{figure}

We perform an analysis for real market data, which consists of a single day of midpoint stock prices of Apple Inc. (NASDAQ:AAPL) on March $6^{th}$, 2014, printed on the millisecond from the NYSE TAQ data set.  Our a priori estimate of the bid-ask spread driving microstructure noise is centered on \$0.1 as with the simulation data. The volatility paths are shown in Figure \ref{fig:log-vol-real}.  In the case where $\xi^2 = 0$, the model estimates the volatility signal to be, on average, higher than in the cases where $\xi^2 > 0$, which is also seen in the posterior means estimates of $\halpha$ in Figure \ref{fig:posterior-parameters-real}. This is consistent with the simulation-study results, where the $\xi^2 = 0$ model attributes microstructure noise to the log-volatility signal.

Figure \ref{fig:posterior-parameters-real} shows the posterior distributions for the model parameters.   Note that the posterior for $\xi^2$ is centered around $8 \cdot 10^{-9}$ -- two orders of magnitude smaller than the prior center of $2.7 \cdot 10^{-7}$. This value of the posterior mean is roughly equivalent to a bid-ask spread of \$0.01, which is reasonable for a highly-traded stock like AAPL.  Furthermore, note that fixing $\xi^2 = 2.5\cdot 10^{-7}$ compared to treaing $\xi^2$ as an unknown parameter, leads to volatility paths that are smoother and have greater coherence in posterior estimates of the model parameters across sampling periods.  This is especially true for $\htheta$, since the closer the log-volatility process is to being discontinuous, the shorter its timescale must be.  We thus see the obvious trade-off when specifying $\xi^2$: if $\xi^2$ is too large, we run the risk of over-smoothing; if $\xi^2$ is too small, we confound the effects of noise with those of the volatility process.

\subsection{Effect of timescale of inertia on estimates of $\halpha$}\label{se:effect-timescale}

To illustrate the discussion in Section \ref{effect-mean-reverting-rate} on the effect of the log-volatility timescale on our method's ability to learn $\halpha$, we examine the posterior uncertainty for $\halpha$ under two scenarios: 1) increasing sample size $N$ by decreasing the sampling period $\Delta$, and 2) increasing $N$ by increasing the observational period $T$ while keeping $\Delta$ constant. The same simulated dataset is used in both 1) and 2), with $1/\htheta = 15 \mbox{ min}$, such that $\htheta \Delta \leq 1$ and the approximation for the posterior variance of $\halpha$ in \eqref{var_alpha_n2} is applicable.
\begin{figure}[h!]
\centering
\begin{tabular}{cc}
			\begin{minipage}{0.45\textwidth}
				\centering
				\includegraphics[width=1\linewidth]{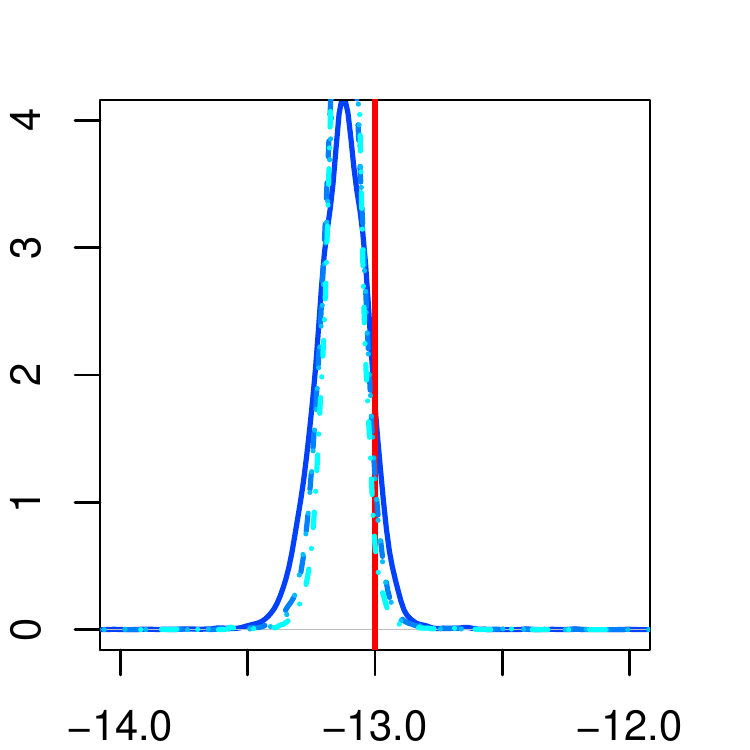}
			\end{minipage} 
			& \begin{minipage}{0.45\textwidth}
				\centering
				\includegraphics[width=1\linewidth]{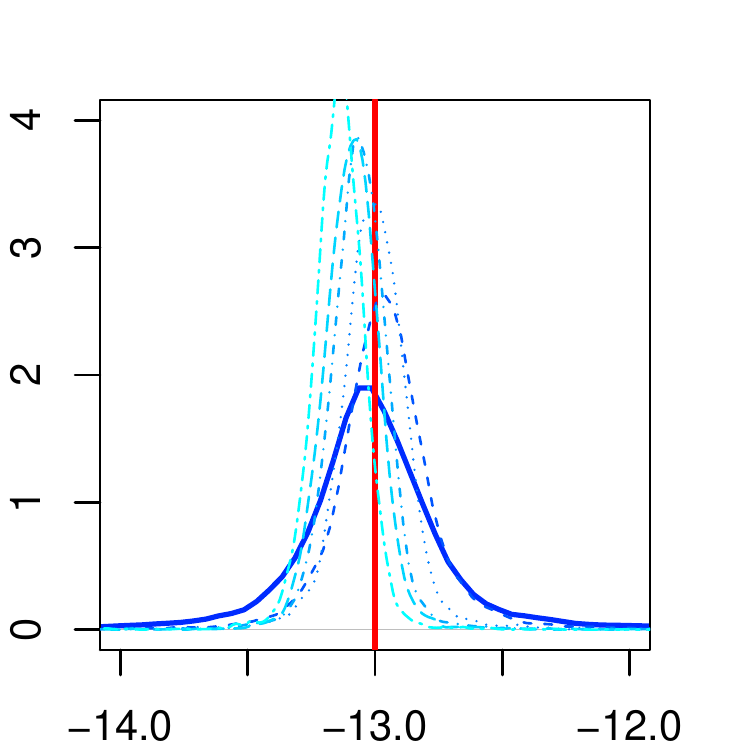}
				\end{minipage} 
\end{tabular}
\caption{Posterior densities of $\halpha$ for two data sets examined under scenarios of increasing sample size $N$ by decreasing $\Delta$ (left) and increasing $N$ by increasing the observational duration $T$ (right). The same data set was used for both cases, where $1/\htheta = 15\, \mbox{min}$ so that $\htheta \Delta \leq 1$ for all posterior samples. The red vertical line signifies the true $\halpha$ value used in the data-generation process. For a fixed $T$ and decreasing $\Delta$ (left) [1 minute \usebox{\legendLineOne}, 30 seconds \usebox{\legendLineTwo}, 15 seconds \usebox{\legendLineThree}, 5 seconds \usebox{\legendLineFour}] the posterior variance of $\halpha$ stays approximately the same. For a fixed $\Delta$ and increasing $T$ [65 minutes \usebox{\legendLineOne}, 130 minutes \usebox{\legendLineTwo}, 195 minutes \usebox{\legendLineThree}, 260 minutes \usebox{\legendLineFour}, 325 minutes \usebox{\legendLineFive}, and 390 \usebox{\legendLineTwo} minutes], the posterior variance of $\halpha$ tends to decrease.}\label{fig:different-phi}
\end{figure}

Under scenario 1), we consider the entire data set and estimate $\halpha$ with $\Delta = 1 \mbox{ min}$, $30 \mbox{ sec}$, $15 \mbox{ sec},$ and $5 \mbox{ sec}$. The posterior distributions for $\halpha$ are shown in the left panel of Figure \ref{fig:different-phi}. We see that the posterior uncertainty for $\halpha$ remains the same with increasing number of intraperiod samples, as suggested by the analysis in Section \ref{effect-mean-reverting-rate}. Under scenario 2), we fix $\Delta = 1 \mbox{ min}$ and increase sample sizes by increasing the observational period $T$, using the first 1/6 (65 min) of the data, the first 2/6 (130 min) of the data, and so on through the entirety of the data (390 min). The right panel of Figure \ref{fig:different-phi} shows the posterior densities for $\halpha$ under this regime. Confirming the discussion in Section \ref{effect-mean-reverting-rate}, we tend to see a decreasing posterior variance for $\halpha$ with increasing observational duration, but not when the sampling frequency increases.  The important takeaway point is that a dataset covering a finite observational period contains a finite amount of information, no matter now finely the observational period is sampled.

%When estimating model parameters, the common intuition is that an increase in sample size leads to a decrease in posterior uncertainty. When dealing with the estimation of stochastic volatility models for high-frequency data, one may be prone to apply this thinking when the sample size is increased by obtaining move frequent price path samples for a fixed observational period. However, as discussed in Section \ref{effect-mean-reverting-rate} and shown in the two scenarios above, in the case where the volatility process has a finite non-zero mean-reversion timescale (as is the case for the OU-process), an increase in the number of intraperiod observations does not add information about the mean-level of the process. Rather, the posterior uncertainty for this model parameter can only be decreased by increasing how long we observe the process. The important takeaway point is that a dataset covering a finite observational period contains a finite amount of information, no matter now finely the observational period is sampled.

\section{Conclusion}

In this paper we have outlined a discrete-time stochastic volatility model for high-frequency data. The model and the algorithm used to estimate it are designed to be coherent across all sampling frequencies. To this end, we elicit priors on the parameters of the continuous-time version of our model and transform them to the discrete-time scale. Both simulation and real data results show that adding the microstructure term in the standard stochastic volatility formulation allows one to fit such models to high-frequency data and extract the log-volatility signal from noisy observations. However, having a good prior estimate of the noise level is an important specification, since attributing some fluctuations in the observed log volatility to microstructure noise has a smoothing effect on the reconstructed log volatility paths. Finally, simulation studies show that the integrated variance estimator derived from our model is well-calibrated and outperforms current kernel-based realized volatility estimators.

\appendix
\section*{Appendix A: Details of the Markov chain Monte Carlo algorithm}\label{ap:mcmc}
We write the full discre-time model 

\begin{align}
	Y_j &= \log(S_j) + \zeta_j  , \nonumber  \\
	\log(S_{j}) &= \mu(\Delta) + \log(S_{j-1}) + \sigma_{j} \epsilon_{j,1}  ,  \label{eq:mod2-appendix}   \\
	\log(\sigma_{j+1}) &= \alpha(\Delta) + \theta(\Delta) \left\{ \log(\sigma_j) - \alpha(\Delta) \right\} + \tau(\Delta) \epsilon_{j,2}, \nonumber
\end{align}
where 
\begin{align*}
\zeta_j &\sim N(0, \xi^2)  ,  &
\left( \begin{matrix} \epsilon_{j,1} \\
                               \epsilon_{j,2} \end{matrix} \right) &\sim 
N \left( \left( \begin{matrix} 0 \\
                                           0 \end{matrix} \right),  
\left( \begin{matrix} 1 & \rho \\
                               \rho & 1 \end{matrix} \right) \right)  , &
\log(\sigma_1) & \sim N \left( \alpha(\Delta) , \frac{\tau(\Delta)^2}{1 - \theta(\Delta)^2} \right)  ,  &   \log(S_0) & \sim N \left( \eta, \kappa^2 \right).
\end{align*}
This model is nonlinear in terms of the volatility due to the formulation of its evolution on the log-scale. To reparameterize the model to be linear in terms of volatility and thereby use the Kalman Filter and Sampler, we take equation \eqref{eq:mod2-appendix} and transform it so that it is linear in terms of $\log(\sigma_j)$, 
\[ 
\log(S_{j}) = \mu(\Delta) + \log(S_{j-1}) + \sigma_{j} \epsilon_{j,1} \quad \leftrightarrow \quad \underbrace{ \log\left[ \left| \log(S_{j}/S_{j-1}) - \mu(\Delta) \right| \right] }_{y_j^*} = \underbrace{  \log(\sigma_j) }_{h_j} + \underbrace{ \log(  \epsilon_{j,1}^2  )/2 }_{\epsilon_{j,1}^{*}}. 
\]
Having defined $y^*$, $h_j$, and $\epsilon_{j,1}^*$, the model becomes linear in the terms involving the volatility:
\begin{eqnarray}
	Y_j &=& \log(S_j) + \zeta_j  ,    \\
	y_j^* &=& h_j + \epsilon^*_{j,1}, \label{eq:yjstar} \\
	h_{j+1} &=& \alpha(\Delta) + \theta(\Delta) \left\{ h_j- \alpha(\Delta) \right\} + \tau(\Delta) \epsilon_{j,2}.  \label{eq:hj},  
\end{eqnarray}
We approximate $\epsilon^*_{j,1}$ as a mixture of normals 
\[ 
	\epsilon^*_{j,1} = \log( \epsilon_{j,1}^2 )/2 \sim \sum_{l=1}^{10} p_l N \left( \frac{m_l}{2}, \frac{v_l^2}{4} \right).
\] 
We can introduce the mixture indicators $\gamma_1, \ldots, \gamma_{n(\Delta)}$ such that
\begin{align*}
\log( \epsilon^2_{j,1} )/2 \mid \gamma_j &\sim N \left( \frac{m_{\gamma_j}}{2}, \frac{v_{\gamma_j}^2}{4} \right)   ,   &   \Pr(\gamma_k = l) = p_l .
\end{align*}
Hence, conditionally on the true prices, the indicators $\gamma_{1}, \ldots, \gamma_{n(\Delta)}$, and the hyperparameters $\mu(\Delta)$, $\alpha(\Delta)$, $\theta(\Delta)$, $\tau(\Delta)$ and $\rho$, we have again a linear state-space model with Gaussian innovations. However, due to correlation of  the innovations $\epsilon_{j,1}$ and $\epsilon_{j,2}$, we need a joint distribution for the transformed and approximated $\epsilon_{j,1}^*$ and $\epsilon_{j,2}$. To this end, we directly follow the approach in \cite{omori2007stochastic}, beginning with known expression 
\begin{eqnarray*}
	p( \epsilon_{j,2}, \epsilon_{j,1}^* | \gamma_j ) &=& p( \epsilon_{j,2} | \epsilon_{j,1}^*, \gamma_j ) p( \epsilon_{j,1}^* | \gamma_j) \\
	&=& p( \epsilon_{j,2} | \underbrace{ d_j \exp( \epsilon_{j,1}^* ) }_{\epsilon_{j,1}}, \gamma_j ) p( \epsilon_{j,1}^* | \gamma_j) \\
	&=& \dNormal{\epsilon_{j,2}}{\rho \tau(\Delta) d_j \exp( \epsilon_{j,1}^* )}{\tau(\Delta)^2 (1-\rho^2) } \dNormal{\epsilon_{j,1}^*}{ \frac{m_{\gamma_j}}{2} }{ \frac{v_{\gamma_j}^2}{4} },
\end{eqnarray*}
where $d_j$ is the sign of $\epsilon_{j,1}$. The nonlinear term $\exp( \epsilon_{j,1}^* )$ is approximated by a linear function, where the constants $(a_{\gamma_j}, b_{\gamma_j})$ are chosen to minimize the expected squared difference between $\exp(\epsilon_{j,1}^*)$ and its approximation, as done in \cite{omori2007stochastic}
\[ 
\exp( \epsilon_{j,1}^* ) | \gamma_j \approx \exp(m_{\gamma_j} /2) (a_{\gamma_j} + b_{\gamma_j}( 2\epsilon_{j,1}^* - m_{\gamma_j} ) ). 
\]
If $z_{j}^*, z_{j} \stackrel{iid}{\sim} N(0,1),$ the joint distribution for the conditional distribution of the pair $(\epsilon_{j,1}^*, \epsilon_{j,2} | \gamma_j)$ can be written as 

\begin{equation}
		\left\{ \left. \left( \begin{array}{c} \epsilon_{j,1}^{*} \\ \epsilon_{j,2} \end{array} \right) \right| \gamma_j \right\} = \left( \begin{array}{c} m_{\gamma_j}/2 \\ d_j \rho\tau(\Delta) \exp(m_{\gamma_j}/2) a_{\gamma_j}  \end{array} \right) + \left( \begin{array}{cc} v_{\gamma_j}/2 & 0 \\ d_j \rho \tau(\Delta) b_{\gamma_j} v_{\gamma_j} \exp(m_{\gamma_j}/2) & \tau(\Delta)\sqrt{1-\rho^2}  \end{array} \right) \left( \begin{array}{c} z_j^* \\ z_j \end{array} \right).
\end{equation}
Rearranging equation \eqref{eq:yjstar} to express $z_j^*$ in terms of $y_j^*, h_j, m_{\gamma_j}$, and $v_{\gamma_j}$, then substituting into equation \eqref{eq:hj} allows us to finally write down the model in the convenient linear state-space form, where the innovations in the state-evolution equations are independent:
\begin{align*}
	Y_j &= \log(S_j) + \zeta_j,    \\
	y_j^* &= h_j + \frac{m_{\gamma_j}}{2} + \frac{v_{\gamma_j}}{2} \,\, z^*_{j}, \\
	h_{j+1} &= \theta_{j}(\Delta) h_j + \alpha_j(\Delta) + \tau(\Delta) \sqrt{1 - \rho^2} z_j,  
\end{align*}
with 
\begin{align*}
	\theta_{j}(\Delta) &= \theta(\Delta) - d_j \rho \tau(\Delta) b_{\gamma_j} v_{\gamma_j} \exp(m_{\gamma_j}/2), \\
	\alpha_{j}(\Delta) &= \alpha(\Delta)( 1 - \theta(\Delta)) + \rho \tau(\Delta) d_j \exp( m_{\gamma_j} / 2) a_{\gamma_j} + d_j \rho \tau(\Delta) b_{\gamma_j} v_{\gamma_j} \exp(m_{\gamma_j} / 2 ) \frac{y_j^* - m_{\gamma_j}/2}{v_{\gamma_j}/2}.
\end{align*}

The full likelihood for the model can be written as 
\begin{align*}
	\MoveEqLeft p(Y_1, \ldots, Y_{n(\Delta)} | h_{1,\ldots,n(\Delta), n(\Delta)+1}, \log( S_{0,\ldots,n(\Delta)} ), \gamma_{1,\ldots,n(\Delta)} \xi, \rho, \tau(\Delta), \theta(\Delta), \alpha(\Delta) ) \propto  \\
	& \,\,\,\,\,\,\,\, \prod_{j=1}^{n(\Delta)} \xi^{-1/2} \expo{ -\frac{1}{2\xi} (Y_j - \log(S_j))^2 } \\
%& [\xi] \\
	&\times \prod_{j=1}^{n(\Delta)} (v_{\gamma_j} / 2)^{-1} \expo{ -\frac{1}{2 v^2_{\gamma_j}/4 } ( y^*_j - h_j - m_{\gamma_j}/2 )^2 } \\
%& [ \gamma_{j}, \log(S_j), \log(\sigma_j), \mu(\Delta)  ]\\
	&\times \prod_{j=1}^{n(\Delta)} \left(\tau(\Delta) \sqrt{1 - \rho^2} \right)^{-1} \expo{ -\frac{1}{2 \tau(\Delta)^2 (1-\rho^2) } \left( h_{j+1} - \theta_j(\Delta) h_j - \alpha_j(\Delta)  \right)^2 } \\
%& [\log(S_j), \log(\sigma_j), \theta(\Delta), \alpha(\Delta), \tau(\Delta), \rho] \\
	&\times p(\log(S_0)) p(\log(\sigma_1)).
\end{align*}
For our MCMC algorithm, we implement a Gibbs sampler where we simulate posterior draws from the full conditional posteriors for each set of parameters in the following steps: 
\begin{enumerate}
	\item \textbf{Sample} $\boldsymbol{\tau^2 (\Delta)}$. With an Inverse-Gamma prior $\tau^2(\Delta) \sim \InvGam{ a_{\tau^2}(\Delta)}{b_{\tau^2}(\Delta) }$, the full conditional posterior for $\tau^2(\Delta)$ is also an Inverse-Gamma with 
	\begin{align*}
		p(\tau^2(\Delta) | - ) &=  \mbox{Inv-Gamma}\left( \tau^2(\Delta) \left| A, B  \right. \right) \\
		A &= a_{\tau^2} + \frac{n(\Delta) + 1}{2} \\
		B &= b_{\tau^2} +\frac{1}{2(1-\rho^2)} \sum_{j=1}^{n(\Delta)}\left( h_{j+1} - \theta_j(\Delta) h_j - \alpha_j(\Delta)  \right)^2 + \frac{1-\theta(\Delta)^2}{2} (h_1 - \alpha(\Delta))^2.
	\end{align*}

	\item \textbf{Sample} $\boldsymbol{\theta (\Delta)}.$ The full conditional posterior for $\theta(\Delta)$ is given by
	\begin{align*}
		\MoveEqLeft p(\theta(\Delta) | - ) \propto \\
 		& p(\theta(\Delta)) \left[  \prod_{j=1}^{n(\Delta)} \frac{1}{\tau(\Delta) \sqrt{1-\rho^2} } \expo{ (h_{j+1} - \theta_j(\Delta) h_j - \alpha_j(\Delta) )^2 } \right] \\
		& \times \sqrt{\frac{1 - \theta(\Delta)^2}{\tau^2(\Delta)}} \expo{ -\frac{1 - \theta(\Delta)^2}{2\tau^2(\Delta) } (h_1 - \alpha(\Delta))^2 }. 
	\end{align*}
	With a normal prior for $\theta(\Delta)$, the product term 
        \[ p(\theta(\Delta)) \left[  \prod_{j=1}^{n(\Delta)} \frac{1}{\tau(\Delta) \sqrt{1-\rho^2} } \expo{ (h_{j+1} - \theta_j(\Delta) h_j - \alpha_j(\Delta) )^2 } \right] \]
        can be reduced to a Normal-kernel form. This can be used as an efficient proposal distribution in a Metropolis-Hasting step, with the rest of the likelihood used to reject or accept the proposal. 

	\item \textbf{Sample} $\boldsymbol{\alpha (\Delta)}.$ The full conditional posterior for $\alpha(\Delta)$ is given by
	\begin{align*}
		\MoveEqLeft p(\alpha(\Delta) | - ) \propto \\
 		& p(\alpha(\Delta)) \left[  \prod_{j=1}^{n(\Delta)} \frac{1}{\tau(\Delta) \sqrt{1-\rho^2} } \expo{ (h_{j+1} - \theta_j(\Delta) h_j - \alpha_j(\Delta) )^2 } \right]\\
		& \times \sqrt{\frac{1 - \theta(\Delta)^2}{\tau^2(\Delta)}} \expo{ -\frac{1 - \theta(\Delta)^2}{2\tau^2(\Delta) } (h_1 - \alpha(\Delta))^2 }. 
	\end{align*}
	Just as above, with a normal prior for $\alpha(\Delta)$, the product term 
\[p(\alpha(\Delta)) \left[  \prod_{j=1}^{n(\Delta)} \frac{1}{\tau(\Delta) \sqrt{1-\rho^2} } \expo{ (h_{j+1} - \theta_j(\Delta) h_j - \alpha_j(\Delta) )^2 } \right] \]
can be reduced to a Normal-kernel form and used an efficient proposal in a Metropolis-Hastings step.

	\item \textbf{Sample} $\boldsymbol{\xi}.$ With an Inverse-Gamma prior on $\xi$ such that $\xi \sim \InvGam{a_\xi}{b_\xi}$, the full conditional posterior for $\xi$ is also Inverse-Gamma
\[ p(\xi | - ) = \mbox{Inverse-Gamma}\left( \xi \left|  a_\xi + n(\Delta)/2 , b_\xi + \frac{1}{2}\sum_{j=1}^{n(\Delta)} ( Y_j - \log(S_j) )^2 \right. \right). \]

	\item \textbf{Sample} $\boldsymbol{\mu(\Delta)}.$ For sampling $\mu(\Delta)$, it is convenient to consider the state-space model in terms of the latent log-prices $\log(S_j)$. The parameter $\mu(\Delta)$ appears only in the evolution of the log-price process. Conditional on all other parameters, we have 
\[ \log(S_j) = \log(S_{j-1}) + \mu(\Delta) + \sigma_j \epsilon_{j,1}, \]
so that the full conditional for $\mu(\Delta)$ is 
\[ p(\mu(\Delta) | - ) \propto p(\mu(\Delta)) N\left( \left( \sum_{j=1}^{n(\Delta)} \frac{\log(S_j/S_{j-1}}{\sigma_j^2} \right) \left( \sum_{j=1}^{n(\Delta)} \frac{1}{\sigma_j^2} \right)^{-1}, \left( \sum_{j=1}^{n(\Delta)} \frac{1}{\sigma_j^2} \right)^{-1} \right). \]

Because each $y_j^*$ is dependent on $\mu(\Delta)$ we must be careful to re-define $y_j^*$ after this sample.
	\item \textbf{Sample} $\boldsymbol{\gamma_{1,\ldots, n(\Delta)}}$. Since each $\gamma_j$ can take on a finite number of values, for each $j$ we sample the discrete posterior where 
\[ p(\gamma_j = l | - ) \propto p(\gamma = l) (v_{l} / 2)^{-1} \expo{ -\frac{1}{2 v^2_{l}/4 } ( y^*_j - h_j - m_{l}/2 )^2 }. \]

	\item \textbf{Sample the latent log-volatilities}. Conditional on all other parameters, the portion of the state-space model where $h_j$ appears is comprised of the linear system
	\begin{align*}
		y_j^* &= h_j + \frac{m_{\gamma_j}}{2} + \frac{v_{\gamma_j}}{2} \,\, z^*_{j}, & z^*_{j} \sim N(0,1) \\
	h_{j+1} &= \theta_{j}(\Delta) h_j + \alpha_j(\Delta) + \tau(\Delta) \sqrt{1 - \rho^2} z_j, & z_j \sim N(0,1) \\	
		p(h_1) &= \dNormal{h_1}{\alpha(\Delta)}{\frac{\tau(\Delta)^2}{1 - \theta(\Delta)^2}}
	\end{align*}
We can efficiently obtain posterior samples for $h_{1, \ldots, n(\Delta), n(\Delta)+1}$ using the Kalman Forward Filter and Backward Sampler. 

	\item \textbf{Sampling the latent log-prices}.  Conditional on all other parameters, the portion of the state-space model where $\log(S_j)$ appears is comprised of the linear system
\begin{align*}
	Y_j &= \log(S_j) + \zeta_j, & \zeta_j \sim N(0,\xi^2) \\
	\log(S_{j}) &= \mu(\Delta) + \log(S_{j-1}) + \sigma_{j} \epsilon_{j,1}, & \epsilon_{j,1} \sim N(0,1)  \\
	p(\log(S_0)) &= \dNormal{\log(S_0)}{\eta}{\kappa^2}.
\end{align*}
As before, we sample $\log(S_j)$ using the Kalman Forward Filter and Backward Sampler. Because each $y_j^*$ is dependent on $\log(S_j)$ we must not forget to re-define $y_j^*$ after this sample.
\end{enumerate}

\bibliography{master-bibliography}

\begin{thebibliography}{31}
\providecommand{\natexlab}[1]{#1}
\providecommand{\url}[1]{\texttt{#1}}
\expandafter\ifx\csname urlstyle\endcsname\relax
  \providecommand{\doi}[1]{doi: #1}\else
  \providecommand{\doi}{doi: \begingroup \urlstyle{rm}\Url}\fi

\bibitem[Ait-Sahalia et~al.(2011)Ait-Sahalia, Mykland, and Zhang]{ait2011ultra}
Yacine Ait-Sahalia, Per~A Mykland, and Lan Zhang.
\newblock Ultra high frequency volatility estimation with dependent
  microstructure noise.
\newblock \emph{Journal of Econometrics}, 160\penalty0 (1):\penalty0 160--175,
  2011.

\bibitem[Alizadeh et~al.(2002)Alizadeh, Brandt, and Diebold]{alizadeh2002range}
Sassan Alizadeh, Michael~W Brandt, and Francis~X Diebold.
\newblock Range-based estimation of stochastic volatility models.
\newblock \emph{The Journal of Finance}, 57\penalty0 (3):\penalty0 1047--1091,
  2002.

\bibitem[Andersen and Bollerslev(1997)]{andersen1997intraday}
Torben~G Andersen and Tim Bollerslev.
\newblock Intraday periodicity and volatility persistence in financial markets.
\newblock \emph{Journal of empirical finance}, 4\penalty0 (2):\penalty0
  115--158, 1997.

\bibitem[Andersen et~al.(1999)Andersen, Bollerslev, and
  Lange]{andersen1999forecasting}
Torben~G Andersen, Tim Bollerslev, and Steve Lange.
\newblock Forecasting financial market volatility: Sample frequency vis-a-vis
  forecast horizon.
\newblock \emph{Journal of Empirical Finance}, 6\penalty0 (5):\penalty0
  457--477, 1999.

\bibitem[Andersen et~al.(2001)Andersen, Bollerslev, Diebold, and
  Labys]{andersen2001distribution}
Torben~G Andersen, Tim Bollerslev, Francis~X Diebold, and Paul Labys.
\newblock The distribution of realized exchange rate volatility.
\newblock \emph{Journal of the American statistical association}, 96\penalty0
  (453):\penalty0 42--55, 2001.

\bibitem[Barndorff-Nielsen and Shephard(2002)]{barndorff2002estimating}
Ole~E Barndorff-Nielsen and Neil Shephard.
\newblock Estimating quadratic variation using realized variance.
\newblock \emph{Journal of Applied Econometrics}, 17\penalty0 (5):\penalty0
  457--477, 2002.

\bibitem[Barndorff-Nielsen et~al.(2008)Barndorff-Nielsen, Hansen, Lunde, and
  Shephard]{barndorff2008designing}
Ole~E Barndorff-Nielsen, Peter~Reinhard Hansen, Asger Lunde, and Neil Shephard.
\newblock Designing realized kernels to measure the ex post variation of equity
  prices in the presence of noise.
\newblock \emph{Econometrica}, 76\penalty0 (6):\penalty0 1481--1536, 2008.

\bibitem[Black(1976)]{black1976pricing}
Fischer Black.
\newblock The pricing of commodity contracts.
\newblock \emph{Journal of financial economics}, 3\penalty0 (1):\penalty0
  167--179, 1976.

\bibitem[Bollerslev(1986)]{bollerslev1986}
Tim Bollerslev.
\newblock Generalized autoregressive conditional heteroskedasticity.
\newblock \emph{Journal of econometrics}, 31\penalty0 (3):\penalty0 307--327,
  1986.

\bibitem[Bollerslev and Zhou(2002)]{bollerslev2002estimating}
Tim Bollerslev and Hao Zhou.
\newblock Estimating stochastic volatility diffusion using conditional moments
  of integrated volatility.
\newblock \emph{Journal of Econometrics}, 109\penalty0 (1):\penalty0 33--65,
  2002.

\bibitem[Brandt and Diebold(2003)]{brandt2003no-arb}
Michael~W Brandt and Francis~X Diebold.
\newblock A no-arbitrage approach to range-based estimation of return
  covariances and correlations.
\newblock Technical report, National Bureau of Economic Research, 2003.

\bibitem[Carter and Kohn(1994)]{carter1994gibbs}
Chris~K Carter and Robert Kohn.
\newblock On gibbs sampling for state space models.
\newblock \emph{Biometrika}, 81\penalty0 (3):\penalty0 541--553, 1994.

\bibitem[Casella and Berger(2002)]{casella2002statistical}
George Casella and Roger~L Berger.
\newblock \emph{Statistical inference}, volume~2.
\newblock Duxbury Pacific Grove, CA, 2002.

\bibitem[Chou et~al.(2010)Chou, Chou, and Liu]{chou2010range}
Ray~Yeutien Chou, Hengchih Chou, and Nathan Liu.
\newblock Range volatility models and their applications in finance.
\newblock In \emph{Handbook of Quantitative Finance and Risk Management}, pages
  1273--1281. Springer, 2010.

\bibitem[Comte and Renault(1998)]{comte1998long}
Fabienne Comte and Eric Renault.
\newblock Long memory in continuous-time stochastic volatility models.
\newblock \emph{Mathematical Finance}, 8\penalty0 (4):\penalty0 291--323, 1998.

\bibitem[Drost and Nijman(1993)]{drost1993aggregation}
Feike~C Drost and Theo~E Nijman.
\newblock Temporal aggregation of garch processes.
\newblock \emph{Econometrica: Journal of the Econometric Society}, pages
  909--927, 1993.

\bibitem[Fr{\"u}hwirth-Schnatter(1994)]{fruhwirth1994data}
Sylvia Fr{\"u}hwirth-Schnatter.
\newblock Data augmentation and dynamic linear models.
\newblock \emph{Journal of time series analysis}, 15\penalty0 (2):\penalty0
  183--202, 1994.

\bibitem[Hansen and Lunde(2006)]{hansen2006realized}
Peter~R Hansen and Asger Lunde.
\newblock Realized variance and market microstructure noise.
\newblock \emph{Journal of Business \& Economic Statistics}, 24\penalty0
  (2):\penalty0 127--161, 2006.

\bibitem[Hansen et~al.(2012)Hansen, Huang, and Shek]{hansen2012realized}
Peter~Reinhard Hansen, Zhuo Huang, and Howard~Howan Shek.
\newblock Realized {GARCH}: a joint model for returns and realized measures of
  volatility.
\newblock \emph{Journal of Applied Econometrics}, 27\penalty0 (6):\penalty0
  877--906, 2012.

\bibitem[Hull and White(1987)]{hull1987pricing}
John Hull and Alan White.
\newblock The pricing of options on assets with stochastic volatilities.
\newblock \emph{The journal of finance}, 42\penalty0 (2):\penalty0 281--300,
  1987.

\bibitem[Hwang et~al.(2013)Hwang, Shin, et~al.]{hwang2013stationary-bootstrap}
Eunju Hwang, Dong~Wan Shin, et~al.
\newblock Stationary bootstrapping realized volatility under market
  microstructure noise.
\newblock \emph{Electronic Journal of Statistics}, 7:\penalty0 2032--2053,
  2013.

\bibitem[Maneesoonthorn et~al.(2014)Maneesoonthorn, Forbes, and
  Martin]{maneesoonthorn2014inference}
Worapree Maneesoonthorn, Catherine~S Forbes, and Gael~M Martin.
\newblock Inference on self-exciting jumps in prices and volatility using high
  frequency measures.
\newblock \emph{arXiv preprint arXiv:1401.3911}, 2014.

\bibitem[Omori et~al.(2007)Omori, Chib, Shephard, and
  Nakajima]{omori2007stochastic}
Yasuhiro Omori, Siddhartha Chib, Neil Shephard, and Jouchi Nakajima.
\newblock Stochastic volatility with leverage: Fast and efficient likelihood
  inference.
\newblock \emph{Journal of Econometrics}, 140\penalty0 (2):\penalty0 425--449,
  2007.

\bibitem[Pigorsch et~al.(2012)Pigorsch, Pigorsch, and
  Popov]{pigorsch2012volatility}
Christian Pigorsch, Uta Pigorsch, and Ivaylo Popov.
\newblock Volatility estimation based on high-frequency data.
\newblock In \emph{Handbook of Computational Finance}, pages 335--369.
  Springer, 2012.

\bibitem[Shirota et~al.(2014)Shirota, Hizu, and Omori]{shirota2014realized}
Shinichiro Shirota, Takayuki Hizu, and Yasuhiro Omori.
\newblock Realized stochastic volatility with leverage and long memory.
\newblock \emph{Computational Statistics \& Data Analysis}, 76:\penalty0
  618--641, 2014.

\bibitem[Stoll(2000)]{stoll2000presidential}
Hans~R Stoll.
\newblock Presidential address: friction.
\newblock \emph{The Journal of Finance}, 55\penalty0 (4):\penalty0 1479--1514,
  2000.

\bibitem[Takahashi et~al.(2009)Takahashi, Omori, and
  Watanabe]{takahashi2009estimating}
Makoto Takahashi, Yasuhiro Omori, and Toshiaki Watanabe.
\newblock Estimating stochastic volatility models using daily returns and
  realized volatility simultaneously.
\newblock \emph{Computational Statistics \& Data Analysis}, 53\penalty0
  (6):\penalty0 2404--2426, 2009.

\bibitem[Venter and de~Jongh(2012)]{venter2012extended}
JH~Venter and PJ~de~Jongh.
\newblock Extended stochastic volatility models incorporating realised
  measures.
\newblock \emph{Computational Statistics \& Data Analysis}, 2012.

\bibitem[Zhang et~al.(2005)Zhang, Mykland, and A{\"\i}t-Sahalia]{zhang2005tale}
Lan Zhang, Per~A Mykland, and Yacine A{\"\i}t-Sahalia.
\newblock A tale of two time scales.
\newblock \emph{Journal of the American Statistical Association}, 100\penalty0
  (472), 2005.

\bibitem[Zhou(1996)]{zhou1996high}
Bin Zhou.
\newblock High-frequency data and volatility in foreign-exchange rates.
\newblock \emph{Journal of Business \& Economic Statistics}, 14\penalty0
  (1):\penalty0 45--52, 1996.

\bibitem[Zumbach(2000)]{zumbach2000pitfalls}
Gilles Zumbach.
\newblock The pitfalls in fitting garch (1, 1) processes.
\newblock In \emph{Advances in Quantitative Asset Management}, pages 179--200.
  Springer, 2000.

\end{thebibliography}
\end{document}